%% file: ms.tex
\pdfoutput=1
\documentclass[twocolumn,times,tighten,numberedappendix,appendixfloats]{aastex6}

\usepackage{epstopdf}	% for pdflatex
\usepackage{graphicx}		% for including figures
\usepackage{latexsym}		% additional math symbols
\usepackage{bm}			% bold math package
\usepackage[english]{babel}
\usepackage{color}		% for colored text and boxes; remove for subm.
\usepackage{booktabs}	% for hlines with space between
\usepackage{multirow}	% for multirow

\definecolor{red}{rgb}{1.0,0.0,0.0}%

\newcommand{\eg}	{e.g.,}%
\newcommand{\ie}	{i.e.,}%
\newcommand{\etal}	{et~al.}%
\newcommand{\HST}	{\emph{HST}}%
\newcommand{\JWST}	{\emph{JWST}}%
\newcommand{\Spitzer}	{\emph{Spitzer}}%
\newcommand{\WISE}	{\emph{WISE}}%
\newcommand{\AV}	{\ensuremath{A_{V}}}%
\newcommand{\AVinput}	{\ensuremath{A_{V,\mathrm{inp}}}}%
\newcommand{\Alam}	{\ensuremath{A_{\lambda}}}%
\newcommand{\betaV}	{\ensuremath{\beta_{V}}}%
\newcommand{\betaVzero}	{\ensuremath{\beta_{V\!,0}}}%
\newcommand{\betalam}	{\ensuremath{\beta_{\lambda}}}%
\newcommand{\betalamzero}{\ensuremath{\beta_{\lambda,0}}}%
\newcommand{\rbeta}[1]{\ensuremath{r_{\beta}(#1,L)}}%
\newcommand{\code}[1]	{\textsf{\small #1}}%		# for in main text
\newcommand{\SB}	{\code{Starburst99}}%
\newcommand{\BC}	{\code{BC03}}%
\newcommand{\scode}[1]	{\textsf{\scriptsize #1}}%	# for in captions
\newcommand{\sSB}	{\scode{Starburst99}}%
\newcommand{\sBC}	{\scode{BC03}}%
\newcommand{\norsfe}	{3000}%	# Number of randomly generated Stochastic Total
\newcommand{\norsfee}	{1000}%	# Number of randomly generated Stochastic SFH3
\newcommand{\norsfes}	{1000}%	# Number of randomly generated Stochastic SFH4
\newcommand{\norsfel}	{1000}%	# Number of randomly generated Stochastic SFH5
\newcommand{\Msun}	{\ensuremath{\textrm{M}_{\odot}}}% # Mass, not Magnitude!

\newlength{\txw}
\newlength{\txh}

\newcommand{\citebox}[1]{\hspace*{1.5em}\parbox[t]{0.45\txw}{\linespread{1.2}\selectfont\hspace*{-1.8em}#1\\[-6pt]}}%

\begin{document}
%%%%%%%%%%%%%%%%%%%%%%%%%%%%%%%%%%%%%%%%%%%%%%%%%%%%%%%%%%%%%%%%%%%%%%%%%%%%

\title{Analysis of the Intrinsic Mid-Infrared \bm$L$ band to Visible--
	Near-Infrared Flux Ratios in Spectral Synthesis Models of Composite 
	Stellar Populations}

\author{Duho Kim\altaffilmark{1}, Rolf A. Jansen\altaffilmark{2}, \&
	Rogier A. Windhorst\altaffilmark{2}}

\affil{School of Earth \& Space Exploration, Arizona State
        University, Tempe, AZ 85287-1404, USA}
\altaffiltext{1}{Duho.Kim@asu.edu}
\altaffiltext{2}{Department of Physics, Arizona State University, Tempe,
	AZ 85287-1504, USA}

\shortauthors{Kim, D., \etal}
\shorttitle{\bm$L$ band to Visible--Near-IR Flux Ratios of Composite Stellar
	Populations}
\journalinfo{{\rm Published on} The Astrophysical Journal}
%*%\submitted{\today}\received{\today}\accepted{\today}\published{\today}

\begin{abstract} We analyze the intrinsic flux ratios of various
visible--near-infrared filters with respect to 3.5\,\micron\ for simple and
composite stellar populations (CSPs), and their dependence on age, metallicity,
and star formation history (SFH). UV/optical light from stars is reddened and
attenuated by dust, where different sightlines across a galaxy suffer varying
amounts of extinction. Tamura \etal\ (2009) developed an approximate method to
correct for dust extinction on a pixel-by-pixel basis, dubbed the ``\betaV''
method, by comparing the observed flux ratio to an empirical estimate of the
intrinsic ratio of visible and $\sim$3.5\,\micron\ data. Through extensive
modeling, we aim to validate the ``\betaV'' method for various filters spanning
the visible through near-infrared wavelength range, for a wide variety of
simple and CSPs. Combining \SB\ and \BC\ models, we built spectral energy
distributions (SEDs) of simple (SSP) and composite (CSP) stellar populations
for various realistic SFHs, while taking metallicity evolution into account. We
convolve various 0.44--1.65\,\micron\ filter throughput curves with each model
SED to obtain intrinsic flux ratios \betalamzero. When unconstrained in
redshift, the total allowed range of \betaVzero\ is 0.6--4.7, or almost a
factor of eight. At known redshifts, and in particular at low redshifts
($z$\,$\lesssim$\,0.01), \betaVzero\ is predicted to span a narrow range of
0.6--1.9, especially for early-type galaxies (0.6--0.7), and is consistent with
observed \betaV\ values. The \betalam\ method can therefore serve as a
first-order dust-correction method for large galaxy surveys that combine \JWST\
(rest-frame 3.5\micron) and \HST\ (rest-frame visible--near-IR) data.
\end{abstract}


\section{Introduction}

Spectral energy distribution (SED) information across a wide frequency range
can be used to infer the nature of the emitting source, and simultaneously the
effect of the intervening interstellar medium (ISM) along the line of sight
(\eg\ Lindblad 1941; Elmegreen 1980; Walterbos \& Kennicutt 1988; Waller \etal\
1992; Witt \etal\ 1992; Calzetti \etal\ 1994; Boselli \etal\ 2003; Deo \etal\
2006). Dust in the ISM of galaxies effectively scatters or absorbs light at
shorter wavelengths, such as ultraviolet (UV) and visible photons, and
reradiates the absorbed energy in the far-infrared (far-IR) (\eg\ Buat \& Xu
1996; Charlot \& Fall 2000; Dale \etal\ 2001; Bell \etal\ 2002; Panuzzo \etal\
2003; Boissier \etal\ 2004; Buat \etal\ 2005). Accordingly, dust transforms the
shape of the galaxy SEDs, making it harder to study the intrinsic properties of
astronomical sources (\eg\ Trumpler 1930; Mathis \etal\ 1977; Viallefond \etal\
1982; Caplan \& Deharveng 1985,1986; Roussel \etal\ 2005; Driver \etal\ 2008).

In order to directly account for the transformation of the SED by dust, it
would be especially helpful to have UV data accompanied by far-IR data
(25--350\,\micron) at the same angular resolution (Calzetti \etal\ 2000; Bell
\etal\ 2002; Boselli \etal\ 2003; Boissier \etal\ 2004; Buat \etal\ 2005). The
Earth's atmosphere, however, is mostly opaque to the far-IR (Elsasser 1938),
necessitating the use of space telescopes like \emph{IRAS} (Neugebauer \etal\
1984), \emph{ISO} (Kessler \etal\ 1996), \Spitzer\ (Werner \etal\ 2004), and
\emph{Herschel} (Pilbratt 2004). Even then, it is challenging to build and
operate far-IR detectors, and for a given telescope aperture, the spatial
resolution is $\sim$40--800 times worse than in the optical (Xu \& Helou 1996;
Price \etal\ 2002). For these reasons, shorter wavelength data have been used
in various ways to correct for extinction by dust in many studies (\eg\
Lindblad 1941; Rudy 1984; Calzetti \etal\ 1994; Petersen \& Gammelgaard 1997;
Meurer \etal\ 1999; Scoville \etal\ 2001; Bell \etal\ 2002; Kong \etal\ 2004;
Ma\'{\i}z-Apell\'{a}niz \etal\ 2004; Calzetti \etal\ 2005; Rela\~{n}o \etal\
2006; Kennicutt \etal\ 2009), each with their own advantages and disadvantages
(see, \eg\ Tamura \etal\ 2009).\\[-7pt]

Tamura \etal\ (2009; hereafter T09) developed a simple, approximate,
dust-correction method, dubbed the ``\betaV'' method, which uses photometry in
two broadband filters at optical ($\lambda$\,$>$\,4000\AA) and mid-infrared
(mid-IR) $L$-band ($\sim$3.5\,\micron) rest-frame wavelengths, respectively.
The \betaV\ method is primarily intended for the study of large numbers of
spatially resolved galaxies at low to intermediate redshifts
(0.1\,$\lesssim$\,$z$\,$\lesssim$\,2), for which multiwavelength observations
are too expensive and approximate dust corrections may still present a marked
improvement over ignoring the effects of extinction altogether, or over
adopting a single number as a canonical extinction value for a given galaxy. In
particular, T09 and Tamura \etal\ (2010) applied the \betaV\
method to one nearby late-type spiral galaxy, NGC\,959, by using $V$-band
images obtained from the ground with the Vatican Advanced Technology Telescope
and 3.6\,\micron\ images from space with the \Spitzer/InfraRed Array Camera
(IRAC), and using ancillary far- and near-UV images from the \emph{Galaxy
Evolution Explorer} (\emph{GALEX}) in order to better distinguish pixels
dominated by younger stellar populations from those dominated by older ones.
From an analysis of the histogram of the observed pixel flux ratios, they
adopted values of 1.10 for ``older'' pixels and 1.32 for ``younger'' pixels for
the extinction-free $V$ to 3.6\,\micron\ flux ratio, \betaVzero. By doing so,
they were able to map the extinction across NGC\,959, and obtain
extinction-corrected images in which they discerned a hitherto unrecognized
stellar bar. The \betaVzero\ values that they used for ``older'' pixels were
consistent with the 0.5\,$\lesssim$\,\betaVzero\,$\lesssim$\,2 derived from the
theoretical SED models for simple stellar populations (SSPs) of Anders \&
Fritze-von~Alvensleben (2003). The \betaVzero\ values for ``younger'' pixels,
on the other hand, were significantly smaller than those expected from theory
(4\,$\lesssim$\,\betaVzero\,$\lesssim$\,7). The authors argued that blending of
light from underlying and neighboring older stellar populations was the likely
origin of this discrepancy. This effect will be resolution dependent and hence
redshift dependent, because poor spatial resolution will mix the light from a
larger area within a galaxy. To address the impact of these issues,
quantitative analysis of both the effect of blending of the light from
different stellar populations and of the effects of spatial resolution on
extinction estimates will be necessary.\\[-7pt]

As part of a larger project to investigate the validity, robustness, and limits
of the \betaV\ method (Jansen \etal\ 2017, in preparation), we construct SED
models for stellar populations with various metallicities, ages, and star
formation histories (SFHs), in order to quantify how the intrinsic optical to
mid-IR flux ratio, \betalamzero, will vary. The result will be relevant,
particularly, for future surveys of intermediate-redshift ($z$\,$\lesssim$\,2)
galaxies that combine images from the \emph{James Webb Space Telescope} (\JWST;
Gardner \etal\ 2006) and the \emph{Hubble Space Telescope} (\HST) at rest-frame
$\sim$3.5\micron\ and visible--near-IR wavelengths, respectively. As the
spatial resolution ($\propto$\,$\lambda/D$) of images from \JWST/NIRCam at
3.5\micron\ will be comparable to that of \HST/ACS\,WFC and WFC3\,UVIS in $V$
to within a factor 2.5, an unprecedented detailed study of galaxies over the
past 4.4\,Gyr, virtually unhampered by extinction, will become possible when
the \betaV\ method proves valid.\\[-7pt]

This paper is organized as follows: in \S2 we explain how we combine the
results of two different stellar population synthesis codes published in the
literature, and how we use our adopted SSP SEDs to generate SEDs of composite
stellar populations (CSPs). In \S3, for different SFHs, we present
\betalamzero\ values obtained by convolving the SEDs with various sets of
filters. In \S4 we analyze and evaluate the \betalam\ method as a
dust-correction method. We briefly summarize in \S5.\\[-7pt]

We adopt the Planck\,2015 (Planck Collaboration \etal\ 2016) cosmology
with ($H_{0}$ = 67.8 km\,sec$^{-1}$\,Mpc$^{-1}$; $\Omega_{m}$ = 0.308;
$\Omega_{\Lambda}$ = 0.692), and we will use AB magnitudes (Oke 1974; Oke \&
Gunn 1983) throughout.

\section{SED Models}

\subsection{Combining the SSP Model SEDs of \SB\ and \BC}

The first step of our computational analysis is to build a large family of SSP
model SEDs. Each SSP represents a single generation of stars of the same age
and chemical composition, with stellar masses that were distributed according
to an adopted initial mass function (IMF; Salpeter 1955; Kroupa 2002; Chabrier
2003) at birth. There are several stellar population synthesis codes available
in the literature (\eg\ Fioc \& Rocca-Volmerange 1997; Leitherer \etal\ 1999;
Bruzual \& Charlot 2003; Maraston 2005; Vazdekis \etal\ 2010). We select \SB\
(Leitherer \etal\ 1999 and V\'{a}zquez \& Leitherer 2005) for very young
($\lesssim$30\,Myr) stellar populations, since the stellar evolutionary tracks
of the Geneva group (Schaller \etal\ 1992; Charbonnel \etal\ 1993; Schaerer
\etal\ 1993a, 1993b; Meynet \etal\ 1994) adopted by \SB\ are optimized for
massive young stars, and include, \eg\ the Wolf-Rayet phase. \SB\ also includes
nebular continuum from the gas enshrouding the stellar populations, as is the
general case for newborn and young stars. On the other hand, the stellar
evolutionary tracks of the Padova group (Girardi \etal\ 2000; Marigo \etal\
2008) are a good match to observations of intermediate- and low-mass stars that
dominate older stellar populations (V\'{a}zquez \& Leitherer 2005). We
therefore adopt the \textsf{\small Padova1994} tracks in \BC\ (Bruzual \&
Charlot 2003) for older ($\gtrsim$100\,Myr) stellar populations. Below, we
describe how we combine the two sets of SED models into a database that can be
used for stellar populations of any age. The age grid of the \BC\ code is fixed
as 221 logarithmic steps between 10$^5$ and 10$^{10}$\,years. In the \SB\ code,
however, the user can specify the age step and scale, so we set 1000
logarithmic time steps between 10$^6$ and 10$^{10}$\,years in order to provide
a finer grid than \BC. For both codes, we adopted a Salpeter (1955) stellar
IMF, with a power-law slope of $-$2.35, and with minimum and maximum stellar
masses of 0.1 and 100\,\Msun. Each code has a different set of metallicities:
$Z$\,=\,0.001, 0.004, 0.008, 0.02, and 0.04 for \SB\ and $Z$\,=\,0.0001,
0.0004, 0.004, 0.008, 0.02, and 0.05 for \BC. We extrapolate the \SB\ SEDs to
match the metallicities of \BC\ using the method described in the following.
For each code, we log-exponentially extrapolated in
$[Z]$\,=\,$\log_{10}(Z/Z_{\odot})$ (where $Z_{\odot}$\,$=$\,0.02) from the two
SEDs that are closest in metallicity to the desired value. For example, to
obtain our $Z$\,=\,0.0004 SED in \SB, we combined SEDs for $Z$\,=\,0.001 and
$Z$\,=\,0.004 in \SB\ with relative weights of 1.40 and $-$0.40, which are
derived from $F_1$\,=\,$(F_2 - w\cdot F_3)/(1 - w)$, where $F_1$, $F_2$, and
$F_3$ denote the logarithm of the wavelength-dependent flux densities of SEDs
with metallicities of $Z_1$, $Z_2$, and $Z_3$ (with $Z_1$$<$$Z_2$$<$$Z_3$), and
where the weight $w$\,=\,$(e^{Z_2}-e^{Z_1})/(e^{Z_3}-e^{Z_1})$. Similarly, when
interpolating $F_2$ into $F_1$ and $F_3$, we use $F_2 = (1-w)\cdot F_1 + w\cdot
F_3$. Note that the SEDs $F_1$, $F_2$, and $F_3$ and metallicities $Z_1$,
$Z_2$, and $Z_3$ are in log scale, with the latter all known (\eg\ 0.0004,
0.001, and 0.004 giving $w$\,$=$\,0.285 in our example).
  %
%%%%%%%%%%%%%%%%%%%%%%%%%%%%%%%%% FIGURE 1 %%%%%%%%%%%%%%%%%%%%%%%%%%%%%%%%%
\noindent\begin{figure}[t]
\centerline{
  \includegraphics[width=0.48\txw]{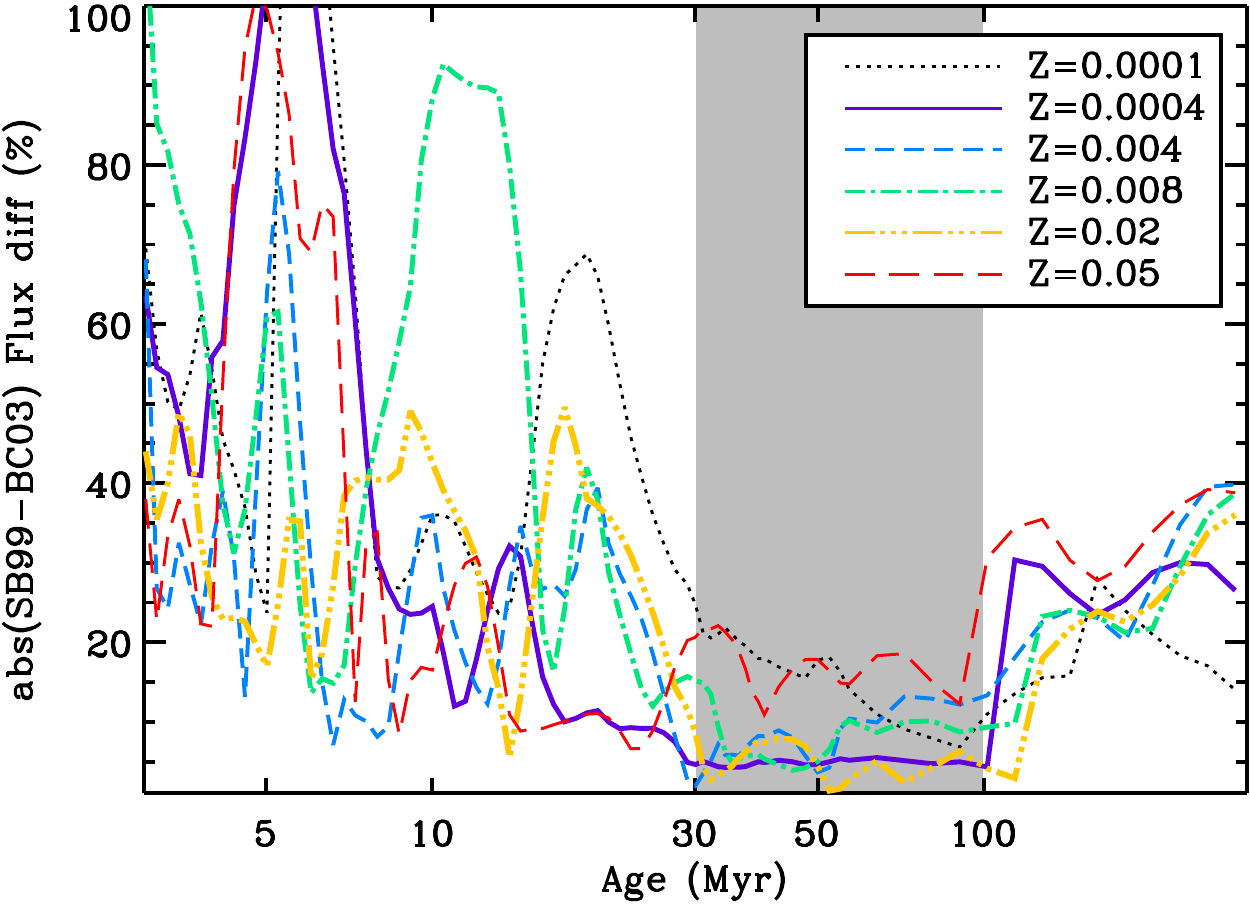}
}
\caption{\noindent\small
Integrated absolute flux difference over the 0.4--3.75\,\micron\ wavelength
range between \sSB\ and \sBC\ SEDs (see Eq.~\ref{eq:absfd}) as a
function of SSP age for various metallicities. There is no specific age where
both codes converge on an identical SED, but the percentage difference reaches
a general minimum between $\sim$30 and $\sim$100\,Myr, where they are mostly
$\lesssim$10$\%$.
\label{fig:figure1}}
\end{figure}
%%%%%%%%%%%%%%%%%%%%%%%%%%%%%%%%%%%%%%%%%%%%%%%%%%%%%%%%%%%%%%%%%%%%%%%%%%%%%
  %
%%%%%%%%%%%%%%%%%%%%%%%%%%%%%%%%%  TABLE 1  %%%%%%%%%%%%%%%%%%%%%%%%%%%%%%%%%
%\placetable{1}
\noindent\begin{table}[b]
\caption{\small Transition age range adopted when combining \sSB\
and \sBC\ stellar population synthesis model SEDs for various metallicities,
$Z$.  Minimum, maximum, and mean relative flux differences of the two SEDs in
the transition age range, and rms thereof are also tabulated.\label{tab:table1}}
\setlength{\tabcolsep}{4.2pt}
\begin{tabular}{cccccc}
\hline\\[-5pt]
\multirow{2}{*}{$Z$}&Age~range&Min.~Diff.&Max.~Diff.&Mean~Diff.&RMS\\
                    &  (Myr)  &   (\%)   &   (\%)   &   (\%)  & (\%)\\[2pt]
\hline\\[-5pt]
0.0001   &  30--100 &  ~~7 &  25  &  17  &  5.0\\
0.0004   &  30--100 &  ~~4 &  ~~5 &  ~~5 &  0.3\\
0.004~~  &  30--100 &  ~~2 &  13  &  ~~7 &  3.0\\
0.008~~  &  30--100 &  ~~4 &  15  &  ~~8 &  3.3\\
0.02~~~~ &  30--100 &  ~~1 &  ~~9 &  ~~5 &  2.1\\
0.05~~~~ &  30--100 &  11  &  22  &  17  &  3.2\\[2pt]
\hline\\[-3pt]
\end{tabular}
\end{table}
%%%%%%%%%%%%%%%%%%%%%%%%%%%%%%%%%%%%%%%%%%%%%%%%%%%%%%%%%%%%%%%%%%%%%%%%%%%%%
  %

For intermediate ages (30\,$\lesssim$\,Age\,$\lesssim$\,100\,Myr), we
interpolate the results of the two codes onto the \BC\ Age grid as follows. To
minimize discontinuities and sudden changes of SEDs for intermediate ages, we
inspected the integrated absolute flux differences between \SB\ and \BC\ SEDs
over the 0.4--3.75\,\micron\ wavelength range (the relevant range for the
\betaV\ method; see T09). This quantity is for a given age given by:
\begin{equation}
\frac{\int_{\lambda_1}^{\lambda_2}{\bigl| F_{SB99}(\lambda)-F_{BC03}(\lambda)\bigr|}
} { \onehalf \left(\int_{\lambda_1}^{\lambda_2}{F_{SB99}(\lambda)} + \int_{\lambda_1}^{\lambda_2}{F_{BC03}(\lambda)}\right) } ,
\label{eq:absfd}
\end{equation}
with ($\lambda_1$,\,$\lambda_2$)\,=\,(0.4,\,3.75)\,\micron, and its dependence
on age and metallicity is shown in Figure~\ref{fig:figure1}.
Table~\ref{tab:table1} lists the minimum, maximum, mean, and rms differences.
At 30\,Myr, 100\% of the \SB\ SED and 0\% of the \BC\ SED is used, and the
contributions are linearly reduced (increased) to 0\% (100\%) at 100\,Myr.
Interpolating using $\log($Age$)$ instead of Age yields nearly identical
results, since the age ranges are relatively narrow.

\begin{figure*}[t]
\centerline{
  \includegraphics[width=0.485\txw]{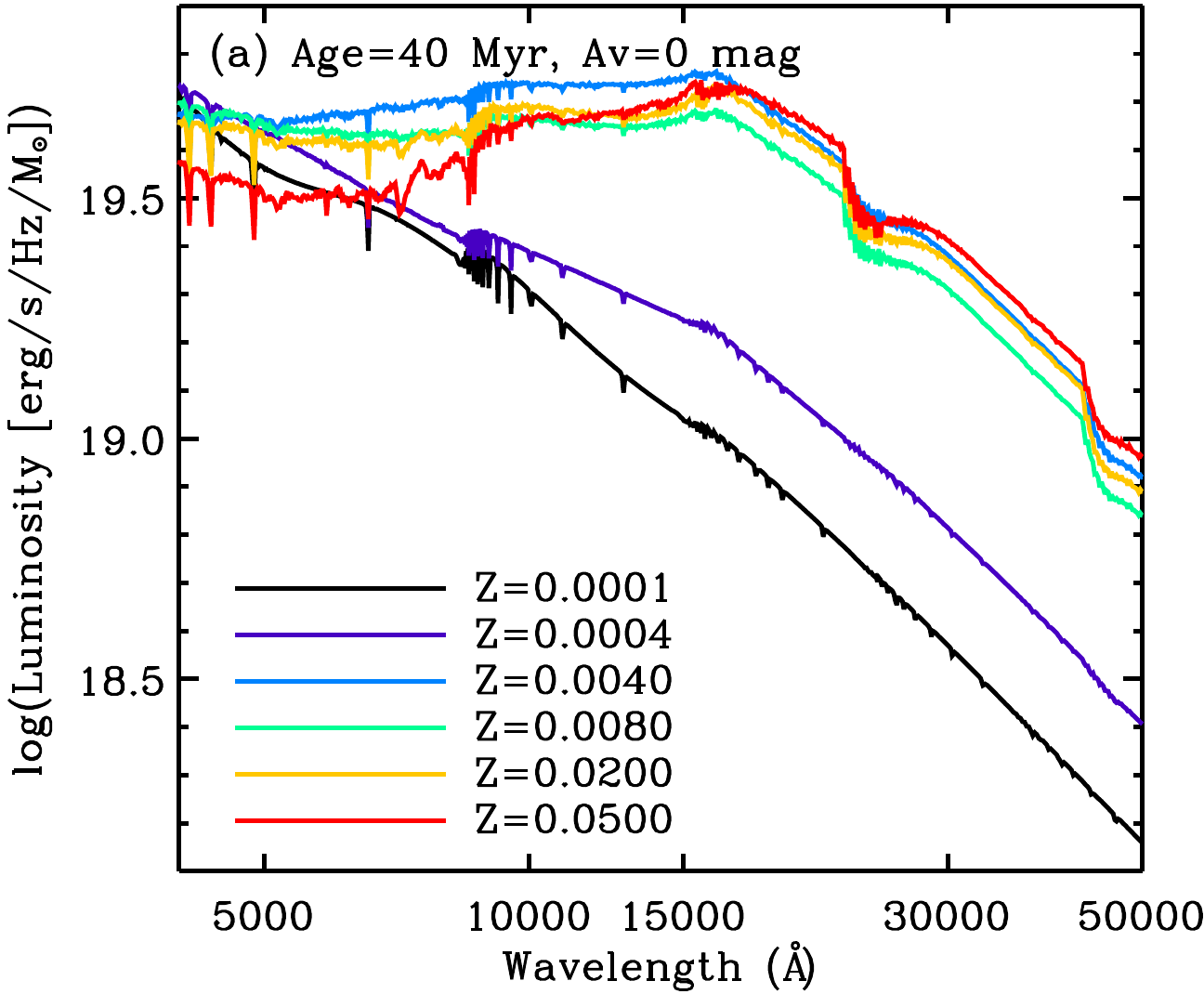}\ \ \ \ \ \ 
  \includegraphics[width=0.485\txw]{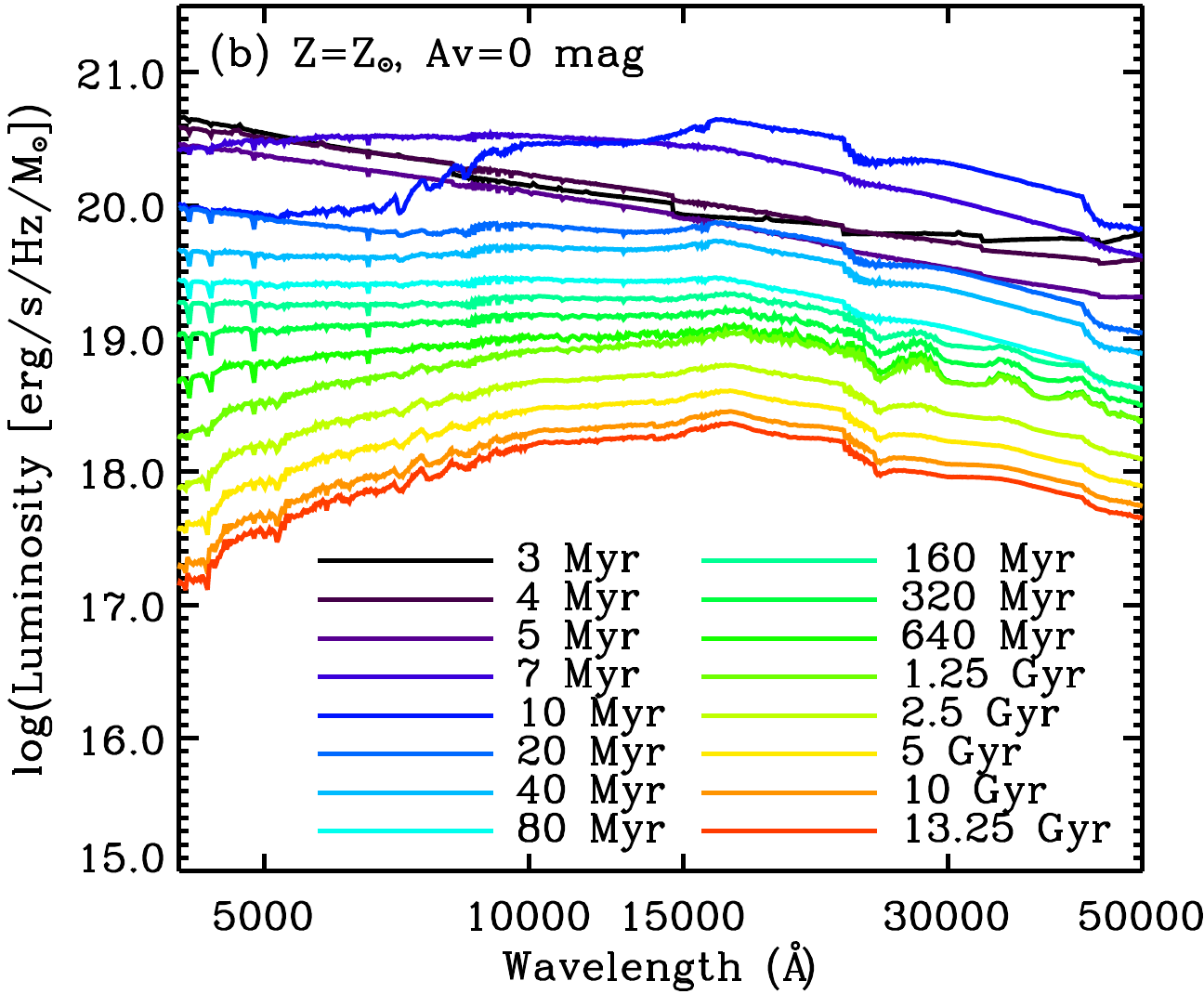}
}
\caption{\noindent\small
Examples of our adopted combined array of SSP SEDs. (a) SEDs for a 40\,Myr old
stellar population, for six different metallicities and no extinction; (b) SEDs
for 16 different ages with fixed solar metallicity ($Z_{\odot}$\,=\,0.02) and
zero extinction.\label{fig:figure2}}
\end{figure*}

Figure~\ref{fig:figure2} shows examples of our combined SSP SEDs for six
metallicities and 16 ages. We use this set of ages and metallicities when we
calculate the $L$ band to visible--near-IR flux ratios of stellar populations
with various SFHs described in \S3. We also use the same parameter set when we
select a manageably small but comprehensive set of SSP SEDs, which we extend
with additional extinction-related parameters in Appendix A.
  %
%%%%%%%%%%%%%%%%%%%%%%%%%%%%%%%%%  FIGURE 2  %%%%%%%%%%%%%%%%%%%%%%%%%%%%%%%%%
\noindent
%%%%%%%%%%%%%%%%%%%%%%%%%%%%%%%%%%%%%%%%%%%%%%%%%%%%%%%%%%%%%%%%%%%%%%%%%%%%%%

\subsection{SFH, Metallicity Evolution, and Construction of CSP SEDs}

Whereas an SSP consists of purely coeval stars, realistic star formation within
galaxies is characterized by stochastic and/or temporally extended, possibly
spatially propagating, episodes of star formation. The CSP of a larger region
within a galaxy therefore effectively records its SFH. If we denote that SFH as
the time-dependent star formation rate, $\psi(t)$, and the SED of an SSP as a
function of its age, $t$, and metallicity, $Z$, by $F_{\lambda,{\rm
SSP}}(t,Z)$, then we can express the SED of a CSP as the superposition of
multiple SSPs:\\[-4pt]
\begin{equation} 
F_{\lambda,{\rm CSP}}(t,Z) = 
 \sum_{t'=t_0}^{t}{F_{\lambda,{\rm SSP}}(t-t'\!,Z(t'))\,\psi(t')\,\delta t'}\,,\ 
\label{eq:fcsp}
\end{equation}
where $F_{\lambda,{\rm SSP}}$ is derived by interpolating our combined SSP SEDs
to generate a set of SEDs with 28 logarithmic steps in metallicity (between
$Z$\,=\,0.0001 and $Z$\,=\,$0.05$) and 1000 logarithmic steps in Age (between
$t$\,=\,1\,Myr and $t$\,=\,13.8\,Gyr), $Z(t')$ is used to explicitly denote the
time-dependence of the metallicity, and $t_0$ denotes the time of the first
episode of star formation. The metallicity at time step $t'$ results from
metal-enriched gas returned to the ISM by stars from previous generations of
stellar populations, and from accretion of gas from other nearby galaxies or
from the IGM and circum-galactic medium (whether more enriched, or diluted) by
the end of time step $(t-\delta t)$. Note that this implies that the
metallicity of the resulting CSP differs from that of the metallicity of (most
of) its constituent SSPs, and that the chemical evolution of the CSP will not
be entirely self-consistent with the choice of the SFH. It does account,
however, for the open-box, rather than closed-box, nature of a galaxy or galaxy
region in an empirical way. We return to the functional form of $Z(t')$ shortly.

%%%%%%%%%%%%%%%%%%%%%%%%%%%%%%%  FIGURE 3  %%%%%%%%%%%%%%%%%%%%%%%%%%%%%%%%%%%
\noindent\begin{figure}[t]
\centerline{
  \includegraphics[width=0.48\txw]{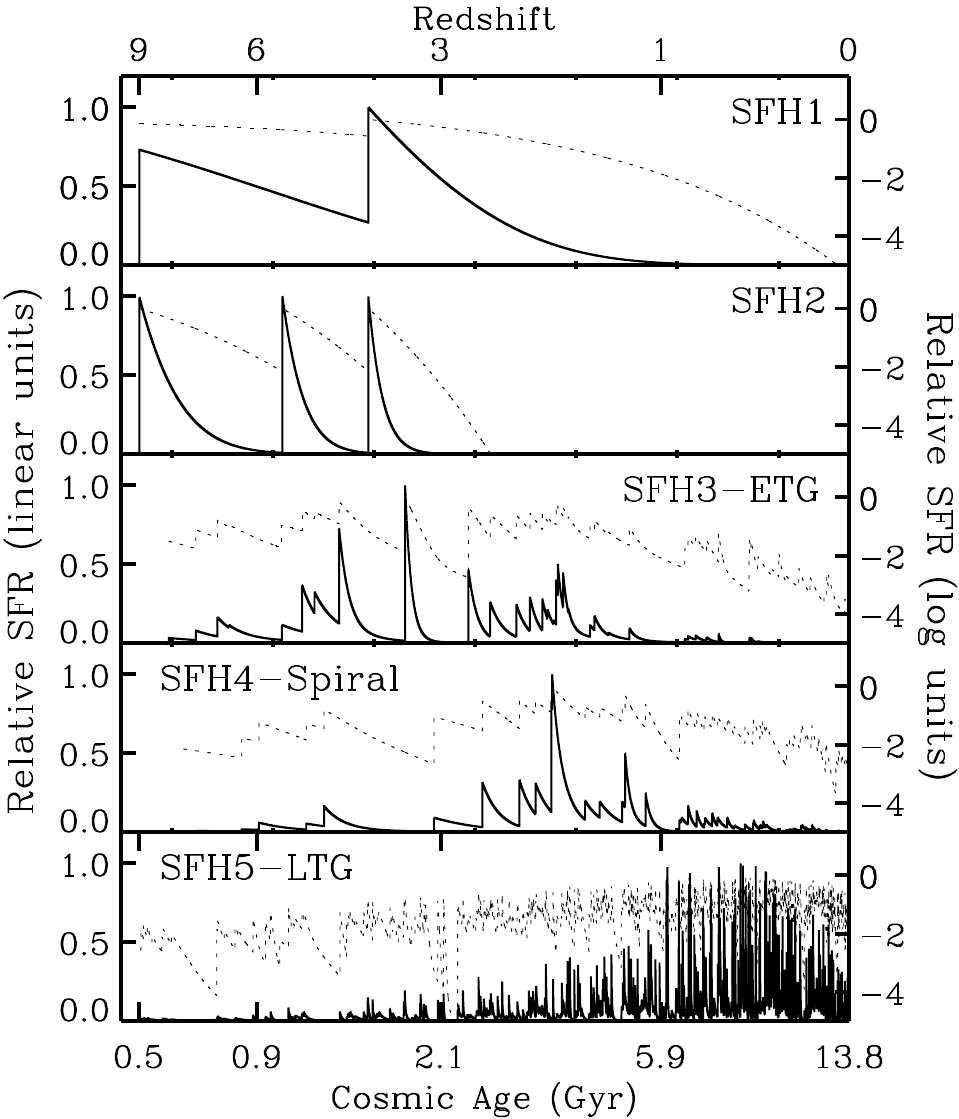}
} 
\caption{\noindent\small
Examples of five arbitrary scenarios of stochastic SFHs. Relative star
formation rates ($\psi$, normalized to the peak of the strongest star formation
episode) are plotted as a function of cosmic time (bottom axis) and redshift
(top axis) in both linear (solid curves; left axis) and logarithmic (dotted;
right axis) units. SFH1 and 2 represent the cases of two and three bursts with
long (1\,Gyr) and shorter (0.5\,Gyr) $e$-folding times, respectively. SFH3, 4,
and 5 represent simulated CSPs of early-type (E,S0), spiral (Sa--Sbc) and
late-type (Sc,Sd) galaxies, respectively.\label{fig:figure3}}
\end{figure}
%%%%%%%%%%%%%%%%%%%%%%%%%%%%%%%%%%%%%%%%%%%%%%%%%%%%%%%%%%%%%%%%%%%%%%%%%%%%%%

%%%%%%%%%%%%%%%%%%%%%%%%%%%%%%%%%  FIGURE 4  %%%%%%%%%%%%%%%%%%%%%%%%%%%%%%%%%
\noindent\begin{figure*}[t]
\centerline{
  \includegraphics[width=0.48\txw]{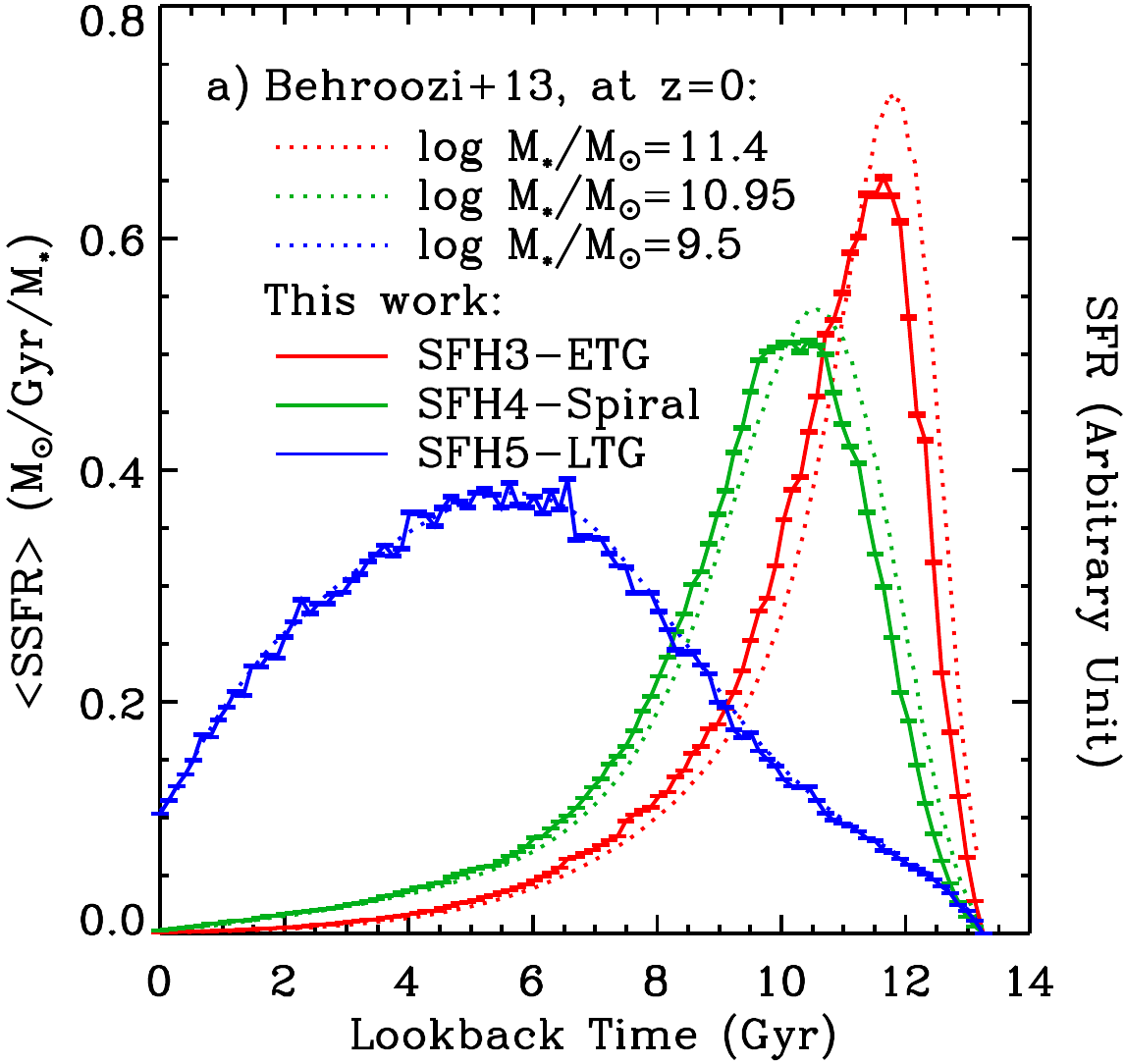} \ \ \ \ \ 
  \includegraphics[width=0.48\txw]{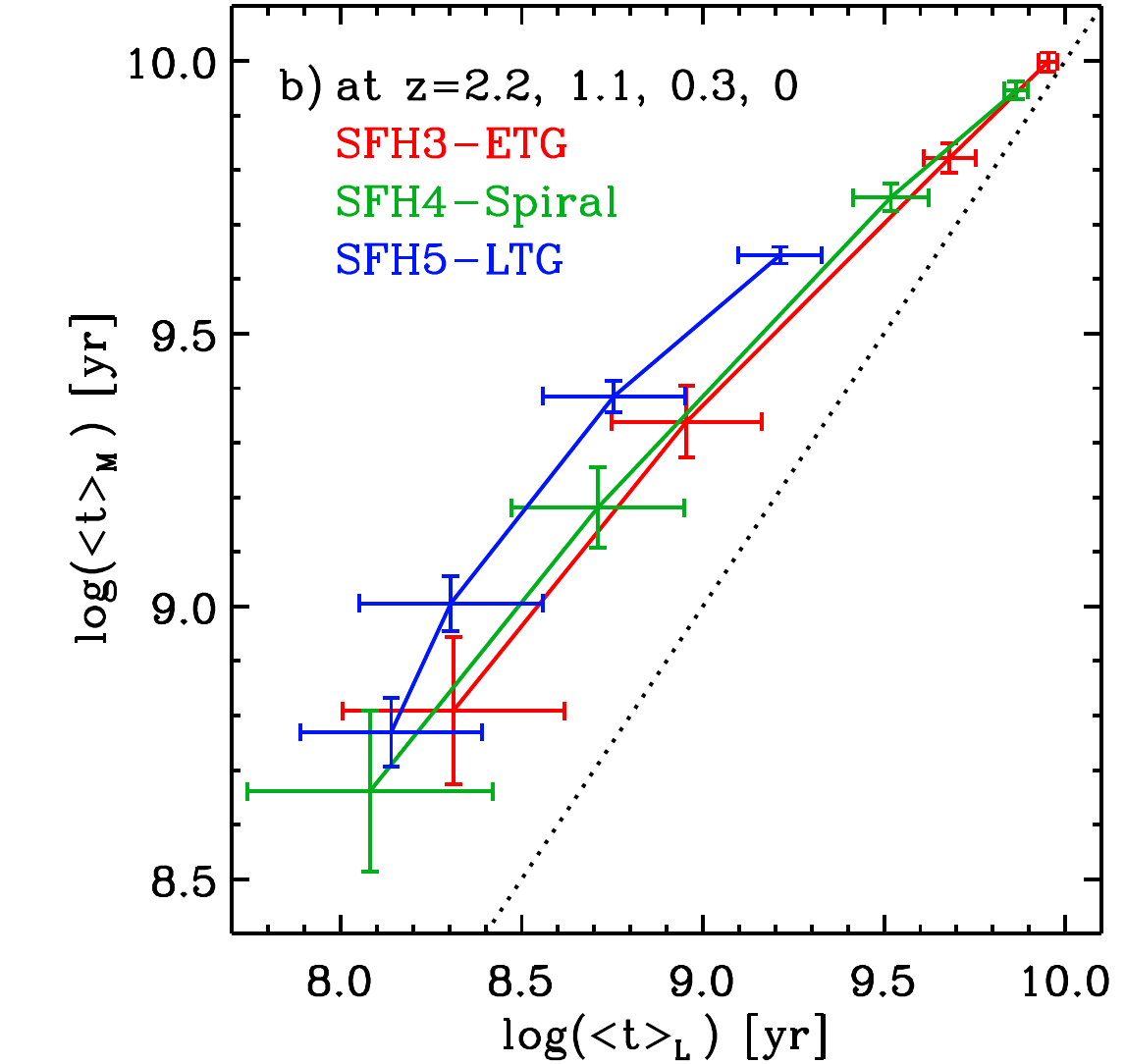}
}
\caption{\noindent\small
(a) Specific SFR versus lookback time for our families of stochastic SFHs. We
show the mean and standard deviation of the mean SFR normalized by the total
stellar mass built up at $z$\,=\,0 for 1000 random realizations each of SFH3,
SFH4, and SFH5 (solid distributions). The dotted curves represent the models of
Behroozi \etal\ (2013) for present-day stellar masses of 2.5$\times$10$^{11}$,
8.9$\times$10$^{10}$, and 3.1$\times$10$^{9}$\,\Msun, representative of
early-type (E), spiral (Sb), and late-type (Sd) galaxies, respectively. (b)
Mean and scatter of the light- and mass-weighted ages of our stochastic SFHs at
redshifts of 0 (upper right), 0.3, 1.1, and 2.2 (lower left). The mass-weighted
ages are always larger than the light-weighted ones, but the gap between them
decreases from redshift $z$\,=\,2 to the present
($z$\,=\,0).\label{fig:figure4}}
\end{figure*}
%%%%%%%%%%%%%%%%%%%%%%%%%%%%%%%%%%%%%%%%%%%%%%%%%%%%%%%%%%%%%%%%%%%%%%%%%%%%%%%

We started by building CSP SEDs for stellar populations with exponentially
declining star formation rates, $\psi(t)$\,=\,$\psi_{0}$\,${e^{-t/\tau}}$ for
various values of the $e$-folding time $\tau$ (100, 250, 500\,Myrs, 1, 2, 5,
and 10\,Gyr, where the latter two approach the case of a \emph{constant} star
formation rate). Note that for these exponentially declining SFHs, which were
merely meant to test the effect of combining SSPs of different ages, we did not
yet take metallicity evolution into account. We then produced SEDs for more
realistic stochastic SFHs, built from multiple partially overlapping
exponentially declining starbursts with onset times $t_{0,i}$, $e$-folding time
$\tau_i$, and peak amplitude $\psi_{0,i}$\ from\\[-10pt]
\begin{equation}
  \psi(t) = \sum_{i}^{n} \psi_{0,i}\, e^{-(t-t_{0,i})/\tau_i} \; .
\label{eq:psi}
\end{equation}
We adopt five different families of stochastic SFHs (five examples of which are
shown in Figure~\ref{fig:figure3}). SFH1 and SFH2 are simple test cases with
$n$\,=\,2 long ($\tau_i$\,=\,1\,Gyr) and $n$\,=\,3 short
($\tau_i$\,=\,100\,Myr) bursts, respectively. For each of SFH3, SFH4, and SFH5
we randomly generated 1000 stochastic composite SFHs by combining multiple
exponentially declining starbursts with different constraints in order to
simulate galaxies of different (Hubble) types and present-day stellar masses.
For SFH3, we restrict $e$-folding times $\tau_i$ to
50\,Myr\,$<$\,$\tau_i$\,$<$\,1\,Gyr, for SFH4
100\,Myr\,$<$\,$\tau_i$\,$<$\,500\,Myr, and for SFH5 to
5\,Myr\,$<$\,$\tau_i$\,$<$\,100\,Myr, representative of likely starbursts in
early-type, spiral, and late-type galaxies. The mean $e$-folding times are
$\sim$300\,Myr, $\sim$250\,Myr, and $\sim$30\,Myr, respectively. We then
estimate the number of starburst episodes required for each SFH family by
dividing the Hubble time (13.8\,Gyr) by these mean $e$-folding times, giving
44, 53, and 427 for SFH3, SFH4, and SFH5.

We furthermore adopt the model SFHs of Behroozi \etal\ (2013) to define
constraints for our mean starburst amplitudes $\psi_{0,i}$ as a function of
time of onset $t_{0,i}$. We chose model SFHs resulting in present-day stellar
masses of 2.5$\times$10$^{11}$, 8.9$\times$10$^{10}$, and
3.1$\times$10$^{9}$\,\Msun\ ($\log(\textrm{M}/\Msun)$ of 11.4, 10.95, and 9.5)
for SFH3, SFH4, and SFH5, respectively, to match the distribution of stellar
masses for Hubble types E, Sb, and Sd found by Gonz{\'a}lez-Delgado \etal\
(2015; hereafter GD15) in the CALIFA IFU survey of 300 local galaxies (see
their Figure~2). Figure~\ref{fig:figure4}(a) compares the mean specific star
formation rates as a function of lookback time for each of our SFH families and
for the corresponding Behroozi \etal\ (2013) models. For SFH3, we find that the
results shown in Figure~\ref{fig:figure4}(a) are sensitive to the choice of
lower bound on $\tau_i$, such that for longer minimum burst durations we would
fail to reproduce the steep rise and relatively fast decline in SF for massive
(early-type) galaxies in the Behroozi \etal\ (2013) models. The resulting SF
peak is reached at $z$\,=\,3.0 for SFH3, at $z$\,=\,2.1 for SFH4, and at
$z$\,=\,0.6 for SFH5.
   %
%%%%%%%%%%%%%%%%%%%%%%%%%%%%%%%%%%  TABLE 2  %%%%%%%%%%%%%%%%%%%%%%%%%%%%%%%%%%
%\placetable{2}
\noindent\begin{table}[!b]
\centering
\caption{\small Log-linear metallicity evolution parameters, fit to the results
of Maiolino \etal\ (2008) combined with our stochastic SFHs. We approximate
that evolution by $\log{Z(z)} = a\cdot z + b$. The rms of the fit in each of
$a$ and $b$ is $<$0.01. We adopt the slopes in parentheses: $-$0.18 for SFH3,
$-$0.30 for SFH4, and $-$0.58 for SFH5.\label{tab:table2}}
\begin{tabular}{ccccccc}
\toprule
\multirow{2}{*}{\bf $\log \mathrm{M} / \mathrm{M}_{\odot}$}	& \multicolumn{2}{c}{SFH3-ETG} & \multicolumn{2}{c}{SFH4-Spiral} & \multicolumn{2}{c}{SFH5-LTG} \\
 \cmidrule(r){2-3}  \cmidrule(r){4-5}  \cmidrule(r){6-7}
		& $a$ & $b$ & $a$ & $b$ & $a$ & $b$	\\[2pt]
\midrule
9.50  & $-$0.37 & 0.25 & $-$0.49 & 0.29 & ($-$0.58) & 0.22 \\
10.95 & $-$0.23 & 0.49 & ($-$0.30) & 0.52 & $-$0.36 & 0.50 \\
11.40 & ($-$0.18) & 0.49 & $-$0.24 & 0.52 & $-$0.30 & 0.51 \\
\bottomrule
\end{tabular}
\end{table}
%%%%%%%%%%%%%%%%%%%%%%%%%%%%%%%%%%%%%%%%%%%%%%%%%%%%%%%%%%%%%%%%%%%%%%%%%%%%%%%

For all CSPs constructed as described above, the normalization of the SFH is
such that $\sum \psi(t') \delta t'$\,=\,M, the total mass in stars formed up to
the present time, which will be higher than the current stellar mass because of
mass returned by stellar winds and supernova explosions, but by no more than
$\sim$30\% (see V\'{a}zquez \& Leitherer 2005). The more recent SF in low-mass
(late-type) galaxies, i.e., downsizing, results in younger ages, especially
light-weighted ages (see Figure~\ref{fig:figure4}(b)). Mass-weighted ages
are always larger than light-weighted ones, but Figure~\ref{fig:figure4}(b)
shows that the gap between them decrease from a factor of 3 at $z$\,=\,2 to
$\sim$10\% at $z$\,=\,0 for SFH3 and SFH4. For SFH5, the difference remains a
factor of $\sim$5 until $z$\,$\sim$\,0.3 before decreasing to a factor of
$\sim$3 at $z$\,$\simeq$\,0.

To account for metallicity evolution and arrive at a functional form for
$Z(t')$ in Eq.~\ref{eq:fcsp}, we combine our stochastic SFHs with the
mass-metallicity relations at $z$\,$\simeq$\,0, 0.3, 1.1, and 2.2 presented by
Maiolino \etal\ (2008). At $z$\,=\,0, we assume stellar masses of
2.5$\times$10$^{11}$, 8.9$\times$10$^{10}$, and 3.1$\times$10$^{9}$\,\Msun\ for
early-type, spiral, and late-type galaxies as above. We find that we can
approximate the metallicity evolution as $\log Z(z) = a\cdot z + b$, and fit
slopes $a$ for each of these present-day stellar masses and SFHs (see
Table~\ref{tab:table2}). We adopt $a$\,=\,$-$0.18, $-$0.30, and $-$0.58 for the
metallicity evolution of SFH3, SFH4, and SFH5, respectively. Note that while we
list fitted values in the final column of Table~\ref{tab:table2} for SFH5 for
each of our three stellar masses, late-type galaxies with M\,$>$\,10$^{10}$ are
extremely rare in nature (\eg\ Kelvin \etal\ 2014).

%%%%%%%%%%%%%%%%%%%%%%%%%%%%%%%%  FIGURE 5  %%%%%%%%%%%%%%%%%%%%%%%%%%%%%%%%%%
\noindent\begin{figure*}[t!]
\centerline{
  \includegraphics[width=\txw]{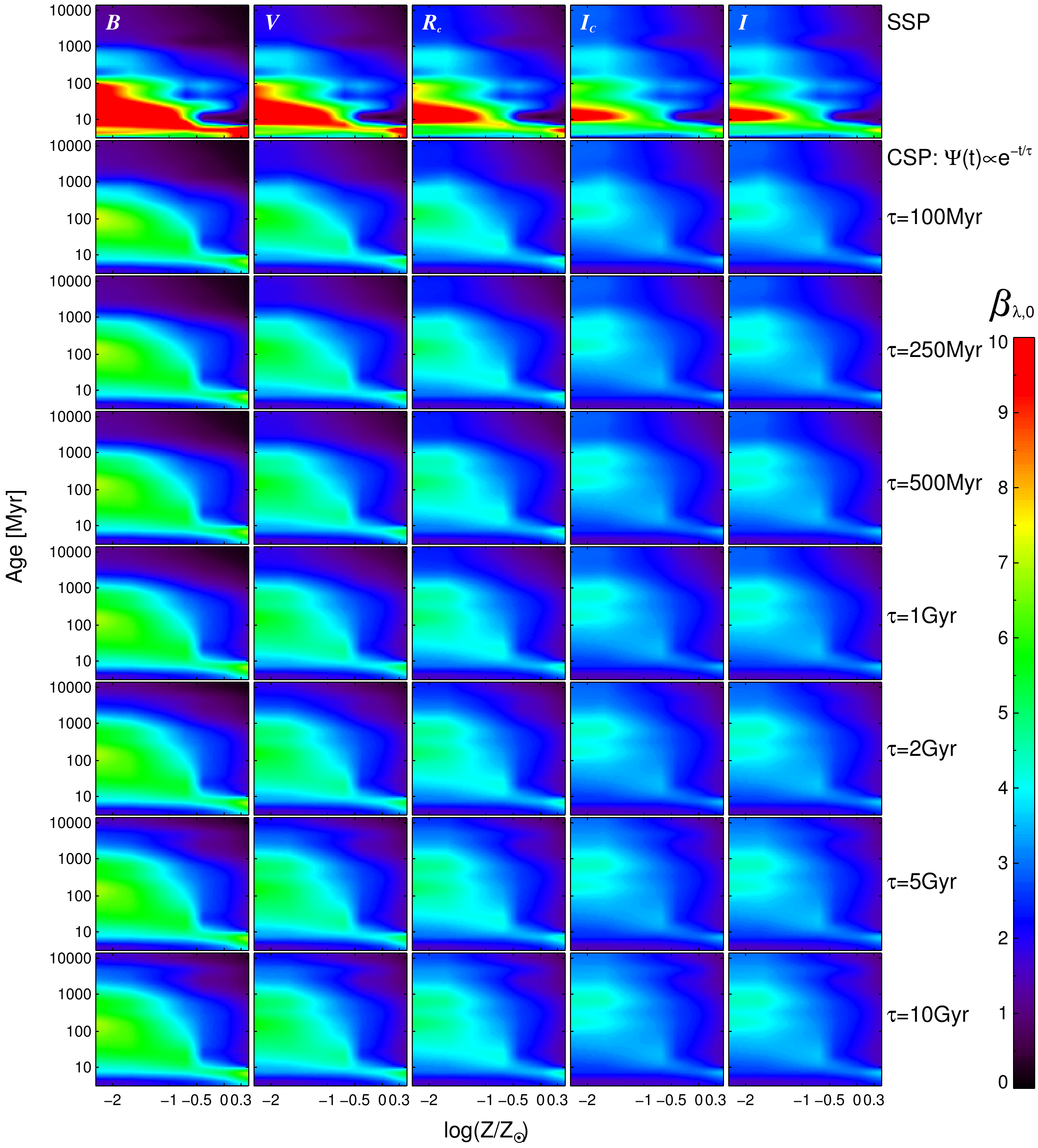}
}
\caption{\noindent\small
Maps of \betalamzero\ values referenced to the Johnson $L$ band as a function
of metallicity ($Z/Z_{\odot})$ and age of the stellar populations for the (left
to right) Johnson $B$, $V\!$, Kron-Cousins-Glass $R_c$, $I_c$, and Johnson $I$
filters for (top to bottom) SSPs, and for seven composite stellar populations
resulting from exponentially declining SFHs with $\tau$\,=\,100, 250, 500\,Myr,
1, 2, 5, and 10\,Gyr. Maps of \betalamzero\ values for our full set of
visible--near-IR filters are available as Figure Set~5 in the electronic
edition of the journal or at the end of this paper.\\[4pt] (The complete figure
set (seven images) is available.)\label{fig:figure5}}
\end{figure*}
%%%%%%%%%%%%%%%%%%%%%%%%%%%%%%%%%%%%%%%%%%%%%%%%%%%%%%%%%%%%%%%%%%%%%%%%%%%%%%
  %
%%%%%%%%%%%%%%%%%%%%%%%%%%%%%%%%%  TABLE 3  %%%%%%%%%%%%%%%%%%%%%%%%%%%%%%%%%
%\placetable{3}
\noindent\begin{table*}
\centering
\caption{\small The mean, standard deviation, minimum and maximum of
\betaVzero\ values inferred for our SSPs and CSPs with exponentially declining
SFHs (2nd column in Figure~\ref{fig:figure5}) for all $Z$, and separately for
$Z$=0.004, $Z$=0.008, and $Z$\,=\,$Z_{\odot}$\,=\,0.02,
respectively.\label{tab:table3}}
\setlength{\tabcolsep}{7pt}
\begin{tabular}{ccccccccccccc}
\toprule
\multirow{2}{*}{SFH} & \multicolumn{3}{c}{0.0001\,$\le$\,$Z$\,$\le$\,0.05} & \multicolumn{3}{c}{$Z$\,=\,0.004} & \multicolumn{3}{c}{$Z$\,=\,0.008} & \multicolumn{3}{c}{$Z$\,=\,0.02} \\
\cmidrule(r){2-4} \cmidrule(r){5-7} \cmidrule(r){8-10} \cmidrule(r){11-13} 
& $\langle$\betaVzero$\rangle$ & Min & Max & 
 $\langle$\betaVzero$\rangle$ & Min & Max & $\langle$\betaVzero$\rangle$ & Min & Max & $\langle$\betaVzero$\rangle$ & Min & Max \\ 
\midrule
SSP  			& 3.84$\pm$3.57 & 0.10 & 23.64 & 3.06$\pm$2.10 & 1.03 & 8.99 & 2.41$\pm$1.65 & 0.73 & 7.79 & 2.06$\pm$1.67 & 0.37 & 7.77 \\
$\tau$=100\,Myr & 2.74$\pm$1.37 & 0.35 & 6.05 & 2.67$\pm$1.22 & 1.04 & 4.71 & 2.13$\pm$0.89 & 0.73 & 3.89 & 1.86$\pm$0.92 & 0.55 & 4.13 \\
$\tau$=250\,Myr & 2.80$\pm$1.36 & 0.35 & 5.88 & 2.75$\pm$1.20 & 1.04 & 4.67 & 2.18$\pm$0.87 & 0.73 & 3.87 & 1.90$\pm$0.90 & 0.55 & 4.11 \\
$\tau$=500\,Myr & 2.85$\pm$1.35 & 0.35 & 5.82 & 2.82$\pm$1.17 & 1.04 & 4.65 & 2.22$\pm$0.85 & 0.73 & 3.86 & 1.93$\pm$0.84 & 0.56 & 4.11 \\
$\tau$=1\,Gyr	& 2.92$\pm$1.33 & 0.36 & 5.78 & 2.88$\pm$1.13 & 1.04 & 4.64 & 2.26$\pm$0.82 & 0.74 & 3.86 & 1.97$\pm$0.84 & 0.56 & 4.10 \\
$\tau$=2\,Gyr 	& 2.98$\pm$1.29 & 0.37 & 5.77 & 2.95$\pm$1.06 & 1.06 & 4.64 & 2.31$\pm$0.76 & 0.77 & 3.86 & 2.02$\pm$0.79 & 0.59 & 4.10 \\
$\tau$=5\,Gyr 	& 3.05$\pm$1.24 & 0.48 & 5.76 & 3.02$\pm$0.98 & 1.24 & 4.64 & 2.37$\pm$0.69 & 0.96 & 3.86 & 2.07$\pm$0.73 & 0.76 & 4.10 \\
$\tau$=10\,Gyr 	& 3.08$\pm$1.22 & 0.57 & 5.75 & 3.05$\pm$0.95 & 1.38 & 4.64 & 2.39$\pm$0.66 & 1.11 & 3.85 & 2.09$\pm$0.70 & 0.89 & 4.10 \\
\bottomrule\\
\end{tabular}
\end{table*}
%%%%%%%%%%%%%%%%%%%%%%%%%%%%%%%%%%%%%%%%%%%%%%%%%%%%%%%%%%%%%%%%%%%%%%%%%%%%%

\section{Intrinsic \betalamzero\ Flux Ratios from Model SEDs}

T09 developed the \betaV\ method to estimate the amount of dust extinction in a
galaxy using images through just two broadband filters: one visible ($V$) and
one in the mid-IR near 3.5\,\micron\ ($L$ band), specifically, \Spitzer/IRAC
3.6\,\micron\ in their case. The $L$ band is where the extinction by dust
reaches a minimum, and where there is still little emission from warm dust,
polycyclic aromatic hydrocarbons (PAHs), and silicates (although there is a
significant C--H stretching PAH feature near 3.3\,\micron; L\'eger \& Puget
1984; Allamandola, Tielens \& Barker 1989). If we have knowledge of the
intrinsic SED of a simple or CSP, then we can calculate the intrinsic flux
ratio \betaVzero\,=\,$f_{V\!,0}/f_{L,0}$. This ratio will be a function of age
$t$, metallicity $Z$, and SFH $\psi(t)$. If \betaVzero\ values were to fall in
a very narrow range for a wide range of such CSP parameters, and if our model
SEDs were accurate representations of observed stellar populations, then
comparing the observed (\betaV) and the intrinsic (\betaVzero) flux ratios
would allow us to infer the missing flux in the $V$ band. We furthermore assume
that the extinction in bandpasses centered near 3.5\,\micron\ ($L$) is
negligible. The extinction in magnitudes is given by $\AV$\,=\,$(m_V -
m_{V\!,0})$, which can be rewritten in terms of the observed $L$-band flux and
the intrinsic $V$-to-$L$-band flux ratio \betaVzero\ as\\[-8pt]
\begin{equation}
  A_V \simeq\ m_V - [-2.5\,\log\left(\betaVzero\times f_L\right) - V_{\rm zp}] \ ,
\end{equation}
where $f_L$\,$\simeq$\,$f_{L,0}$, $V_{\rm zp}$ is the zero-point magnitude for
the $V$ filter, and the $\simeq$ -symbol serves to recall the approximate
nature of this method.

Here, we consider a more general extension of the \betaV\ method, the \betalam\
method, where we compute \betalamzero\ for a large selection of filter pairs in
the rest-frame optical ($\lambda$\,$\sim$\,0.4--1\,\micron) and rest-frame
mid-IR ($\lambda$\,$\sim$\,3.4--3.6\,\micron). In total, 29 visible--near-IR
and 5 mid-IR filter throughput curves were convolved\footnote{While we mean the
product $T(\lambda)\cdot F(\lambda)$ of the filter throughput $T(\lambda)$ and
SED $F(\lambda)$, rather than the convolution operation $T(\lambda)\ast
F(\lambda)$, the term ``convolution'' has become the accepted terminology and
is used throughout.} with the simple and composite SEDs we constructed as
described in \S2 to obtain the \betalamzero\ values for each SED.

We make all models and derived data available on our
website\footnote{http://lambda.la.asu.edu/betav/} as ASCII text tables and as
2D maps of \betalamzero($Z$,$t$) in both PNG\footnote{Portable Network Graphics
(Duce \etal\ 2004 [ISO/IEC~15948:2004]).} and FITS\footnote{Flexible Image
Transport System (Wells \etal\ 1981; Hanisch \etal\ 2001).} format.

%%%%%%%%%%%%%%%%%%%%%%%%%%%%%%%%  FIGURE 6 %%%%%%%%%%%%%%%%%%%%%%%%%%%%%%%%%%%
\noindent\begin{figure*}[ht!]
\centerline{
  \includegraphics[width=\txw]{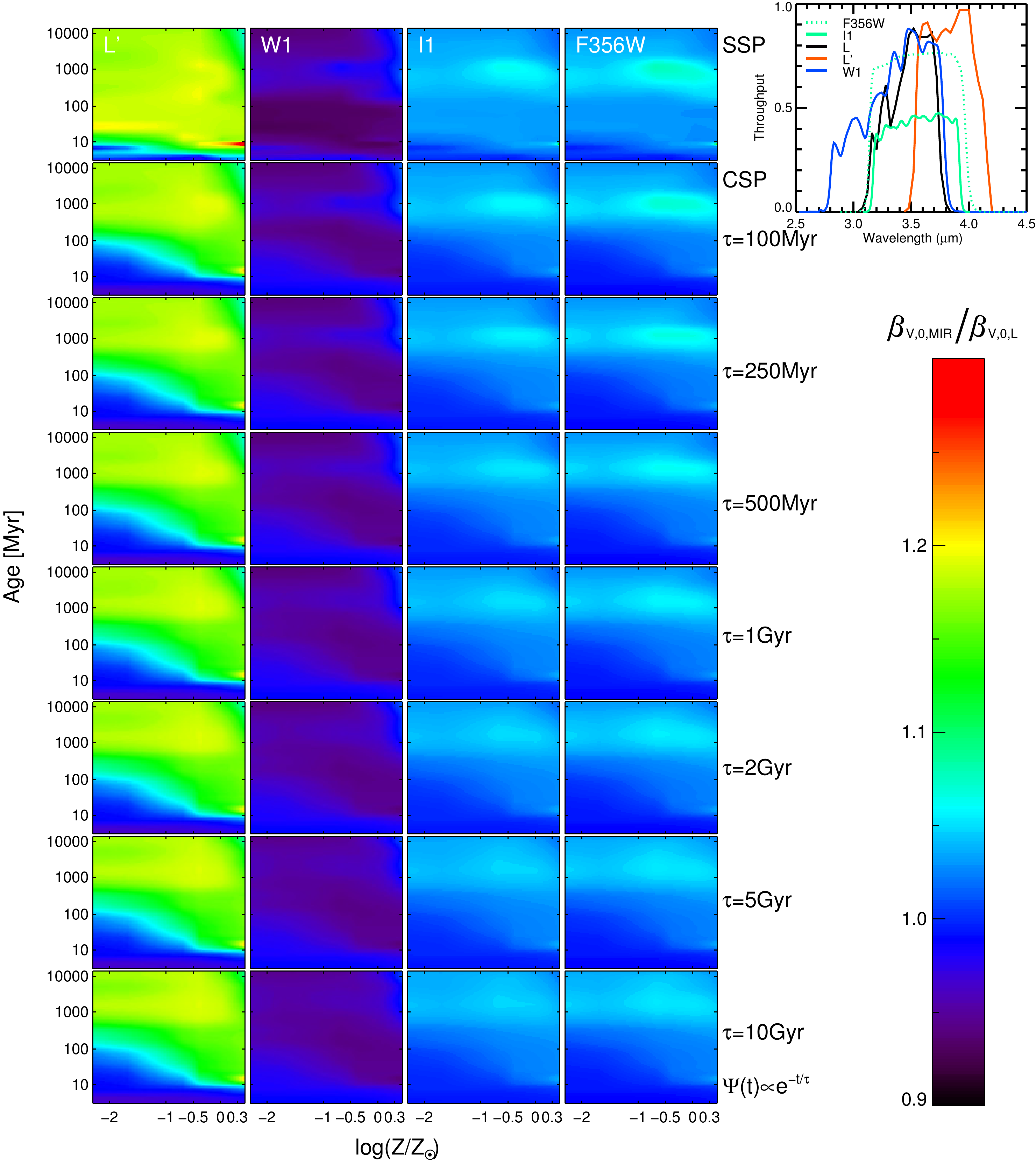}
}
\caption{\noindent\small
Maps of ratios \rbeta{\mathrm{MIR}}\,=\,$F(L)/F(L')$, $F(L)/F(W\textsl{1})$,
$F(L)/F(I\textsl{1})$, and $F(L)/F(\mathrm{F356W})$ as a function of
metallicity ($Z/Z_{\odot})$ and age $t$ of the stellar populations and stellar
population models as shown in Figure~\ref{fig:figure5}. \rbeta{\mathrm{MIR}}
gradually increases as stellar populations age and decreases again for
super-solar metallicity. The color bar in this figure spans a much narrower
range than the one in Figure~\ref{fig:figure5}. For reference, we plot the
throughput curves of each MIR bandpass in the inset at the upper
right.\label{fig:figure6}}
\end{figure*}
%%%%%%%%%%%%%%%%%%%%%%%%%%%%%%%%%%%%%%%%%%%%%%%%%%%%%%%%%%%%%%%%%%%%%%%%%%%%%%
  %
%%%%%%%%%%%%%%%%%%%%%%%%%%%%%%%%%  FIGURE 7 %%%%%%%%%%%%%%%%%%%%%%%%%%%%%%%%%%%
\noindent\begin{figure}[ht!]
\centerline{
  \includegraphics[width=0.485\txw]{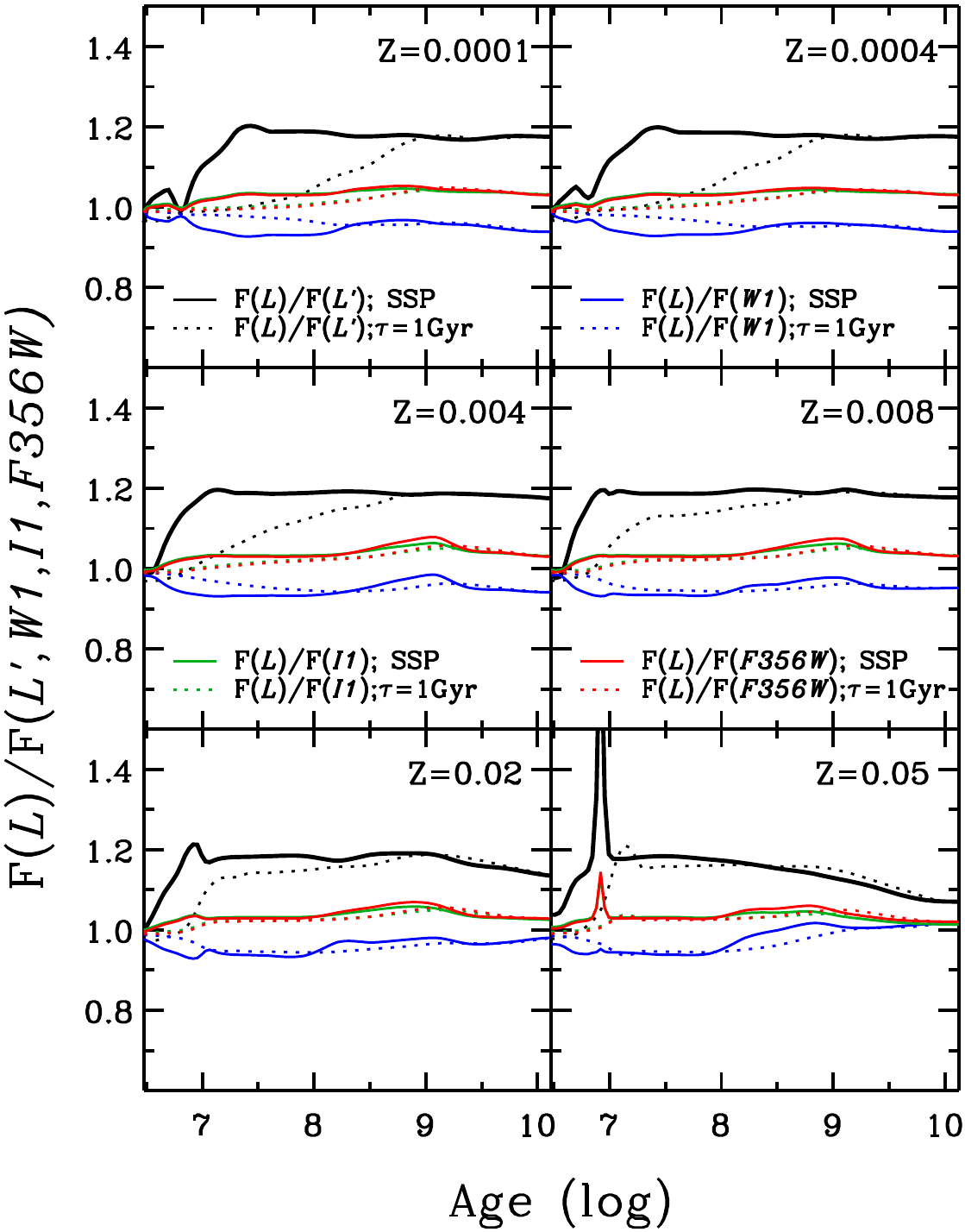}
}
\caption{\noindent\small
Flux ratios between different choices of mid-IR reference filters as a function
of the age of the stellar populations for various metallicities. Solid curves
are for SSPs and dotted curves are for exponentially declining SFHs with a
1\,Gyr $e$-folding time. The ratio generally remains stable in the 0.9--1.2
range for ages larger than a few tens of Myr. While the \betalamzero\ value to
adopt depends on the choice of mid-IR reference filter, it is
\emph{predictable}, so any of the filters considered will be similarly valid
for the application of the \betalam\ method.\label{fig:figure7}}
\end{figure}
%%%%%%%%%%%%%%%%%%%%%%%%%%%%%%%%%%%%%%%%%%%%%%%%%%%%%%%%%%%%%%%%%%%%%%%%%%%%%%
  %
%%%%%%%%%%%%%%%%%%%%%%%%%%%%%%  FIGURE 8  %%%%%%%%%%%%%%%%%%%%%%%%%%%%%%%%%%%%%
\noindent\begin{figure}[ht!]
\centerline{
  \includegraphics[width=0.485\txw]{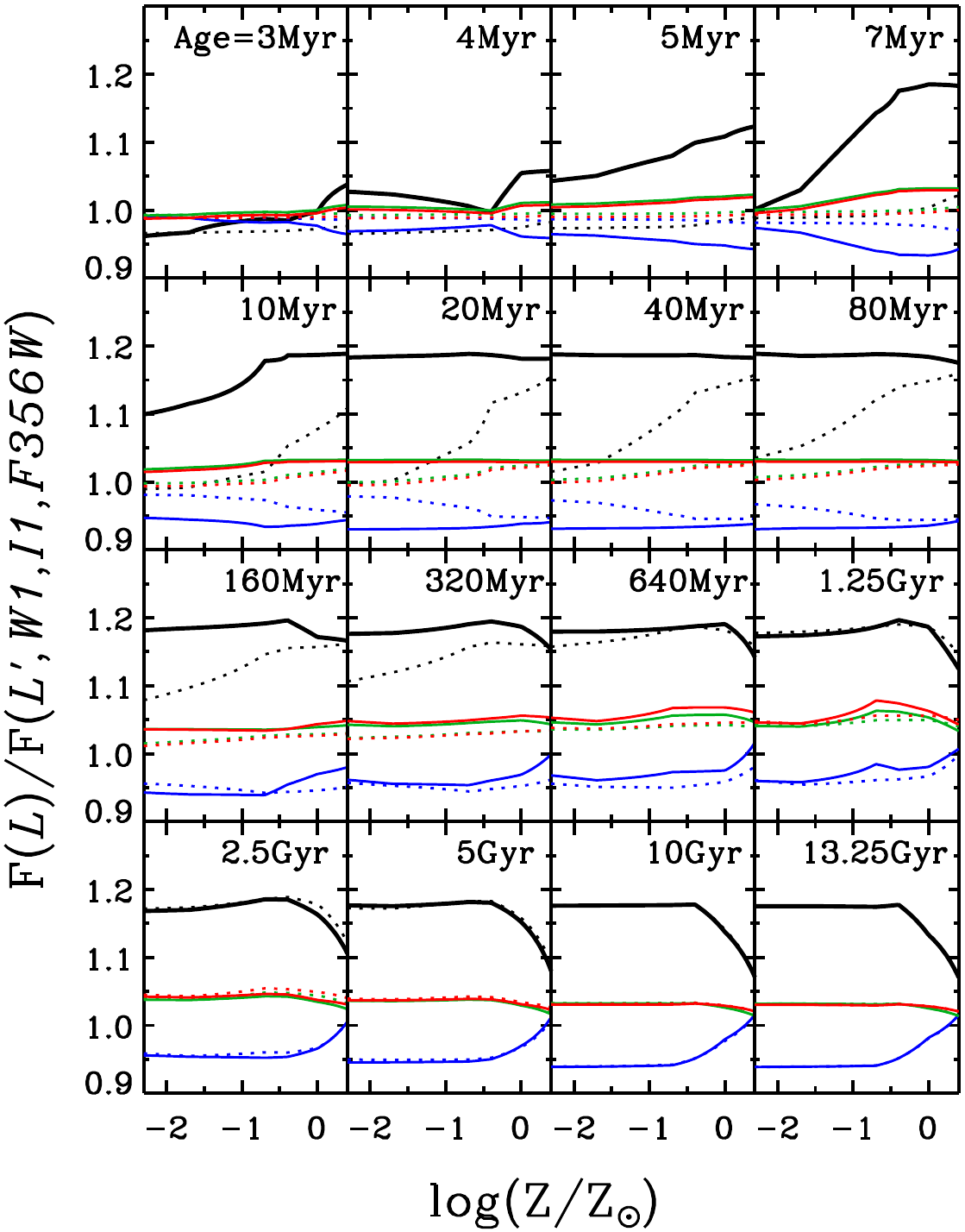}
}
\caption{\noindent\small
Flux ratios between different choices of mid-IR reference filters as a function
of the metallicity of the stellar populations for various ages. Line and color
schemes are the same as in Figure~\ref{fig:figure7}. If we exclude the red
supergiants feature at ages of $\sim$5--10\,Myr and high metallicity, the
choice of mid-IR reference filter affects \betalamzero\ at the $\lesssim$20\%
level, and for a given reference filter, much less than that over most of the
range in age and metallicity.\label{fig:figure8}}
\end{figure}
%%%%%%%%%%%%%%%%%%%%%%%%%%%%%%%%%%%%%%%%%%%%%%%%%%%%%%%%%%%%%%%%%%%%%%%%%%%%%%%
\vspace*{-10pt}

\subsection{SSP and Exponentially Declining SFHs}

Figure~\ref{fig:figure5} depicts an example of 2D maps of \betalamzero\ ratios as
a function of metallicity and stellar population age (or time since the onset
of SF) for the Johnson $B$, $V\!$, Kron-Cousins-Glass $R_c$, $I_c$, and Johnson
$I$ filter (Bessell 1990), when referenced to the Johnson $L$ band (Bessell \&
Brett 1988). From top to bottom, we show the results for SSPs and for seven
exponentially declining SFHs with $\tau$\,=\,100, 250, and 500\,Myr, and 1, 2,
5, and 10\,Gyr. The final row approximates a continuous nearly constant star
formation rate. For each panel, we started with a 2D array of SEDs that has six
rows and 16 columns, corresponding to the metallicity and age values of
Figure~\ref{fig:figure2}, for which we computed \betalamzero\ values by
convolving each of the filter curves with each of the 6$\times$16 SEDs. The
resulting array of \betalamzero\ values was expanded through log--log
cubic-spline interpolation into the finer 100$\times$100 grids of metallicity
and age as shown. SSPs show the highest dispersion and widest range of
\betaVzero\ values (see Table~\ref{tab:table3}). For exponentially declining
SFHs, as $\tau$ increases, the standard deviations and min--max ranges
decrease, while the mean \betaVzero\ values themselves slightly increase
($\lesssim$11\%). The lower metallicity population has higher \betaVzero\
values as well ($\sim$50\% for $Z$\,$=$\,0.0004 compared to $Z$\,$=$\,0.02).

Similarly, in Figure Set~5, we show \betalamzero\ referenced to the same
Johnson $L$ band for the SDSS $g$, $r$, $i$, and $z$ filters (Gunn \etal\
1998), as well as for eight \HST/ACS\,WFC (Ford \etal\ 2003; Avila \etal\ 2015)
and eight \HST/WFC3\,UVIS (Dressel 2015) filters. The choice of filters was
motivated by their common use for \HST\ deep and medium-deep surveys, such as
the Great Observatories Origins Deep Survey (GOODS; Dickinson \etal\ 2003, p.
324; Giavalisco \etal\ 2004), the Cosmic Assembly Near-IR Deep Extragalactic
Legacy Survey (CANDELS; Grogin \etal\ 2011; Koekemoer \etal\ 2011), the
Cosmological Evolution Survey (COSMOS; Scoville \etal\ 2007), the Cluster
Lensing and Supernova Survey (CLASH; Postman \etal\ 2012), and the \HST/WFC3
Early Release Science (ERS) program (Windhorst \etal\ 2011). In Figure Set~5,
moreover, we extend our coverage to the near-IR with four WFC3/IR filters.

%% commented out due to adding these figures at the end of ms.pdf
%% which is not intended to be - DK 040417
%%
%\smallskip
%\input{figset5}
%\smallskip

%%%%%%%%%%%%%%%%%%%%%%%%%%%%%%%%%  TABLE 4  %%%%%%%%%%%%%%%%%%%%%%%%%%%%%%%%%
%\placetable{4}
\noindent\begin{table}[b!]
\caption{\small Comparison of the mean values and standard deviations of
\rbeta{\mathit{MIR}} = $\beta_{\lambda,0}^{\mathit{MIR}}/\beta_{\lambda,0}^{L}$
= $F(L)/F(\mathit{MIR})$, before and after 3$\sigma$ rejection, for our SSPs
and CSPs with exponentially declining SFHs (first and fifth rows in
Figure~\ref{fig:figure6}).\label{tab:table4}}
\centering
\setlength{\tabcolsep}{4.3pt}
\begin{tabular}{ccccc}
\hline\\[-5pt]
\multirow{2}{*}{$\mathit{MIR}$} & \multicolumn{2}{c}{SSP} & \multicolumn{2}{c}{$\tau$=1\,Gyr} \\
& Mean & Robust mean & Mean & Robust mean \\[2pt]
\hline\\[-5pt]
$L'$   & 1.16$\pm$0.05 & 1.18$\pm$0.01 & 1.11$\pm$0.07 & 1.11$\pm$0.08 \\
$W$\textsl{1} & 0.95$\pm$0.02 & 0.95$\pm$0.02 & 0.96$\pm$0.01 & 0.96$\pm$0.02 \\
$I$\textsl{1} & 1.03$\pm$0.01 & 1.04$\pm$0.01 & 1.02$\pm$0.02 & 1.02$\pm$0.02 \\
F356W & 1.03$\pm$0.02 & 1.03$\pm$0.02 & 1.02$\pm$0.02 & 1.02$\pm$0.02 \\[2pt]
\hline
\end{tabular}
\end{table}
%%%%%%%%%%%%%%%%%%%%%%%%%%%%%%%%%%%%%%%%%%%%%%%%%%%%%%%%%%%%%%%%%%%%%%%%%%%%%
  %
One might wonder how sensitive the \betalam\ method is to the exact choice of
the mid-IR reference filter. We thus considered the ground-based Johnson $L$
and $L'$ filters, the space-based \WISE\ $W$\textsl{1} bandpass ($\lambda_{\rm
eff}$\,=\,3.3526\,\micron, $\Delta\lambda$\,=\,0.66\,\micron; Wright \etal\
2010; Jarret \etal\ 2011), the \Spitzer/IRAC $I$\textsl{1} bandpass
($\lambda_{\rm eff}$\,=\,3.550\,\micron, $\Delta\lambda$\,=\,0.75\,\micron;
Fazio \etal\ 2004), and the \JWST/NIRCam (Horner \& Rieke 2004; Gardner \etal\
2006; Rieke 2011) F356W filter ($\lambda_{\rm eff}$\,=\,3.568\,\micron,
$\Delta\lambda$\,=\,0.781\,\micron)\footnote{
http://www.stsci.edu/jwst/instruments/nircam/instrumentdesign/filters/}. In
Figure~\ref{fig:figure6} and Table~\ref{tab:table4}, we compare the ratios of
\betalamzero\ computed relative to each mid-IR (MIR) band and to the $L$ band:
\rbeta{\mathrm{MIR}} = $\beta_{\lambda,0}^{\mathit{MIR}}/\beta_{\lambda,0}^{L}$
= $F(L)/F(\mathrm{MIR})$.
The ratios get closer to unity and the range in values decreases for the
exponentially declining SFHs with $\tau$\,=\,1\,Gyr. The \betalamzero\ values
referenced to the $I\textsl{1}$ and F356W filters are similar to those
referenced to the $L$ band, whereas \betalamzero\ values referenced to $L'$ and
$W\textsl{1}$ were higher and lower, respectively. The \rbeta{L'} values tend
to be higher than 1 (\ie\
$\beta_{\lambda,0}^{L'}$\,$>$\,$\beta_{\lambda,0}^{L}$), because we are
sampling the Rayleigh-Jeans tail of the stellar SED and the central wavelength
of the $L'$ filter is longer than that of the $L$ filter. Conversely, the
central wavelength of the $W$\textsl{1} bandpass lies shortward of that of $L$,
resulting in \rbeta{W\textsl{1}} smaller than 1. Except for a difference in the
overall throughput, the $I$\textsl{1} and F356W bandpasses have very similar
shapes and similar central wavelengths to the $L$ filter, as shown in the inset
at the upper right side of Figure~\ref{fig:figure6}.

In Figure~\ref{fig:figure7} we plot \rbeta{L'}\ as a function of age for six
different metallicities for SSPs (black solid curves) and for exponentially
declining SFHs (black dotted curves). At low metallicities, \rbeta{L'}\ is seen
to be remarkably independent of age for ages older than 10\,Myr for SSPs. For
CSPs with an exponentially declining SFR and an $e$-folding time of 1\,Gyr,
\rbeta{L'} stabilizes around 500--600\,Myr. At solar metallicity and above, and
particularly for SSPs, sharp features become noticeable around 10\,Myr of age
in both Figure~\ref{fig:figure6} and \ref{fig:figure7}. These features correspond
to the appearance and demise of red supergiants (RSGs; Walcher \etal\ 2011).
The strength or absence of these RSG features at low metallicity remains
uncertain (see, \eg\ discussions in Cervi{\~n}o \& Mas-Hesse 1994; Leitherer
\etal\ 1999; V\'{a}zquez \& Leitherer 2005). The flux ratios for different
mid-IR filters are similar at early ages ($\sim$3\,Myr), but diverge until an
age of $\sim$10\,Myr before stabilizing again for ages larger than 1\,Gyr in
the case of super-solar metallicities. This pattern is more clearly shown in
Fig~\ref{fig:figure8}, which plots \rbeta{\mathit{MIR}} as a function of
metallicity at 16 different ages. \rbeta{\mathit{MIR}} is insensitive to
metallicity at most stellar population ages (the curves in
Fig~\ref{fig:figure8} are nearly flat for most ages $\gtrsim$10\,Myr and for
most sub-solar metallicities). If we exclude the RSG feature at high
metallicity, the choice of mid-IR filter affects \betalamzero\ at the
$\lesssim$20\% level overall, while reference filter-dependent variations with
respect to the $L$ band are much smaller than that over most of the range in
age and metallicity. One can gauge the impact of a given choice of mid-IR
reference filter using Figs.~\ref{fig:figure6}--\ref{fig:figure8}.

%%%%%%%%%%%%%%%%%%%%%%%%%%%%%%%%  FIGURE 9  %%%%%%%%%%%%%%%%%%%%%%%%%%%%%%%%%
\noindent\begin{figure}[t!]
\centerline{
  \includegraphics[width=0.48\txw]{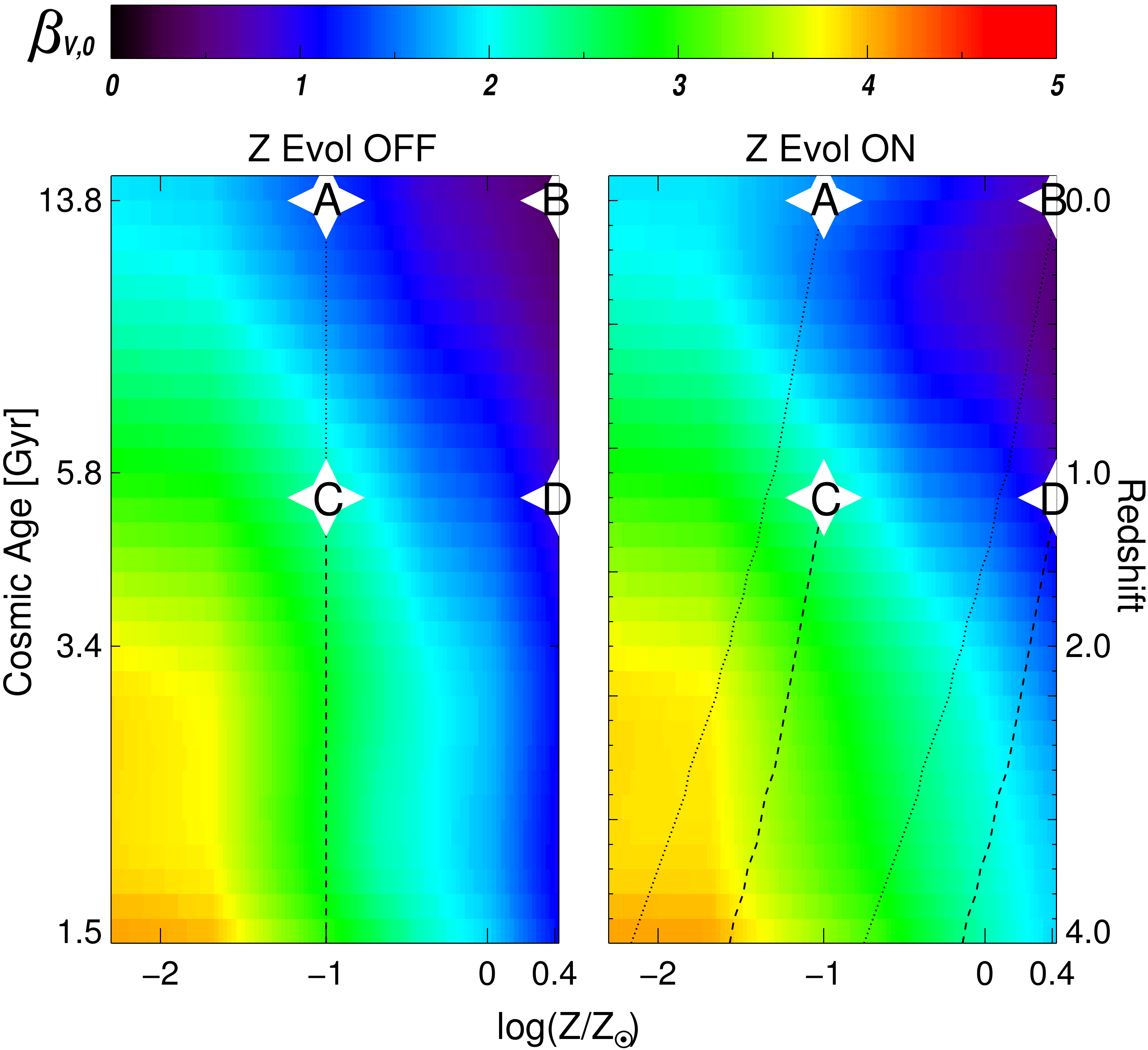}
} 
\caption{\noindent\small
Comparison of \betaVzero\ values when metallicity evolution \emph{is} (right)
and \emph{is not} (left) taken into account for the CSPs characterized by SFH4
(spiral galaxies). For four specific realizations of SFH4, we show the
metallicity tracks as a function of cosmic age of the SSP SEDs, which are
stacked to generate CSP SEDs at that age. The endpoints ($z$,\,$Z$) for tracks
A, B, C, and D are (0.0,\,0.002), (0.0,\,0.05), (1.0,\,0.002), and
(1.0,\,0.05), respectively. When we include metallicity evolution, we find a
stronger dependence of \betaVzero\ on redshift, with higher \betaVzero\ values
for the progenitors of present-day galaxies than in the no-evolution
case.\label{fig:figure9}}
\end{figure}
%%%%%%%%%%%%%%%%%%%%%%%%%%%%%%%%%%%%%%%%%%%%%%%%%%%%%%%%%%%%%%%%%%%%%%%%%%%%%%

\subsection{Stochastic Multiburst SFHs}

The \betaV\ method was originally developed as an approximate dust extinction
correction method for large surveys of spatially resolved galaxies with
unresolved stellar populations. The light from the unresolved stellar
populations will have more complex SFHs than SSP or exponentially declining
SFHs (Gerola \& Seiden 1978; Kauffmann \etal\ 2006; da Silva \etal\ 2012).
Hence, building SEDs for complex SFHs with stochastic starbursts and inspecting
the resulting \betalamzero\ values will be essential to determine whether the
\betaV\ method (or its extension, the \betalam\ method) is applicable. As
mentioned in \S\,2.2, metallicity evolution is taken into account when we build
the SED arrays for stochastic SFHs, which moves the metallicity range toward
higher values when cosmic age increases and redshift decreases. When
metallicity evolution is taken into account, initial metallicities for cosmic
ages less than 1.5\,Gyr ($z$\,$>$\,4) may be below the lower limit
($Z$\,=\,0.0001) of our SSP model SEDs. There, our CSPs will have an artificial
lower limit in metallicity of $Z$\,=\,0.0001, so we will not consider redshifts
$z$\,$>$\,4. Figure~\ref{fig:figure9} shows the difference between \betaVzero\
values when metallicity evolution \emph{is} (right) and \emph{is not} (left)
applied for the CSPs characterized by SFH4 (spiral galaxies; see
Figure~\ref{fig:figure3} and Figure~\ref{fig:figure4}(a)). We adopt a
metallicity evolution of $-$0.30 per unit $z$, appropriate for SFH4 (see
Table~\ref{tab:table2}). To illustrate the effect of metallicity evolution for
four specific realizations of SFH4, we show the metallicity tracks as a
function of cosmic age of the SSP SEDs, which are stacked to generate CSP SEDs
at that age. The effects of metallicity evolution become evident at
$z$\,$\gtrsim$\,1 (cosmic age $\lesssim$5.8\,Gyr). For the progenitors of
galaxies at a given redshift, significantly higher values of \betaVzero\ must
be assumed at longer lookback times than in the no-evolution case, so that
there is a stronger dependence of \betaVzero\ on redshift.

\begin{figure*}[t]
\centerline{
 \parbox[b][0.440\txh][t]{0.03\txw}{\textbf{(a)}} \ 
   \includegraphics[height=0.45\txh]{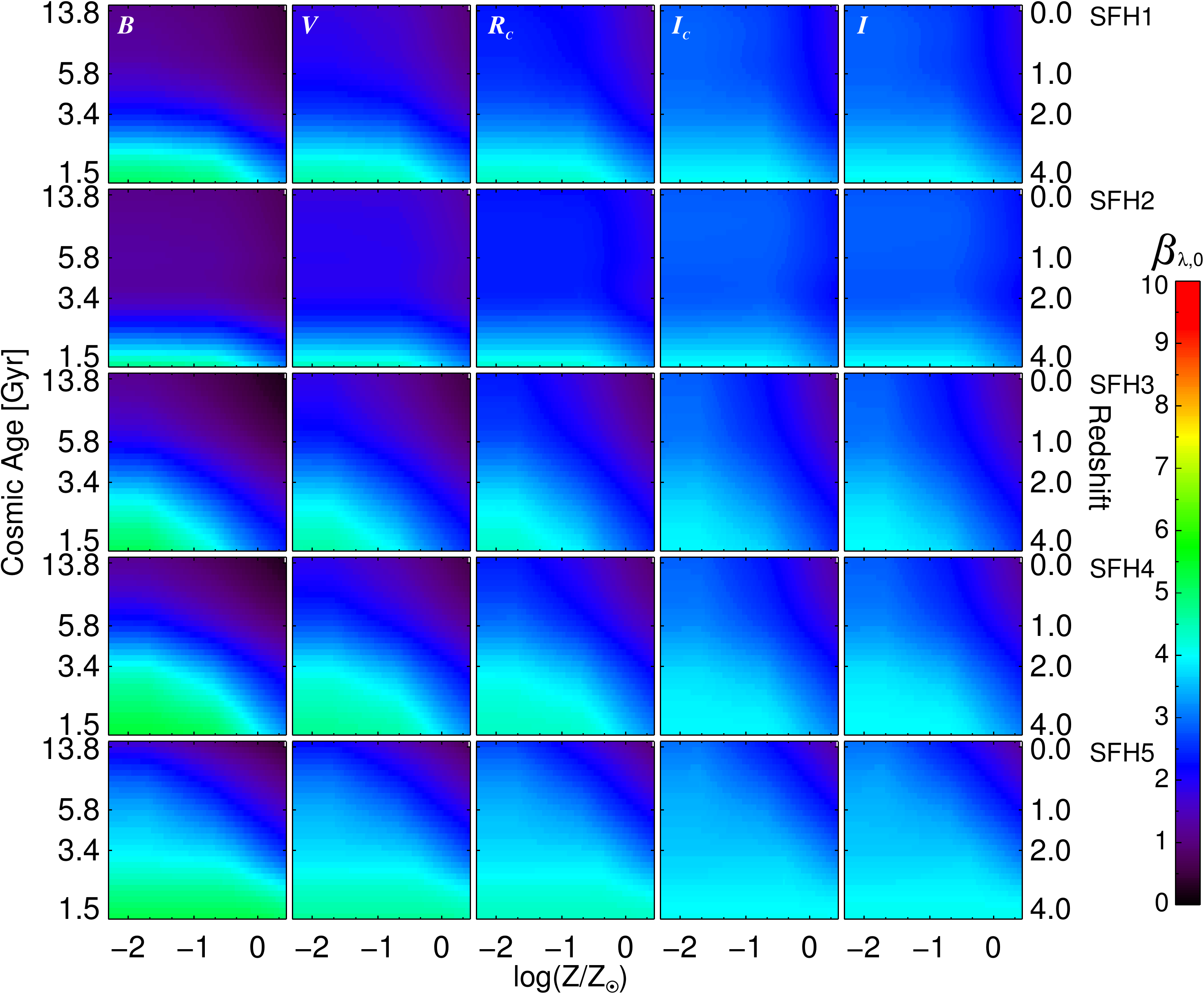}
}
\centerline{
 \parbox[b][0.440\txh][t]{0.03\txw}{\textbf{(b)}} \
   \includegraphics[height=0.45\txh]{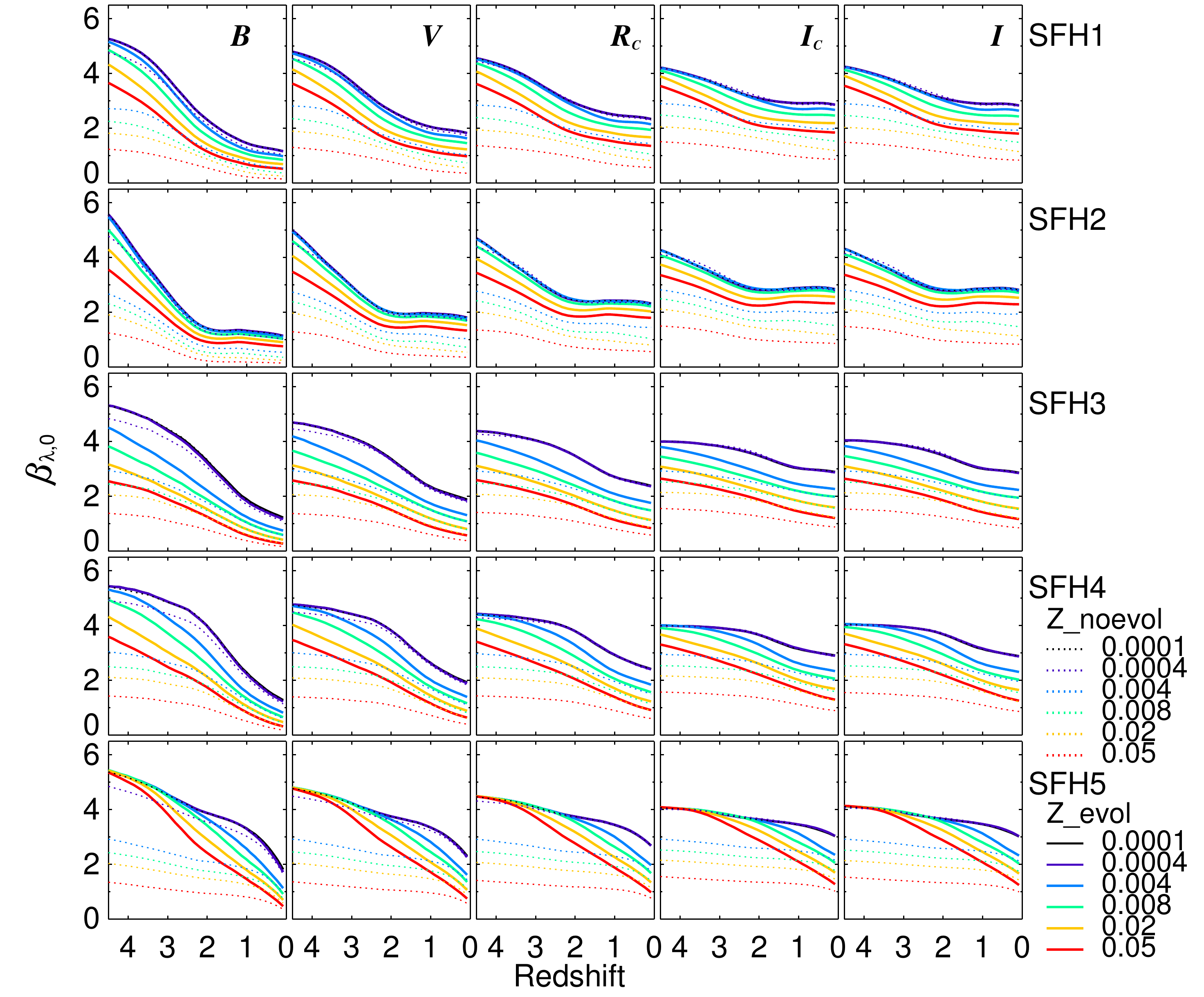}
}
\caption{\noindent\small
(a) Maps of \betalamzero\ values referenced to the Johnson $L$ band as a
function of metallicity ($Z/Z_{\odot})$ and cosmic age for the (left to right)
Johnson $B$, $V\!$, Kron-Cousins-Glass $R_c$, $I_c$, and Johnson $I$ filters
for (top to bottom) stochastic (multiburst) SFHs (see Figure~\ref{fig:figure3})
with metallicity evolution taken into account. (b) \betalamzero\ profiles of
stellar populations with various present-day ($z$$=$0) metallicities that have
different stochastic SFHs as a function of redshift. For comparison, we
overplot the no-evolution case (dotted curves). Each map and each curve in the
panels for SFH3, SFH4, and SFH5 represents the median \betalamzero\ values of
1000 randomly generated SFHs in that family. Maps and profiles of \betalamzero\
values for our full set of visible--near-IR filters are available as Figure
Set~10 in the electronic edition of the journal or at the end of this
paper.\\[4pt] (The complete figure set (seven images) is
available.)\label{fig:figure10}}
\end{figure*}

In Figure~\ref{fig:figure10} we present maps of \betalamzero\ as a function of
cosmic age and metallicity, and 1D profiles of \betalamzero\ values as a
function of redshift. The \betalamzero\ values without metallicity evolution
are overplotted in each 1D profile panel for comparison. Each curve in the
panels for SFH3, SFH4, and SFH5 represents the median \betalamzero\ values of
1000 randomly generated stochastic SFHs in that family. Comparing the solid to
the dotted profiles in Figure~\ref{fig:figure10}(b), we find that the
\betalamzero($z$) values are higher when metallicity evolution is taken into
account, and progressively more so for higher present-day metallicity values
(at low present-day metallicities the difference must necessarily be small).
The spread in \betalamzero\ at a given redshift due to metallicity differences
tends to be smaller than in the no-evolution case. While the range in
\betalamzero\ values at $z$\,$\simeq$\,0 is comparable in both the evolution
and no-evolution cases for each of SFH3, SFH4 and SFH5, this range becomes
narrower toward higher redshifts for SHF4, and especially for SFH5. For these
two SFH families, more of the stars formed relatively recently, with much (or
even most) of the metallicity evolution taking place in the redshift range of
interest (see our adopted slopes in Table~\ref{tab:table2}). The range of
\betaVzero\ for SFH3, SFH4, and SFH5 changes from 2.4--4.7 to 0.6--2.3 as the
redshift decreases from 4 to 0. If we impose a minimal constraint on the
redshift range without assuming any particular SFH family, we find allowed
ranges for \betaVzero\ of 0.6--3.4 when 0\,$<$\,$z$\,$<$\,1, and 0.90--3.85
when 1\,$<$\,$z$\,$<$\,2. When the redshift is known, we find narrower ranges
of 0.57--2.30, 0.70--2.93, 0.90--3.35, 1.17--3.58, and 1.52--3.85 for
\betaVzero\ at $z$=0, 0.5, 1.0, 1.5, and 2.0, respectively. In \S\,3.3 we will
show that these \betalamzero\ ranges are further reduced once we can also
constrain the range of likely SFHs and metallicities. For our full set of
visible--near-IR filters, we refer the reader to Figure Set~10.

%% commented out due to adding these figures at the end of ms.pdf
%% which is not intended to be - DK 040417
%\medskip
%\input{figset10}
%\medskip

%%%%%%%%%%%%%%%%%%%%%%%%%%%%%%%%%  FIGURE 10   %%%%%%%%%%%%%%%%%%%%%%%%%%%%%
\noindent
\begin{figure}[t]
\centerline{
  \includegraphics[width=0.48\txw]{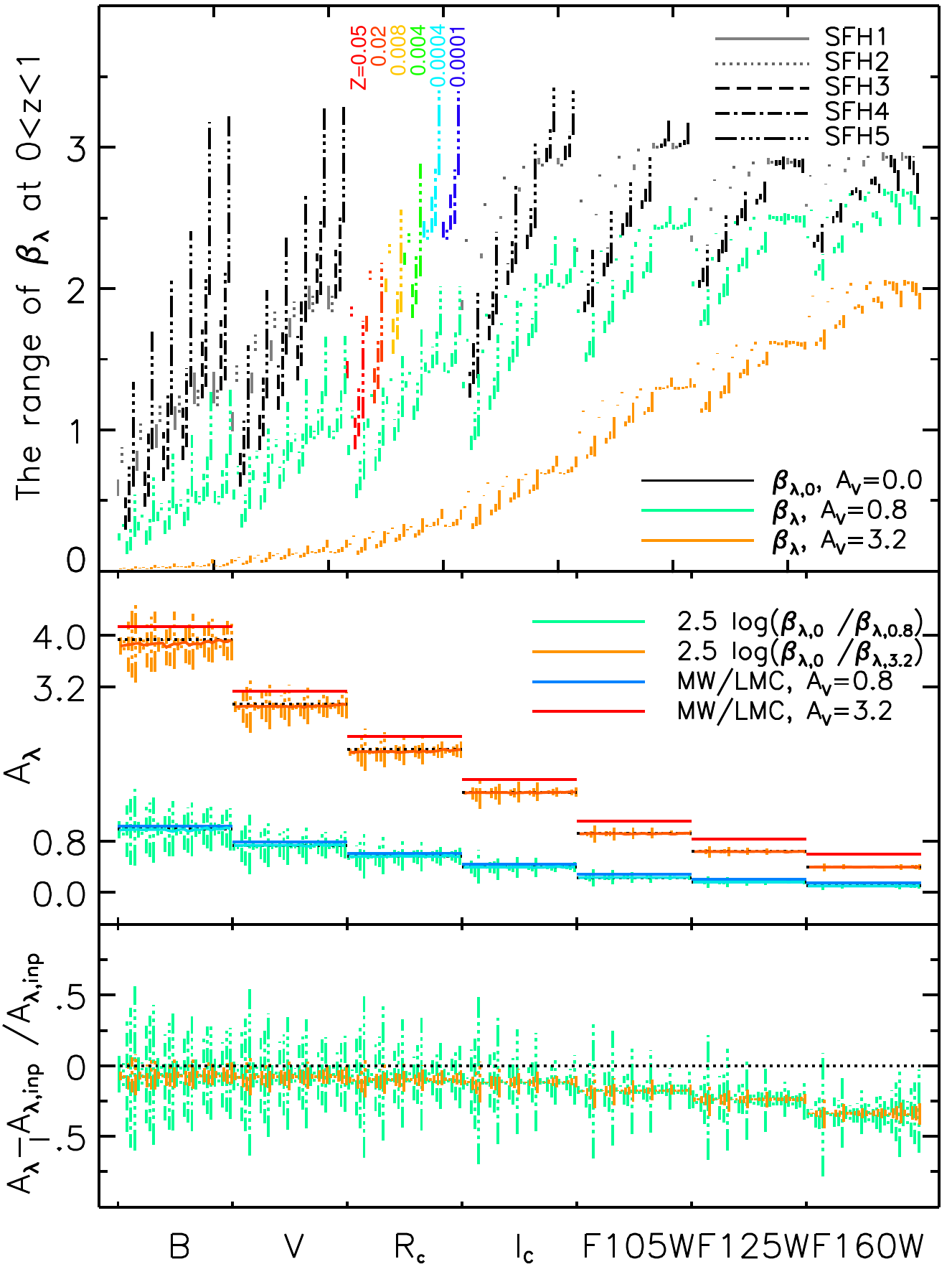}
}
\caption{\noindent\small
(top) The $\pm$1\,$\sigma$ ranges of \betalam\ values, referenced to $L$, for
CSPs at 0\,$<$\,$z$\,$<$\,1 for different choices of the optical--near-IR
filters (bottom axis) for various metallicities, SFHs, and dust extinction
values. (middle) Comparison between dust extinction values calculated using the
ratio of \betalamzero\ and \betalam\ values (center values are connected), and
dust extinction values, $A_{\lambda,\mathrm{inp}}$, using the MW/LMC extinction
law (red and blue horizontal lines). Dotted black lines indicate
$A_{\lambda,\mathrm{inp}}$ values with $-$0.05 and $-$0.19\,mag offsets for
\AV\,=\,0.8 and \AV\,=\,3.2, respectively (see text), which arise from
neglecting the residual extinction in the $L$ filter. (bottom) Normalized
ranges of recovered \Alam\ with respect to $A_{\lambda,\mathrm{inp}}$. For high
extinction values (orange), the \betalam\ method can recover
$A_{\lambda,\mathrm{inp}}$ to better than $\sim$20\% for individual resolved
galaxies when the redshift is only minimally constrained and when allowing a
wide range in SFHs and metallicity. For large samples of galaxies, the \betaV\
method recovers the \emph{mean} extinction to better than
10\%.\label{fig:figure11}}
\end{figure}

%%%%%%%%%%%%%%%%%%%%%%%%%%%%%%%%%%%%%%%%%%%%%%%%%%%%%%%%%%%%%%%%%%%%%%%%%%%%%%
   %
In Figure~\ref{fig:figure11} (top panel) we show the dependence of the
$-$1\,$\sigma$ to $+$1\,$\sigma$ range of \betalamzero\ and \betalam\ values on
the bandpass (wavelength) when the redshift is only minimally constrained to
fall in the 0\,$<$\,$z$\,$<$\,1 interval. The scatter in \betalamzero\ was
derived from the same 1000 randomly generated SFHs per SFH family as used for
Figure~\ref{fig:figure10}. The visible filters show significant ranges of allowed
\betalamzero\ values, especially for SFHs with recent SF such as SFH5. The
near-IR filters show much narrower ranges of \betalamzero. As these filters are
closer to the $L$ band, this can be understood as sampling the light from
similar stellar populations.
   %
%%%%%%%%%%%%%%%%%%%%%%%%%%%%%%%%%  FIGURE 11  %%%%%%%%%%%%%%%%%%%%%%%%%%%%%%%%%
\noindent
%%%%%%%%%%%%%%%%%%%%%%%%%%%%%%%%%%%%%%%%%%%%%%%%%%%%%%%%%%%%%%%%%%%%%%%%%%%%%%

In the middle panel of Figure~\ref{fig:figure11}, we compare the range of
extinction values, \Alam, recovered from the \betalam\ method:\\[-4pt]
\begin{equation}
    A_{\lambda} = 2.5 \log{(\beta_{\lambda,0} / \beta_{\lambda})} \ ,
\label{eq:alam}
\end{equation}
\noindent to the extinction values $A_{\lambda,\mathrm{inp}}$ imposed as\\[-4pt]
\begin{equation}
   F_{\lambda,\mathrm{ext}} = 
       F_{\lambda,\mathrm{int}} \cdot 10^{-0.4\cdot A_{\lambda,\mathrm{inp}}}
\label{eqn:ainput}
\end{equation}
\noindent where $F_{\lambda,\mathrm{int}}$ and $F_{\lambda,\mathrm{ext}}$ are
intrinsic and attenuated fluxes in each bandpass. These input extinction values
are indicated by the horizontal red (\AV\,=\,3.2\,mag) and blue
(\AV\,=\,0.8\,mag) lines. There are small systematic offsets of the median
recovered \Alam\ (indicated by the dark orange and cyan horizontal lines) from
the $A_{\lambda,\mathrm{inp}}$ values (red and blue horizontal lines). In $V$,
these offsets are $-$0.05 and $-$0.24\,mag for $A_V$\,=\,0.8 and
$A_V$\,=\,3.2\,mag, respectively, of which $-$0.05 and $-$0.19\,mag results
from the assumption of the \betalam\ method that the extinction in the $L$ band
is negligible. The relative error due to this assumption becomes more obvious
in the near-IR, where the extinction values are also small. In fact, in the
near-IR filters, this can account for the full offset observed. The additional
$\sim$0.05\,mag offset observed for \AV\,=\,3.2\,mag originates from the
non-normal distribution of \Alam\ values (we plot the center value of the
range, not the mid-point).

In order to determine whether the near-IR filters are a better choice for the
\betalam\ technique over the optical filters, we compare their normalized
scatter of \Alam\ values in the bottom panel of Figure~\ref{fig:figure11}. After
normalization, the level of scatter is consistent from the optical through the
near-IR filters, but the assumption of negligible extinction in $L$ leads to a
progressively worse underestimation of \Alam\ in a relative sense. To have a
sufficient handle on the extinction over a wide range of extinction values, it
is therefore recommended to use the bluest filter available. From this panel we
also see that the minimum scatter in the recovered \Alam\ values consistently
occurs for the lowest metallicity values and for SFHs characterized by little
recent star formation (\eg\ SFH3).
  %
%%%%%%%%%%%%%%%%%%%%%%%%%%%%%%%%%  FIGURE 12  %%%%%%%%%%%%%%%%%%%%%%%%%%%%%%%%%
\noindent\begin{figure*}[t]
\centerline{
\parbox[b][0.345\txh][t]{0.015\txw}{\textbf{(a)}}\hspace*{-0.015\txw} 
  \includegraphics[width=0.48\txw]{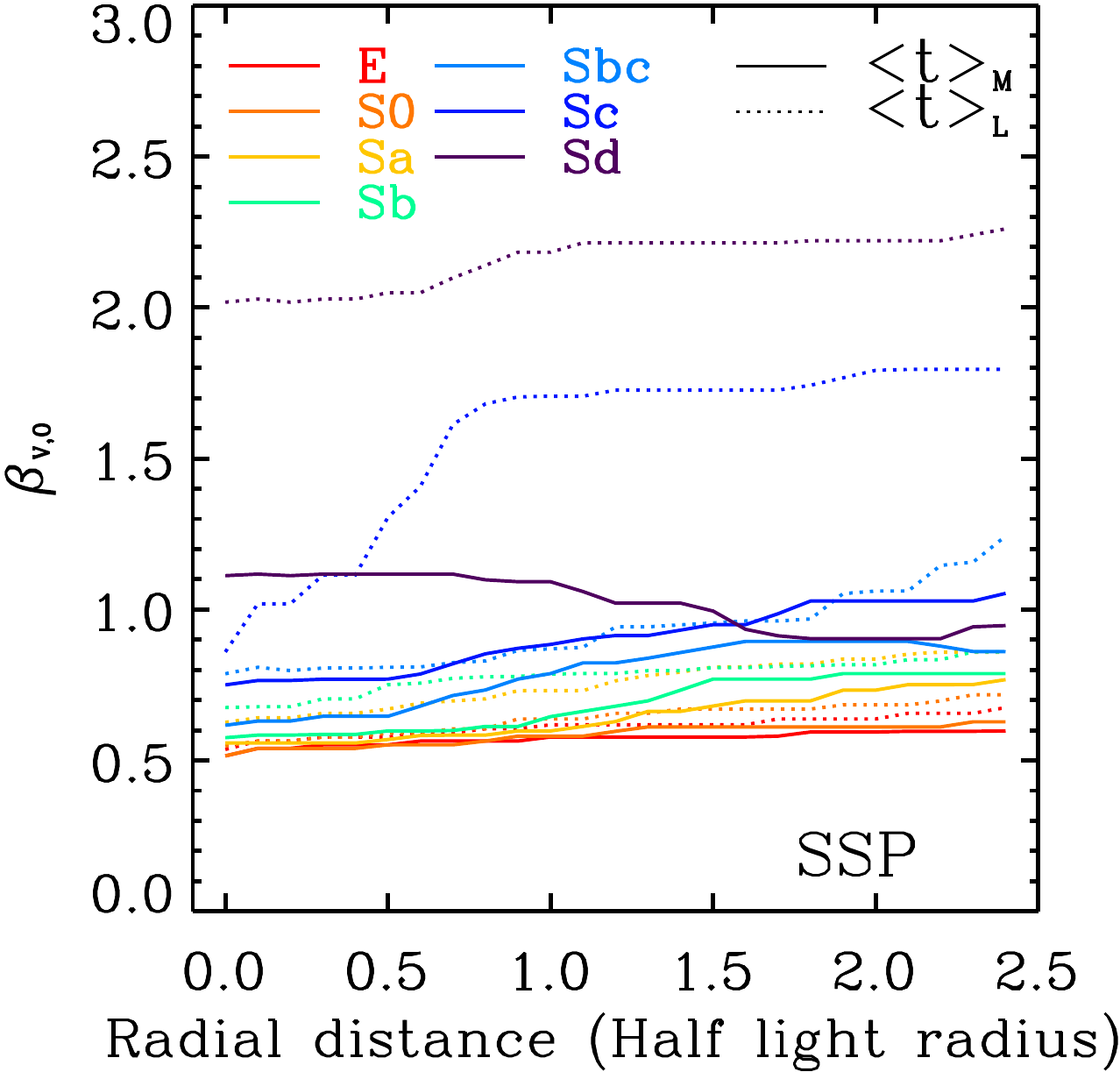} \ \ 
\parbox[b][0.345\txh][t]{0.015\txw}{\textbf{(b)}}\hspace*{-0.015\txw}
  \includegraphics[width=0.48\txw]{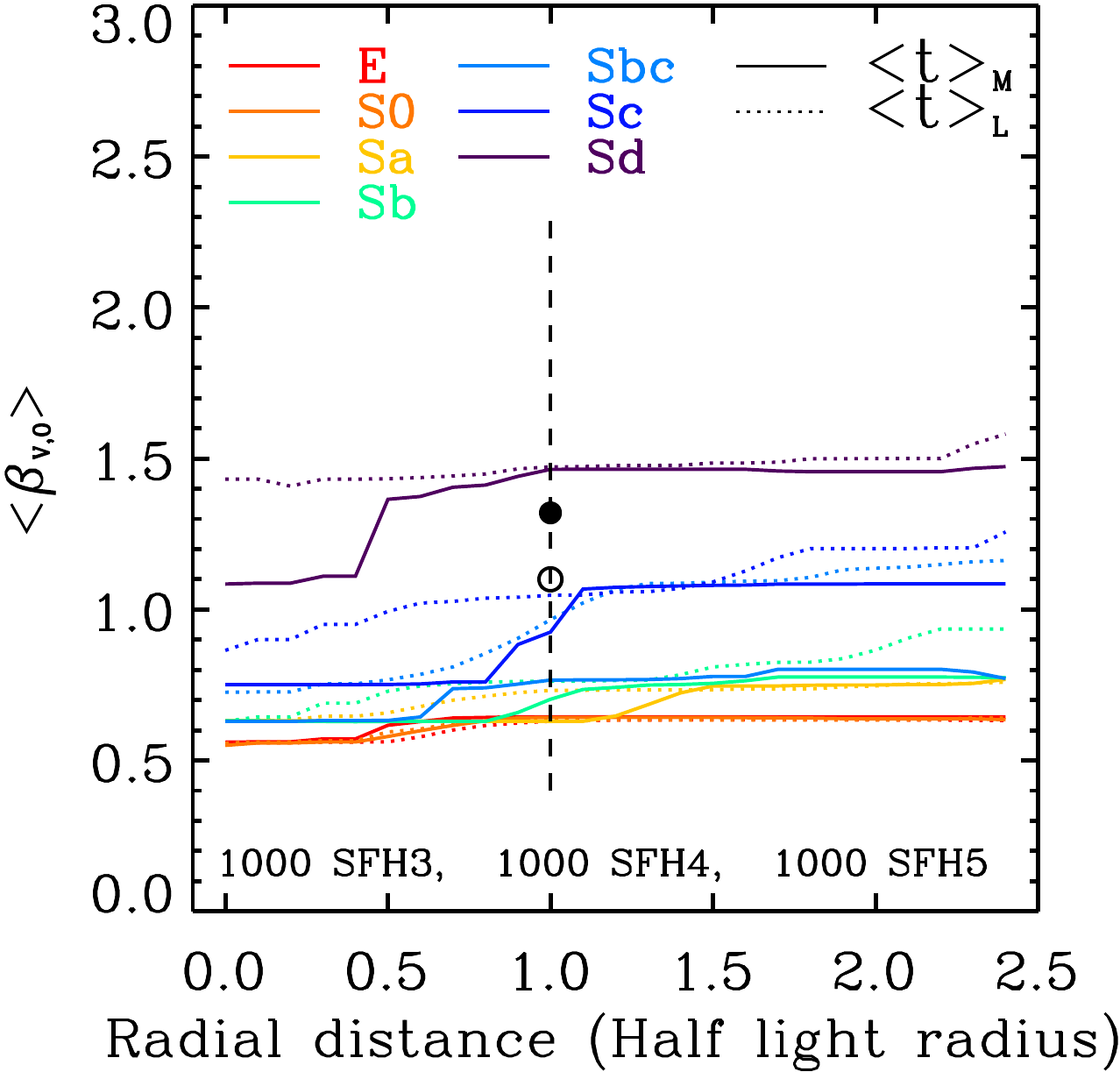}
}
\caption{\noindent\small
\betaVzero\ as a function of radius for galaxies of different Hubble type. (a)
\betaVzero\ profiles derived from an array of SSP SEDs, and (b) mean
$\langle$\betaVzero$\rangle$ profiles resulting from \norsfe\ arrays of
stochastic SFH SEDs that we randomly generated using mass- (solid) and
light-weighted (dotted) age and mass-weighted metallicity values from
Gonz{\'a}lez Delgado \etal\ (2015; GD15). The solid (open) black circle
indicates the value adopted by T09 for younger (older) stellar populations
within NGC\,959 (Sdm). For both SSP and stochastic SFH SEDs, we plot
boxcar-smoothed radial profiles of \betaVzero\ and of average \betaVzero\
values, respectively (see text). The \betaVzero\ values vary little with radius
for early-type galaxies for both SSPs and stochastic SFHs. For late-type
galaxies, the presence of a bulge or older population would reduce the
\betaVzero\ values within $R_e$.\label{fig:figure12}}
\end{figure*}
%%%%%%%%%%%%%%%%%%%%%%%%%%%%%%%%%%%%%%%%%%%%%%%%%%%%%%%%%%%%%%%%%%%%%%%%%%%%%%%

\subsection{\betalamzero\ for Galaxies of Different Hubble Type at
	$z$\,$\simeq$\,0}

Nearby galaxies have been characterized by their morphologies and studied
separately ever since Hubble (1926) first classified them based on their
morphologies (\eg\ de Vaucouleurs 1959; Odewahn 1995; van den Bergh 1998; Buta
\etal\ 2007). Morphology can be estimated from single-band imaging data, and is
known to closely correlate with both structural properties of galaxies and
their SFHs (\eg\ Humason \etal\ 1956; de~Jong 1996; Jansen \etal\ 2000ab;
Windhorst \etal\ 2002; Conselice \etal\ 2004; Taylor \etal\ 2005; Taylor-Mager
\etal\ 2007; Hoyos \etal\ 2016). Examining the relation between \betalamzero\
values and the morphology of galaxies will therefore be useful.

To address this, we first generate multiple arrays of SEDs with different SFH
criteria: \norsfee\ SFH3-ETG, \norsfes\ SFH4-Spiral, and \norsfel\ SFH5-LTG
(see \S2.2). GD15 presented integral field spectroscopy of 300 nearby galaxies
of various Hubble types, and derived radial profiles of stellar population
properties such as age, metallicity, and \AV. We adopt the mean of their
weighted age and metallicity values for different Hubble types as a function of
galactic radii. For spiral galaxies, we adopted the results of their face-on
sample rather than edge-on sample, to minimize the effect of overlapping
stellar populations and high mid-plane extinction. We then derive \betaVzero\
profiles for different Hubble types by comparing age and metallicity values of
GD15 and our SEDs.
  %
%%%%%%%%%%%%%%%%%%%%%%%%%%%%%%%%%%  TABLE 5  %%%%%%%%%%%%%%%%%%%%%%%%%%%%%%%%%%
%\placetable{5}
\noindent\begin{table}[b!]
\centering
\caption{\small \betaVzero\ values at the half-light radii, $\mathrm{R}_e$, of
galaxies of different Hubble type, assuming either SSPs or CSPs with stochastic
SFHs. The indicated ranges of \betaVzero\ for each model are derived from the
uncertainties in the $\langle t\rangle_{\mathrm{L}}$, $\langle
t\rangle_{\mathrm{M}}$, and $\langle Z\rangle_{\mathrm{M}}$ values. For each
type of SFH, we compare mass- and luminosity-weighted \betaVzero\ values and
ranges.\label{tab:table5}}
\setlength{\tabcolsep}{4.3pt}
\begin{tabular}{cccccc}
\toprule
Hubble& \multicolumn{2}{c}{SSP} & \multicolumn{3}{c}{Stochastic SFHs} \\
        \cmidrule(r){2-3}  \cmidrule(r){4-6}
Type  & $\langle t\rangle_{\mathrm{L}}$ & $\langle t\rangle_{\mathrm{M}}$ &
	$\langle t\rangle_{\mathrm{L}}$ & $\langle t\rangle_{\mathrm{M}}$ &
	Family \\[2pt]
\midrule
E0  & 0.60$^{+0.16}_{-0.14}$      & 0.57$^{+0.14}_{-0.13}$         &
	0.64$^{+0.04}_{-0.07}$      & 0.65$^{+0.03}_{-0.07}$ & SFH3 \\
S0  & 0.63$^{+0.14}_{-0.13}$            & 0.58$^{+0.12}_{-0.11}$      &
	0.64$^{+0.03}_{-0.07}$      & 0.64$^{+0.03}_{-0.07}$ & SFH3 \\
Sa  & 0.74$^{+0.17}_{-0.17}$ & 0.61$^{+0.17}_{-0.16}$ &
	0.74$^{+0.14}_{-0.11}$ & 0.65$^{+0.14}_{-0.03}$ & SFH4 \\
Sb  & 0.79$^{+0.08}_{-0.09}$      & 0.65$^{+0.11}_{-0.10}$    &
	0.76$^{+0.05}_{-0.03}$ & 0.72$^{+0.06}_{-0.09}$ & SFH4 \\
Sbc & 0.93$^{+0.50}_{-0.13}$ & 0.78$^{+0.17}_{-0.17}$ &
	1.02$^{+0.29}_{-0.27}$ & 0.77$^{+0.26}_{-0.12}$ & SFH4 \\
Sc  & 1.65$^{+0.49}_{-0.71}$    & 0.88$^{+0.22}_{-0.18}$    &
	1.09$^{+0.37}_{-0.31}$ & 1.06$^{+0.30}_{-0.31}$ & SFH5 \\
Sd  & 2.17$^{+0.38}_{-0.62}$ & 1.03$^{+0.30}_{-0.17}$ &
	1.43$^{+0.44}_{-0.36}$ & 1.39$^{+0.43}_{-0.34}$ & SFH5 \\
\bottomrule
\end{tabular}
\end{table}
%%%%%%%%%%%%%%%%%%%%%%%%%%%%%%%%%%%%%%%%%%%%%%%%%%%%%%%%%%%%%%%%%%%%%%%%%%%%%%%%

At each discrete Age step in Table~\ref{tab:tableA1} for cosmic times later
than 545\,Myr ($z$\,$<$\,9; see Figure~\ref{fig:figure3}), we calculated
mass-weighted ages, $\langle t\rangle$$_{\mathrm{M}}$, light-weighted ages,
$\langle t\rangle$$_{\mathrm{L}}$, and mass-weighted metallicities, $\langle
Z\rangle$$_{\mathrm{M}}$, for consistency with GD15 as
\begin{eqnarray}
\langle t \rangle_{\mathrm{L}} &=& {\sum_{t'=t_0}^{t} \log t'\mathcal{F} (t')
\psi(t') \delta t'} \Big/{\sum_{t'=t_0}^{t} \mathcal{F}(t')\,\psi(t')\,\delta
t'},\quad\ \\	
\langle t \rangle_{\mathrm{M}} &=& {\sum_{t'=t_0}^{t} \log t'\,\psi(t')\,\delta
t'} \Big/ {\sum_{t'=t_0}^{t} \psi(t')\,\delta t'},\\		
\langle Z \rangle_{\mathrm{M}} &=& {\sum_{t'=t_0}^{t} \log
Z(t',Z_0)\,\psi(t')\,\delta t'} \Big/ {\sum_{t'=t_0}^{t} \,\psi(t')\,\delta t'}
\end{eqnarray}

\noindent where $\mathcal{F}$ is the flux at 5635\,\AA, and $\psi(t)$ denotes
the SFR, $Z_0$ is the metallicity at $z$\,=\,0, and metallicity is a function
of cosmic time and $Z_0$, \ie\ $Z(t',Z_0)$.

Figures~\ref{fig:figure12}(a) and (b) show smoothed median radial profiles of
\betaVzero\ for galaxies modeled by SSP and stochastic SFHs, respectively, for
different Hubble types. For stochastic SFHs, we used the SEDs of SFH3-ETG for E
and S0, SFH4-Spiral for Sa--Sbc and SFH5-LTG for Sc and Sd Hubble types,
respectively. $\langle$\betaVzero$\rangle$ in Figure~\ref{fig:figure12}(b) is
the average \betaV\ of 1000 randomly generated SFHs for each stochastic SFH
family. In both panels, each profile was boxcar smoothed using a radial filter
with a width corresponding to one half-light radius. Galaxies of E and S0 type
have an almost constant \betaVzero\ value with increasing galactic radius for
both SSPs and stochastic SFHs. Later types have higher and more fluctuating
\betaVzero\ values and show a stronger dependence on radius.
Table~\ref{tab:table5} lists representative mass- and light-weighted
\betaVzero\ values at the half-light radii ($\mathrm{R}_e$) for each Hubble
type for both SSPs and CSPs resulting from stochastic SFHs. We derive
uncertainties in the \betaVzero\ values from the 1\,$\sigma$ ranges presented
in GD15 (their Figures~8, 13, and 18) by randomly varying the age and
metallicity values in our models within these allowed ranges. The total range
of luminosity-weighted \betaVzero\ values for nearby galaxies thus derived is
0.57--1.87 (the $\langle t\rangle_{\mathrm{L}}$ column for stochastic SFHs).
For a galaxy of known morphological type, however, the likely range in
\betaVzero\ is only $\sim$$\pm$10\% with respect to the mean value for E, S0,
Sa, and Sb galaxies, although late-type spiral and Magellanic-type galaxies
display a wider $\pm$30--50\% range. Figure~\ref{fig:figure12}(b) also shows
that the mean \betaVzero\ values beyond $\mathrm{R}_e$ in the case of
(observed) light-weighted ages are similar, but systematically higher (by up to
$\sim$8--20\% at 2.5\,$\mathrm{R}_e$) than the (intrinsic) mass-weighted ages.

\section{Discussion}

We have generalized the \betaV\ method to a \betalam\ method by placing the
method on a more robust footing that does not require manual estimation of the
intrinsic flux ratios from pixel histograms and that provides evaluation of the
associated uncertainties for a large number of optical--near-IR and mid-IR
filters. Stars with young ages and low metallicities emit light with SEDs that
have high \betalamzero\ values, and as stars age and as stars become more
metal-rich, the \betalamzero\ values decrease and stabilize. T09 also observed
this trend in their 2D map of \betaVzero\ values (their Figure~2), which was
based on SSPs and the model SEDs from Anders \& Fritze-von~Alvensleben (2003).

\subsection{Application to Discrete Sets of Filters at 0\,$<$\,$z$\,$<$\,1.9}

While using near-IR filters results in a narrower \betalamzero\ range over the
redshift range 0\,$<$\,$z$\,$<$\,1 (see Figure~\ref{fig:figure11}), after
normalization, the relative dispersion in the simulated \Alam\ values is
comparable in the optical and near-IR filters. One should be cautious about
using near-IR bandpasses at low redshift, however, since the offset due to the
residual extinction in the mid-IR reference filter becomes more noticeable (see
Figure~\ref{fig:figure11}, bottom panel). Nonetheless, \HST/WFC3 near-IR
filters may be used to sample visible rest-frame wavelengths up to
$z$\,$\sim$\,1.9. In that case, one should use the filter nearest the
corresponding rest-frame wavelength when applying the \betalam\ method. For
instance, at $z$\,$\simeq$\,1.9, the \HST/WFC3\,IR F160W filter would sample
rest-frame $V$, and the \JWST/MIRI F1000W bandpass would sample rest-frame
3.5\,\micron, so one can use the \betaVzero\ values presented in this paper and
on our website\footnotemark[2].

Although the value of \betaVzero\ will depend on the detailed shape of the
throughput curves of the filters sampling the rest-frame $V$ and 3.5\,\micron\
light, Figs.~\ref{fig:figure7} and \ref{fig:figure8} show that such dependence
for even quite mismatched filters (\eg\ $L'$ and \WISE\ $W$\textsl{1}) affects
\betaVzero\ at the $\lesssim$20\% level. For bandpasses that are better matched
to the $L$ band, such as \Spitzer/IRAC $I$\textsl{1} and \JWST/NIRCam F356W,
the differences tend to be at the $\lesssim$5\% level, except for very young
($<$\,10\,Myr) stellar ages and for high metallicities ($Z$\,$>$\,0.02). A
similar caution applies at redshifts where the bandpass sampling rest-frame $V$
light differs very much from that of the $V$ filter at $z$\,$\sim$\,0.
Nonetheless, the dependence of the accuracy of the \betalam\ technique on the
choice of filters in both the optical--near-IR and mid-IR is weak.

\subsection{Application to Galaxy Samples Using Realistic SFHs}

We have extended our models from SSP to more complex SFHs. In real galaxies,
the observed \betalam\ flux ratios are also affected by spatial resolution, as
light from a broad region within a galaxy would result from a combination of
various stellar populations, each resulting from different complex stochastic
SFHs. For example, Galliano \etal\ (2011) obtained a $\sim$50\% difference in
inferred dust mass when they used integrated SEDs with different spatial
resolutions. The sharp features in a map of \betalamzero\ values of SSPs in
Figure~\ref{fig:figure6} at young ($\lesssim$100\,Myr) ages are smeared out
when a galaxy has a more complex SFH, such as an exponentially declining SFH.
In addition, we adopt five families of stochastic SFHs that are composed of
multiple exponentially declining star formation episodes, and probe variations
of \betalamzero\ values over cosmic history. For a minimally constrained
redshift range, 0\,$<$\,$z$\,$<$\,1, the variation in \betalam\ is dominated by
SFH5, representing late-type galaxies. The other SFH families show more modest
variations for a given metallicity and choice of optical--near-IR filter (see
Figs.~\ref{fig:figure10} and \ref{fig:figure11}). Metallicity evolution is
taken into account and makes a noticeable difference (Figs.~\ref{fig:figure9}
and \ref{fig:figure10}(b)), predicting higher \betalamzero\ values than in the
no-evolution case, especially at higher redshifts and for the
higher-metallicity stellar populations at those redshifts.

We first consider the applicability of the \betalam\ method for a galaxy sample
with a large effectively unconstrained redshift range
(0\,$\lesssim$\,$z$\,$\lesssim$\,4). The \betaVzero\ values could span 0.6--4.7
in that case (see Figure~\ref{fig:figure10} and \S3.2). For the minimally
constrained 0\,$<$\,z\,$<$\,1 redshift range, the scatter in recovered \Alam\
is $\sim$0.2\,mag in $V$ and decreases with increasing wavelength, but varies
little with the amount of input extinction imposed (Figure~\ref{fig:figure11},
middle panel). The normalized difference of the recovered and input extinction,
$(\Alam-A_{\lambda,\mathrm{inp}})/A_{\lambda,\mathrm{inp}}$, on the other hand,
is $\sim$23\% and $\sim$16\% for \AV\,=\,0.8 and \AV\,=\,3.2\,mag, respectively
(see bottom panel of Fig~\ref{fig:figure11}). Note, however, that this reflects
what one would expect for individual galaxies (or regions therein). The mean
difference of the recovered and input extinction values is close to 0.0,
suggesting that the \betalam\ method can accurately recover the mean extinction
for a larger sample of galaxies spanning a range of redshifts, SFHs, and
metallicities, and also of a population of galaxies of a given type within a
narrow redshift range. The highly attenuated case (\AV\,=\,3.2) has a smaller
scatter, because the extinction dominates the shape of the SED more than any
other factors, like stellar age or metallicity. For normal (\AV\,$\lesssim$\,1)
galaxies, we recommend using extra information, such as galaxy size and
magnitudes, or if available, a photometric redshift estimate (or a
spectroscopic redshift), for selecting the corresponding \betalamzero\ value.

\subsection{Dependence of \betalamzero\ on Hubble Type and Weighting}

\begin{figure}[t!]
\centerline{
  \includegraphics[width=0.475\txw]{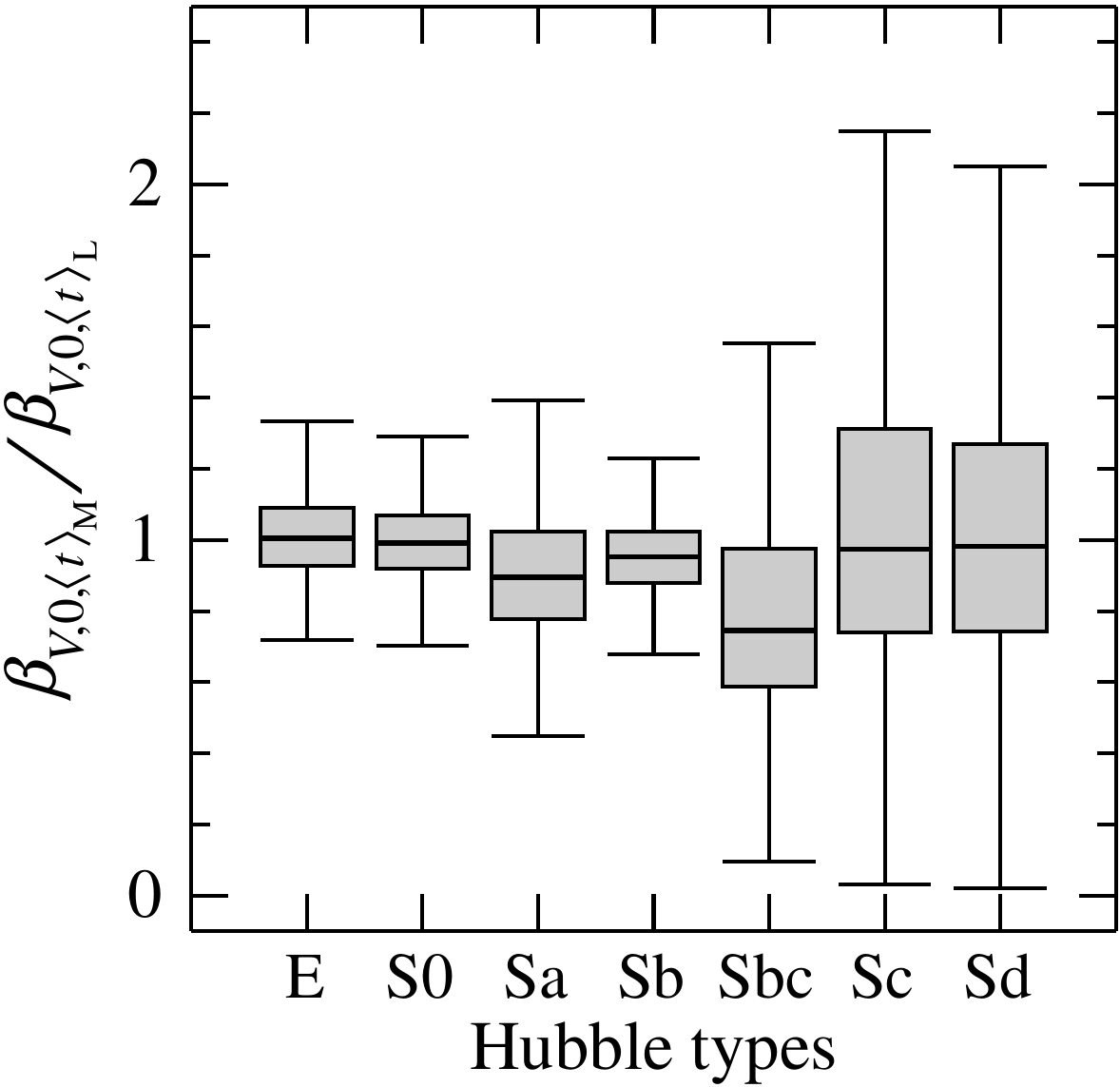}
}
\caption{\noindent\small
Ratio of \betaVzero\ values for mass- and light-weighted ages for different
Hubble types. For each Hubble type, we randomly generated 1000 \betaVzero\
values each for both mass- and light-weighted ages, adopting the mean and
1\,$\sigma$ values at $R_e$ listed in Table~\ref{tab:table5}. The horizontal
lines inside the boxes represent the median ratios. The gray boxes indicate the
quartile range, while the error bars contain 99.7\% of the distribution. The
weighting method does not affect the inferred \betaVzero\ value for a given
Hubble type in a systematic manner.\label{fig:figure13}}
\end{figure}

Because one can classify the morphology of a galaxy with a single-band image
(with the caveat of a slight rest-frame wavelength dependence of the morphology
for different Hubble types, as found by Windhorst \etal\ 2002), studying the
relation between galactic morphology and \betalamzero\ values can be useful,
since the \betalam\ method is specifically designed as a dust-correction
technique when a very limited number of filters are available. We used the
SFH3--5 criteria (see \S2.2) to generate multiple arrays of SEDs combined with
the stellar population parameters from GD15 to obtain the \betaVzero\ values
for different Hubble types (see Figure~\ref{fig:figure12}(b)). For a given
galaxy type, \betaVzero\ values vary little as a function of galactic radius.
Later Hubble types typically have higher \betaVzero\ values (see also
Table.~\ref{tab:table5}), as their stellar populations tend to be younger and
metallicities lower than earlier types (see Figure~17 in GD15). We also
investigate the effect of different weighting methods. The ratios between
\betaVzero\ values resulting from using mass- and light-weighted ages are shown
in Figure~\ref{fig:figure13} as a function of Hubble type. We find no strong
dependence of the median ratios of \betaVzero\ values with different weighting
methods on Hubble type. Also indicated in Figure~\ref{fig:figure13} are the
quartile ranges and the $\pm$3\,$\sigma$ range of these ratios. Light-weighted
\betaVzero\ values are unlikely to differ by more than 10\% (30\%) from
mass-weighted ones for early-type (late-type) galaxies.
   %
%%%%%%%%%%%%%%%%%%%%%%%%%%%%%%%%%  FIGURE 13  %%%%%%%%%%%%%%%%%%%%%%%%%%%%%%%%
\noindent
%%%%%%%%%%%%%%%%%%%%%%%%%%%%%%%%%%%%%%%%%%%%%%%%%%%%%%%%%%%%%%%%%%%%%%%%%%%%%%%%
   %
%%%%%%%%%%%%%%%%%%%%%%%%%%%%%%%%%  FIGURE 14  %%%%%%%%%%%%%%%%%%%%%%%%%%%%%%%%%
\noindent\begin{figure}[t!]
\centerline{
  \includegraphics[width=0.48\txw]{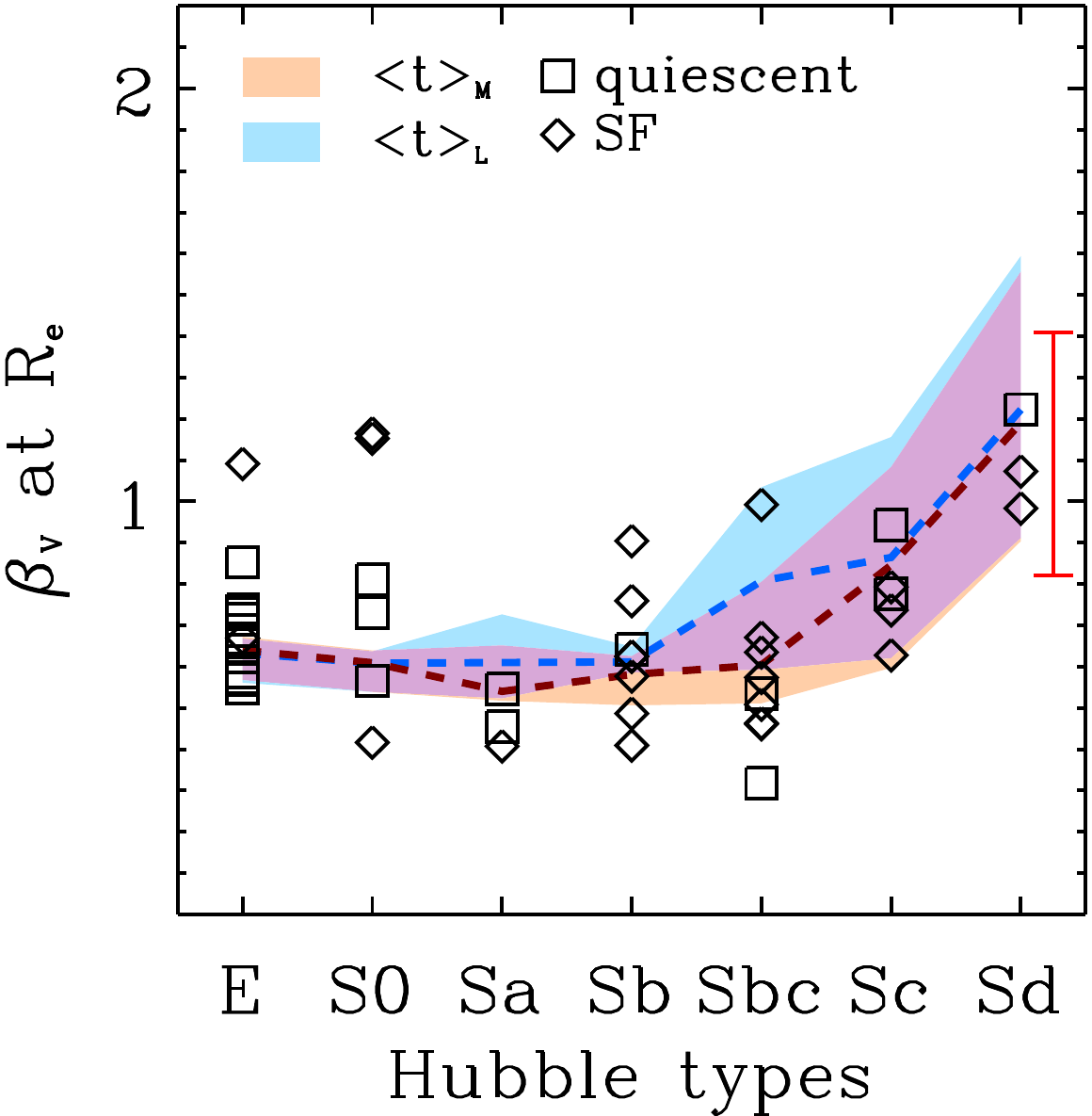}
}
\caption{\noindent\small
\betaV\ values at the half-light radii, $R_e$, as a function of Hubble type
(dashed lines) using mass- (dark brown), or light-weighted (blue) ages and \AV\
values. The shaded regions are the $\pm$1\,$\sigma$ ranges of \betaV\ values.
We overplot the \betaV\ values of nearby galaxies from Brown \etal\ (2014). The
dashed curves represent median trends. The red error bar indicates the range of
\betaV\ values observed by T09 in NGC\,959 (Sdm).\label{fig:figure14}}
\end{figure}
%%%%%%%%%%%%%%%%%%%%%%%%%%%%%%%%%%%%%%%%%%%%%%%%%%%%%%%%%%%%%%%%%%%%%%%%%%%%%%%

To test the reliability of the \betalam\ method, in Figure~\ref{fig:figure14}
we compare the inferred \betaV\ values with observed \betaV\ values for nearby
galaxies. We used $\langle$\betaVzero$\rangle$ values from
Figure~\ref{fig:figure12}(b) at the half-light radius and the allowed
1\,$\sigma$ range therein computed by varying ages and metallicities within the
1\,$\sigma$ observed ranges of GD15 (see their Figs.~8, 11, and 18), coupled
with the mean \AV\ values of the corresponding Hubble types at the half-light
radii from GD15 ($\langle$\AV$\rangle$=0.01, 0.06, 0.22, 0.25, 0.27, 0.26, and
0.18 for E, S0, Sa, Sb, Sbc, Sc, and Sd galaxies) to derive the predicted
ranges for the \betaV\ values one would observe for each of the Hubble types.
The dark brown (blue) dashed line and the orange (blue) shaded region indicate
the median and 1\,$\sigma$ range of the \betaV\ values when using mass-
(light-) weighted ages, and the violet region represents the region of overlap.
The \betaV\ values of 23 nearby ($z$\,$\lesssim$\,0.05) quiescent (limited star
formation) and 28 normally star-forming galaxies from Brown \etal\ (2014) are
overplotted. Galaxies with ``peculiar'' morphologies, and galaxies with
``SF/AGN'', or ``AGN'' BPT classes are excluded. Brown \etal\ (2014) combined
multiple spectra and broadband data of nearby galaxies, and carefully performed
aperture corrections in order to construct a template library. We used their
aperture-corrected SDSS $g$, $r$ band and the \Spitzer/IRAC Channel~1
($I$\textsl{1}) magnitudes to derive the \betaV\ values. The SDSS $g$ and $r$
band magnitudes are converted into $V$-band magnitudes using the transformation
equation from Jester \etal\ (2005). We take the \Spitzer\ $I$\textsl{1}
magnitudes as proxy for $L$-band magnitudes, since Figure~\ref{fig:figure7} and
\ref{fig:figure8} show that the resulting \betaVzero\ are at most
$\lesssim$10\% higher and show little structure. Our model and observation
agree well, except in a few cases for star forming galaxies with Hubble type E
and S0, which seem to be outliers (see \S\,2.1.4 of Brown \etal\ 2014). Last,
we note that the 0.82--1.41 range of \betaV\ values observed by T09 for Sdm
galaxy NGC\,959 is in good agreement with the trend in values inferred for the
galaxy sample of Brown \etal\ (2014), as well as with the range predicted by
our models in Figure~\ref{fig:figure14}, and that their empirical values
adopted for the intrinsic \betaVzero\ ratios are consistent with those
predicted in the present study (\eg\ Figure~\ref{fig:figure12}(b) and
Table~\ref{tab:table5} (Sd)). We conclude that our model \betalamzero\ values
agree with the observations for at least nearby ($z$\,$\simeq$\,$0$) galaxies.
  %
%%%%%%%%%%%%%%%%%%%%%%%%%%%%%%%%%  FIGURE 15  %%%%%%%%%%%%%%%%%%%%%%%%%%%%%%%%%
\noindent\begin{figure*}[t]
\centerline{
\parbox[b][0.355\txh][t]{0.015\txw}{\textbf{(a)}}\hspace*{-0.015\txw} 
  \includegraphics[width=0.475\txw]{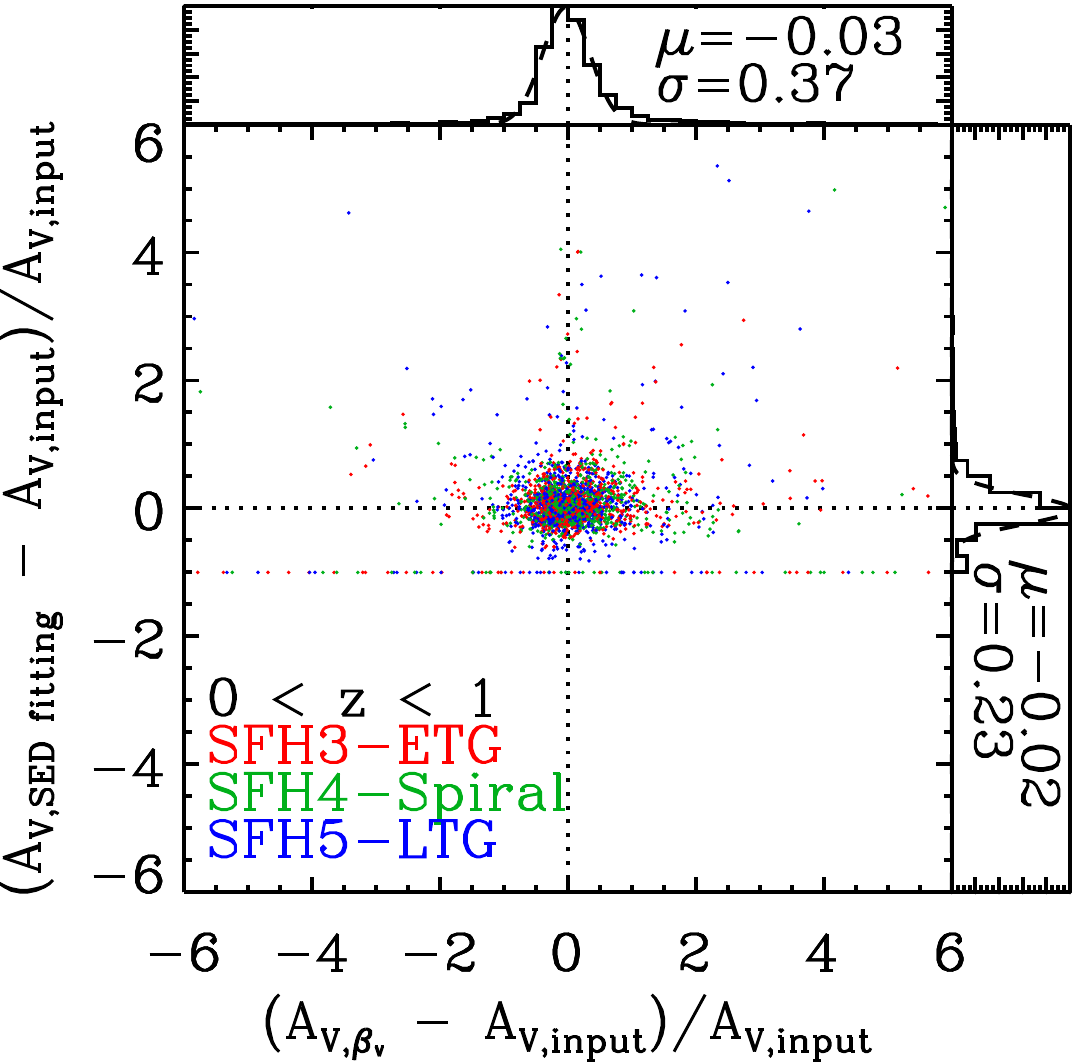}\hspace*{0.025\txw} 
\parbox[b][0.355\txh][t]{0.015\txw}{\textbf{(b)}}\hspace*{-0.015\txw}
  \includegraphics[width=0.475\txw]{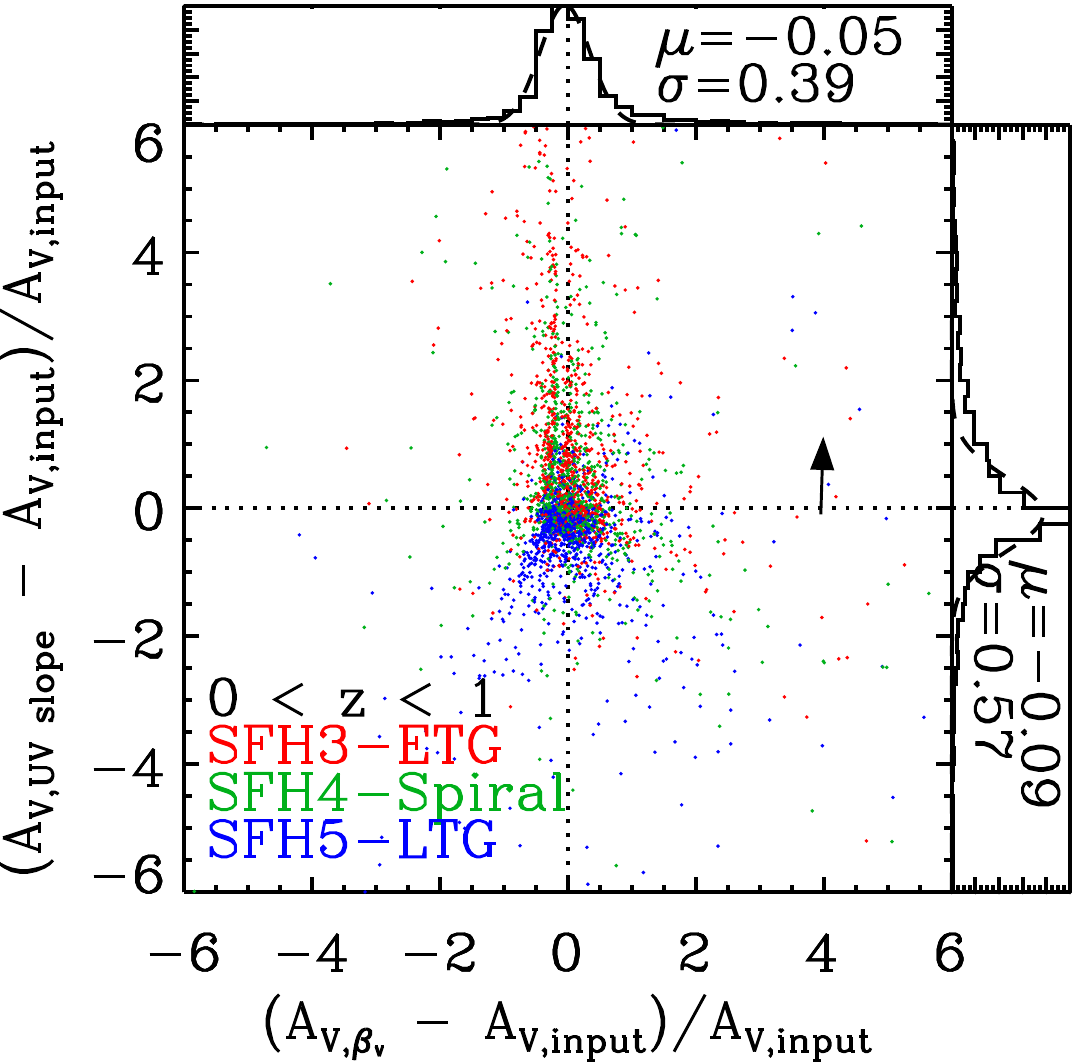}
}
\caption{\noindent\small
Comparison of the \betaV\ method to other dust-correction methods. (a)
Comparison of the normalized difference between recovered and input extinction
values in $V$ for both the SED-fitting and \betaV\ methods. The \betaV\ method
shows a comparable average offset and scatter around the true value of \AV\ to
the SED-fitting method, although it requires less information. To derive \AV\
for our families of stochastic SFHs, we used the 0\,$<$\,$z$\,$<$\,1 average
values of \betaVzero\ of 1.40, 1.48, and 1.78 for SFH3, SFH4, and SFH5,
respectively. (b) As (a) for both the UV-slope and \betaV\ methods. The \AV\
values recovered by the UV-slope method are systematically slightly lower than
the \AVinput\ values ($\mu$\,=\,$-$0.10\,mag; for a MW/LMC extinction law as in
Calzetti \etal 1994), and the scatter is larger than for the \betaV\ method
($\sigma$\,=\,0.60 vs.\ 0.38\,mag). The black arrow shows how the offset would
change if we adopted a Calzetti \etal\ (2000) extinction law
($\mu$\,=\,$+$1.10).\label{fig:figure15}}
\end{figure*}
%%%%%%%%%%%%%%%%%%%%%%%%%%%%%%%%%%%%%%%%%%%%%%%%%%%%%%%%%%%%%%%%%%%%%%%%%%%%%%%
   %
%%%%%%%%%%%%%%%%%%%%%%%%%%%%%%%%%  FIGURE 16  %%%%%%%%%%%%%%%%%%%%%%%%%%%%%%%%
\noindent\begin{figure*}[t]
\centerline{
\parbox[b][0.455\txh][t]{0.015\txw}{\textbf{(a)}}\hspace*{-0.015\txw} 
  \includegraphics[width=0.475\txw]{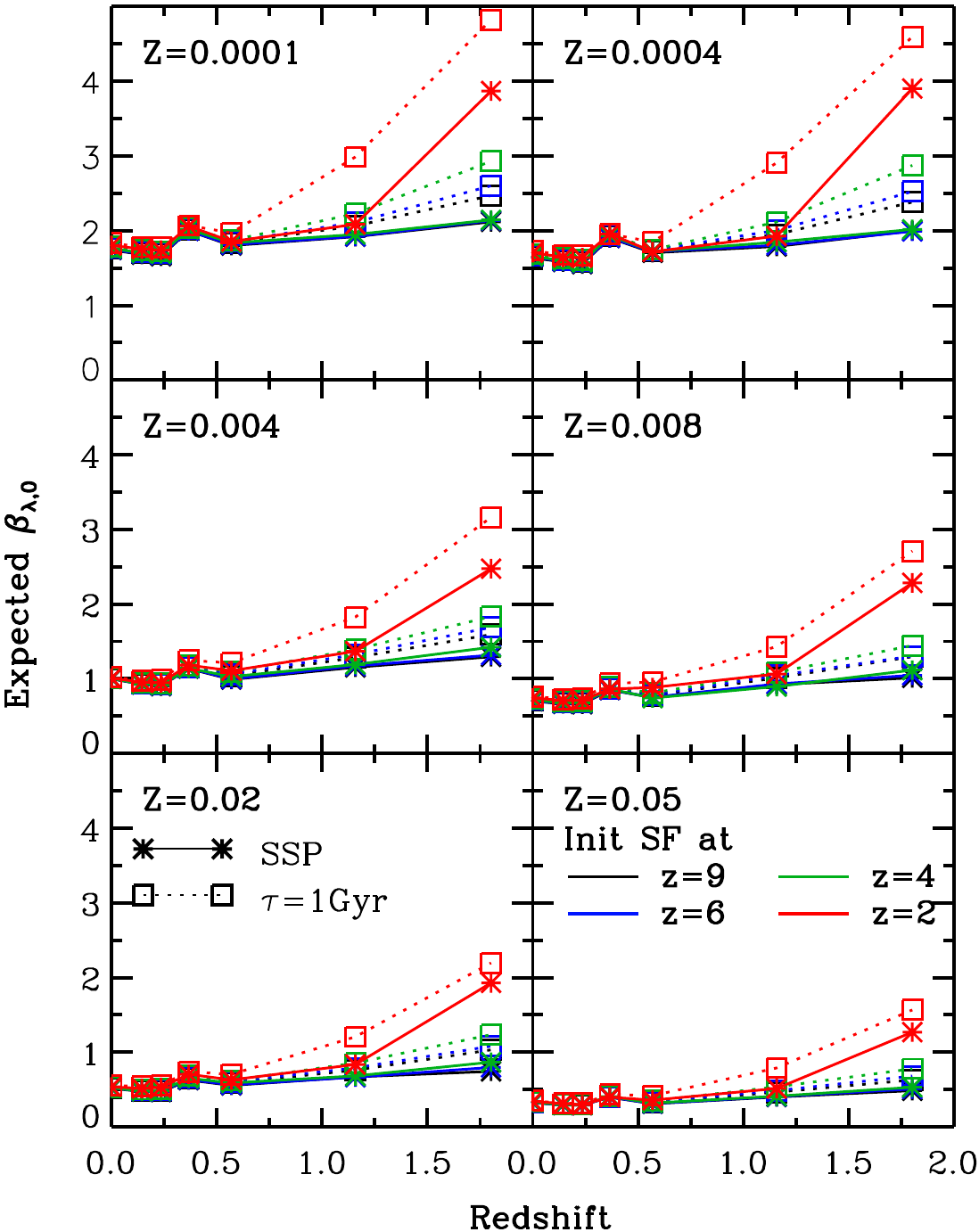}\hspace*{0.025\txw}
\parbox[b][0.455\txh][t]{0.015\txw}{\textbf{(b)}}\hspace*{-0.015\txw}
  \includegraphics[width=0.475\txw]{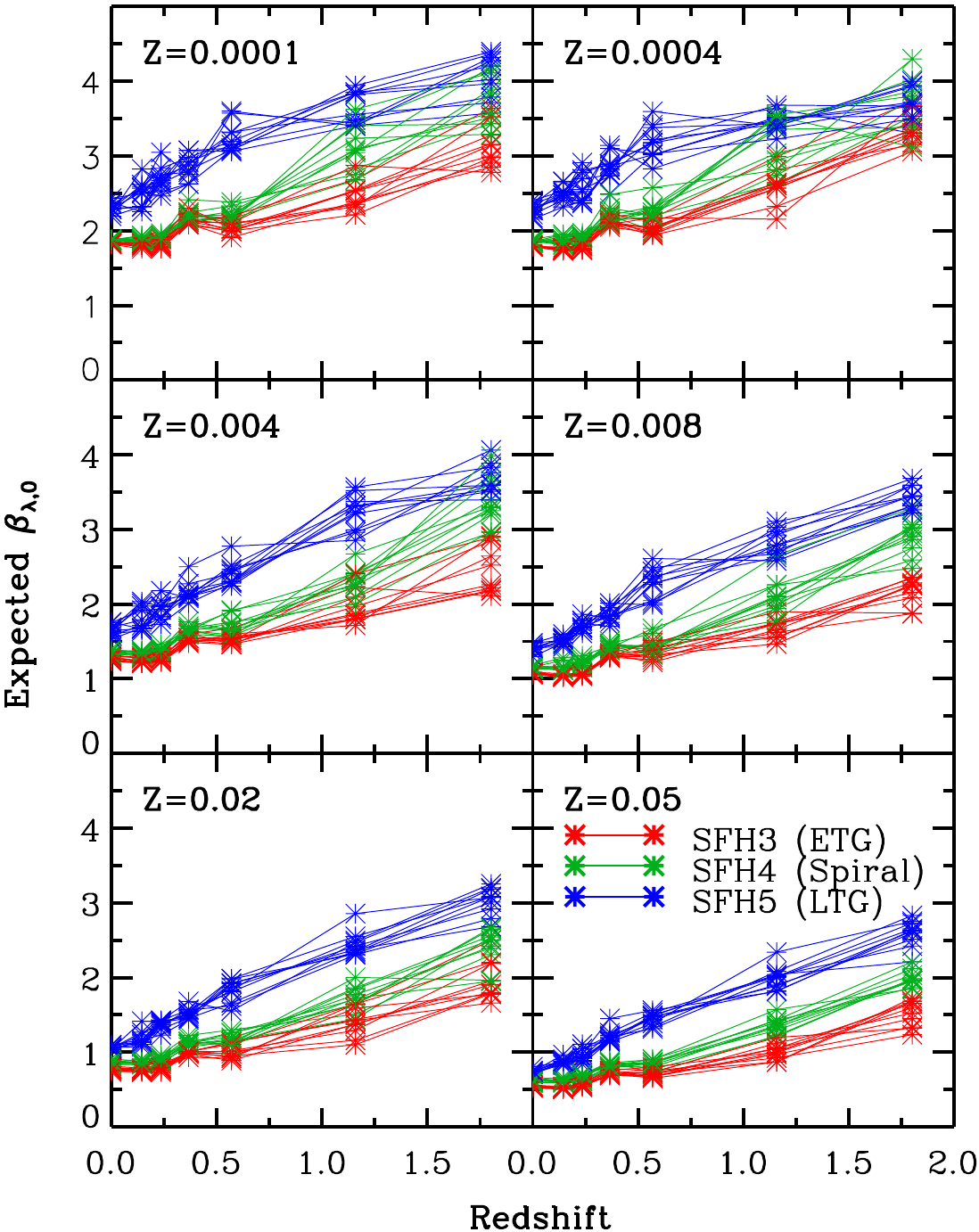}
}
\caption{\noindent\small
Expected intrinsic flux ratios \betalamzero\ as a function of redshift for six
metallicities and various SFHs, for redshifts where well-matched pairs of
optical or near-IR \HST\ and mid-IR \JWST\ filters correspond to rest-frame $V$
and 3.5\,\micron, respectively. (a) \betalamzero\ for SSPs (solid) and
exponentially declining SFHs with an $e$-folding time of 1\,Gyr. Tracks are
color-coded (as indicated in the lower right panel) according to the redshift
of onset of the instantaneous (SSP) or extended star formation. At a fixed
metallicity value and starburst onsets well before $z$\,=\,2, \betalamzero\
remains nearly constant over the entire 0\,$\le$\,$z$\,$\le$\,2 redshift range.
(b) The same as (a) for 10 examples of each of our three stochastic SFH
families. Except for the two very lowest metallicities, \betalamzero\ increases
with redshift in a nearly linear fashion, with little overlap between the
tracks for the three SFH families.\label{fig:figure16}}
\end{figure*}
%%%%%%%%%%%%%%%%%%%%%%%%%%%%%%%%%%%%%%%%%%%%%%%%%%%%%%%%%%%%%%%%%%%%%%%%%%%%%%%%

\subsection{Comparison with Other Methods}

We next compare the \betaV\ method to other established methods of correcting
galaxy SEDs for attenuation by cosmic dust. In Figure~\ref{fig:figure15}(a), we
compare the relative differences between recovered and input extinction values
(normalized to the input value, \AVinput) for (multiband) SED fitting and the
\betaV\ method. When we impose minimal constraints on the allowed redshift
range (0\,$<$\,$z$\,$<$\,1), the SED-fitting method provides estimates that are
closer to the true input value, \AVinput, on average ($\mu$\,=\,$-$0.02 vs.\
$-$0.03\,mag), and with a smaller scatter ($\sigma$\,=\,0.23 vs.\ 0.37\,mag).
The small fraction of data points for the SED-fitting method at a normalized
difference of exactly $-$1 result where the best fit was for \AV\,=\,0\,mag
(out of the finite set of discrete extinction values), when any non-zero
extinction was imposed. These data points do not significantly broaden the
distribution. The \betaV\ method shows a somewhat larger scatter, but not by
much, considering that the \betaV\ method requires significantly less
information than the SED-fitting method to correct for dust extinction. In
order to make this comparison, we construct a large set of randomly generated
SEDs for stochastic SFHs representing early-type, spiral, and late-type
galaxies (our SFH families SFH3, SFH4, and SFH5; see \S\,2.2), sampled at
random redshifts in the range 0\,$<$\,$z$\,$<$\,1, and with random amounts of
extinction (0\,$\leq$\,\AV\,$\leq$\,2\,mag) applied and characterized by a
randomly selected extinction law (Milky Way/Large Magellanic Cloud; MW/LMC,
SMC, or Calzetti). To apply the \betaV\ method and recover \AV, we assume that
the appropriate SFH family to use would be known \emph{a priori} from the
morphological type of an observed galaxy, and only minimally restrict the
redshift to the same 0\,$<$\,$z$\,$<$\,1 range. The average values of
\betaVzero\ in this redshift range are 1.40, 1.48, and 1.78 for SFH3, SFH4, and
SFH5, respectively. Using Eq.~\ref{eq:alam}, we then infer (recover) for each
model SED an extinction value \AV\ from the appropriate mean \betaVzero\ value
and the observed ($V$\,$-$\,$L$) color (expressed as a flux ratio, \ie\
\betaV). As templates for the SED-fitting method, we used the same set of 724
unique and distinguishable SSP SEDs (see Appendix~A), either used as is, or
incorporated into CSP SEDs for seven exponentially declining SFHs, or into our
suite of stochastic SFHs (\S~2.2) with 300 random realizations for each of the
SFH3, SFH4, and SFH5 families. The comparison should therefore not be biased by
differences in the adopted set of input galaxy SED templates. Using the full
rest-frame 0.4\,$\leq$\,$\lambda$\,$\leq$\,3.75\,\micron\ portion of each of
the SEDs, we perform the SED fit to characterize both stellar population
parameters and recover \AV. This represents a \emph{best-case} scenario, the
results of which are shown in Figure~\ref{fig:figure15}(a). If, as would be the
case for a more realistic galaxy survey, the SED is sampled through multiple
filters, each resulting in a single flux density data point, the comparison is
unlikely to be more favorable for the SED-fitting method, although the allowed
redshift range may be narrowed for galaxies displaying a strong 4000\,\AA\
break.

In Figure~\ref{fig:figure15}(b) we compare the relative differences between
recovered and input extinction values for the UV continuum slope method and the
\betaV\ method. Calzetti \etal\ (1994) studied the effect of dust extinction on
the UV continuum in spectra of local starburst galaxies, and provided a
relation between the UV power-law index $\beta$ and the optical depth $\tau$
(their Eq.~4). For a \emph{best-case} scenario, we use their fitting windows,
sampling the 1268--2580\,\AA\ wavelength range while excluding the interval
that might be affected by a 2175\,\AA-bump (if present), and adopt a MW/LMC
extinction law to match Calzetti \etal\ (1994). The \betaV\ method
underestimates the extinction values in this test by $\sim$5\% (compared to 9\%
for the UV continuum slope method) and shows a smaller scatter than the
UV-slope method: $\sigma$\,=\,0.39\,mag for the \betaV\ method versus 0.57\,mag
for the UV-slope method. Note, however, that if we were to adopt a Calzetti
\etal\ (2000) extinction law instead, this would result in recovered extinction
values for the UV-slope method that are systematically \emph{higher} than the
input values by an amount indicated by the black arrow in
Figure~\ref{fig:figure15}(b), whereas the mean difference $\mu$ would change by
only $+$0.04 for the \betaV\ method. We also note that the UV-slope method
shows a stronger dependence of the recovered \Alam\ on SFH (galaxy type) than
does the \betaV\ method.

Our tests therefore show that the \betaV\ method can indeed be used to infer
\AV\ and correct for dust extinction to a level of fidelity that is comparable
to that of more established methods, with fewer or more readily obtainable
data. In particular, we note that \emph{systematic} offsets in the recovered
extinction values are \emph{no worse} than for these two commonly used methods.
These properties makes the \betaV\ (\betalam) method a prime candidate for
application to large imaging surveys with \JWST\ of fields already observed
with \HST\ in the visible--near-IR.

\subsection{\betalamzero\ for Galaxies Observed with \HST\ and \JWST}

For a $z$\,$\simeq$\,0.3 galaxy, rest-frame optical images from \HST/ACS WFC or
WFC3 UVIS (F606W, ..., F850LP) and rest-frame $\sim$3.5\,\micron\ mid-IR images
from \JWST/NIRCam (F410M, F444W, F480M) will have resolutions of
$\sim$0$\farcs$06--0$\farcs$09 and $\lesssim$0$\farcs$19, resolving regions of
$\lesssim$875\,pc in size. At $z$\,$\sim$\,2, \HST/WFC3 IR F160W and \JWST/MIRI
F1000W sample rest-frame $V$ and 3.5\,\micron, but the resulting resolution of
$\sim$0$\farcs$48 would not allow resolving regions smaller than $\sim$4\,kpc.
The combination of these two space telescopes will make studies of galaxies
over billions of years possible in unprecedented detail. Specifically for this
purpose, we therefore generate model SEDs and calculate \betalamzero\ values
for a set of discrete redshifts at which well-matched \HST\ and \JWST\ filter
pairs sample rest-frame $V$ and 3.5\,\micron\ emission. At redshifts $z$\,=\,0,
0.14, 0.24, and 0.37, well-matched \HST\ and \JWST/NIRCam filter pairs are
(F555W, F356W), (F606W, F410M), (F625W, F444W), and (F775W, F480M),
respectively. At $z$\,=\,0.57, 1.16, and 1.8, suitable \HST\ and \JWST/MIRI
filter pairs are (F814W, F560W), (F110W, F770W), and (F160W, F1000W). In
Figure~\ref{fig:figure16}(a) we show the resulting tracks of \betalamzero\
versus redshift for six different metallicities and both SSP and exponentially
declining SFHs with an $e$-folding time of 1\,Gyr, and do so for four different
redshifts of onset of either the instantaneous or extended star formation. For
SF onset times well before $z$\,$\sim$\,2, the tracks remain nearly constant
over the entire redshift range at a given metallicity, although the mean
\betalamzero\ values decrease systematically with increasing metallicity
($\langle$\betalam$\rangle$$_{z<1.2}$\,$\simeq$\,1.9, 1.8, 1.0, 0.8, 0.5, and
0.4 for $Z$\,=\,0.0001, 0.0004, 0.004, 0.008, 0.02, and 0.05, respectively). We
similarly derive \betalamzero\ for these \HST\ and \JWST\ filter pairs for our
three families of stochastic SFHs. In Figure~\ref{fig:figure16}(b) we show
\betalamzero\ versus redshift for the same six metallicity values for 10
examples of each SFH family. The \betalamzero\ tracks for each SFH family
increase systematically and almost linearly with increasing redshift. The
slopes of those tracks differ for each family, SFH5 (late-type) having the
steepest slope and SFH3 (early-type) having the shallowest. Only at the two
lowest metallicities do the late-type tracks start to deviate from their linear
increase with redshift to overlap the tracks of spiral (at $z$\,$>$\,1) and
even early-type galaxies (for $z$\,$\gtrsim$\,1.5). For higher metallicities
there appears to be little overlap between the tracks for the three SFH
families. With some constraints on the stellar metallicity, SFH, and redshift,
the \betalam\ method can be a powerful tool to correct the spatially resolved
images for large samples of galaxies observed with \HST\ and \JWST\ at moderate
redshifts for the effects of extinction by dust.

\medskip

\section{Summary}

We combined SSP SEDs from the \SB\ and \BC\ codes for young and old stellar
ages, respectively, and generated arrays of CSP SEDs for a large variety of
exponentially declining and stochastic SFHs by stacking SSP SEDs as a function
of the adopted time-dependent SFRs. For our large suite of models with
stochastic SFHs, we take the average metallicity evolution as a function of
cosmic age into account. We calculated the intrinsic flux ratios \betalamzero\
between rest-frame visible--near-IR and rest-frame mid-IR 3.5\,\micron\ on a
grid of six metallicities and 16 ages of stellar populations for 13 different
SFH families, for 29 visible--near-IR filters, and 5 mid-IR filters with
central wavelengths near 3.5\,\micron. We find that taking metallicity
evolution into account results in significantly higher \betalamzero\ values at
higher redshifts and for higher-metallicity stellar populations at those
redshifts. We also provide the range of \betaVzero\ and \betaV\ for different
Hubble types of nearby galaxies, and confirm that our models agree with the
observed \betaV\ values. We demonstrated that the \betaV\ method can infer \AV\
to a level of fidelity that is comparable to that of more established methods.
We conclude that the \betaV\ method and its extension, the \betalam\ method
presented here, are valid as a first-order dust-correction method, when using
the morphology and size of a galaxy as broad a priori constraints to its SFH
and redshift, respectively. The \betalam\ method will be applicable to large
samples of galaxies for which resolved imagery is available in one rest-frame
visible--near-IR filter and one rest-frame $\lambda$\,$\sim$\,3.5\,\micron\
bandpass. We make our CSP synthesis models and all our results available via a
dedicated website in the form of FITS and PNG maps, and ASCII data tables of
\betalamzero\ values as a function of age and metallicity.

\acknowledgments

Acknowledgements. This work was funded by NASA/ADAP grant NNX12AE47G (PI:
R.\,A.~Jansen). R.A.W. acknowledges support from NASA JWST grants NAGS-12460
and NNX14AN10G. We thank Claus Leitherer for answering questions about the \SB\
code and Rosa M. Gonz{\'a}lez Delgado and Rub{\'e}n Garc{\'{\i}}a-Benito for
sending us the data of the plots in their paper (GD15). We thank the anonymous
referee for constructive comments that helped us to significantly improve this
paper.

\vspace{4mm}

\software{\BC, \SB, and IDL}

\section{References}

{\footnotesize
\noindent
\citebox{Allamandola, L.J., Tielens, A.G.G.M., \& Barker, J.R. 1989, ApJS, 71, 733}
\citebox{Anders, P., \& Fritze-von~Alvensleben, U. 2003, A\&A 401, 1063}
\citebox{Avila, R., \etal\ 2015, “ACS Instrument Handbook”, Version 14.0 (Baltimore: STScI)}
\citebox{Behroozi, P.S., Wechsler, R.H., \& Conroy, C. 2013, ApJ 770, 57}
\citebox{Bell, E.F., Gordon, K.D., Kennicutt, R.C. Jr., \& Zaritsky, D. 2002, ApJ 565, 994}
\citebox{Bessell, M.S. 1979, PASP 91, 589}
\citebox{Bessell, M.S. 1990, PASP 102, 1181}
\citebox{Bessell, M.S. 2005, \araa, 43, 293}
\citebox{Bessell, M.S., \& Brett, J.M. 1988, PASP 100, 1134}
\citebox{Boissier, S., Boselli, A., Buat, V., Donas, J., \& Milliard, B. 2004, A\&A 424, 465}
\citebox{Boselli, A., Gavazzi, G., \& Sanvito, G. 2003, A\&A 402, 37}
\citebox{Brown, M.J.I., Moustakas, J., Smith, J.-D.T., \etal\ 2014, ApJS 212, 18}
\citebox{Bruzual, G., \& Charlot, S. 2003, MNRAS 344, 1000 (\textsf{BC03})}
\citebox{Buat, V., \& Xu, C. 1996, A\&A 306, 61}
\citebox{Buat, V., Iglesias-P\'{a}ramo, J., Seibert, M., \etal\ 2005, ApJ 619, L51}
\citebox{Buta, R.J., Corwin, H.G., \& Odewahn, S.~C. 2007, The de Vaucouleurs Altlas of Galaxies (Cambridge: Cambridge Univ. Press)}
\citebox{Calzetti, D., Kinney, A.L., \& Storchi-Bergmann, T. 1994, ApJ 429, 582}
\citebox{Calzetti, D., Armus, L., Bohlin, R.C., \etal\ 2000, ApJ 533, 682}
\citebox{Calzetti, D., Kennicutt, R.C. Jr., Bianchi, L., \etal\ 2005, ApJ 633, 871}
\citebox{Caplan, J., \& Deharveng, L. 1985, A\&AS 62, 63}
\citebox{Caplan, J., \& Deharveng, L. 1986, A\&A 155, 297}
\citebox{Cervi{\~n}o, M., \& Mas-Hesse, J.M. 1994, A\&A 284, 749}
\citebox{Chabrier, G. 2003, PASP 115, 763}
\citebox{Charbonnel, C., Meynet, G., Maeder, A., Schaller, G., \& Schaerer, D.\ 1993, \aaps, 101, 415}
\citebox{Charlot, S., \& Fall, S.M. 2000, ApJ 539, 718}
\citebox{Cohen, M., Wheaton, W.A., \& Megeath, S.T.\ 2003, AJ 126, 1090} 
\citebox{Colina, L., \& Bohlin, R.~C. 1994, AJ 108, 1931}
\citebox{Conselice, C.J., Grogin, N.A., Jogee, S., \etal\ 2004, ApJL 600, L139}
\citebox{Dale, D.A., Helou, G., Contursi, A., Silbermann, N.A., \& Kolhatkar, S. 2001, ApJ 549, 215}
\citebox{da Silva, R.L., Fumagalli, M., \& Krumholz, M.\ 2012, ApJ 745, 145}
\citebox{de Jong, R.S. 1996, A\&A 313, 377}
\citebox{Deo, R.P., Crenshaw, D.M., \& Kraemer, S.B. 2006, AJ 132, 321}
\citebox{de Vaucouleurs, G. 1959, Handbuch der Physik 53, 275}
\citebox{Dickinson, M., Giavalisco, M., \& GOODS Team 2003, The Mass of Galaxies at Low and High Redshift, 324}
\citebox{Dressel, L., 2015. “Wide Field Camera 3 Instrument Handbook, Version 7.0” (Baltimore: STScI)}
\citebox{Driver, S.P., Popescu, C.C., Tuffs, R.J., \etal\ 2008, ApJ 678, L101}
\citebox{Duce, D.A., Adler, M., Boutell, T., et~al.\ 2004, 
  ISO/IEC~15948:2004 (\url{http://www.libpng.org/pub/png/spec/iso/}
  and W3C REC-PNG-20031110 \url{https://www.w3.org/TR/PNG/})}
\citebox{Elmegreen, D.M. 1980, ApJS 43, 37}
\citebox{Elsasser, W.M. 1938, ApJ 87, 497}
\citebox{Fazio, G.G., Hora, J.L., Allen, L.E., \etal\ 2004, ApJS 154, 10}
\citebox{Fioc, M., \& Rocca-Volmerange, B. 1997, A\&A 326, 950}
\citebox{Fitzpatrick, E.L. 1999, PASP 111, 63}
\citebox{Ford, H.C., Clampin, M., Hartig, G.F., \etal\ 2003, Proc. SPIE 4854, 81}
\citebox{Galliano, F., Hony, S., Bernard, J.-P., \etal\ 2011, A\&A 536, A88} 
\citebox{Gardner, J.~P., Mather, J.C., Clampin, M., \etal\ 2006, Space Science Reviews, 123, 485}
\citebox{Gerola, H., \& Seiden, P.E.\ 1978, ApJ 223, 129}
\citebox{Giavalisco, M., Ferguson, H.~C., Koekemoer, A.~M., \etal\ 2004, ApJL 600, L93}
\citebox{Girardi, L., Bressan, A., Bertelli, G., \& Chiosi, C. 2000, A\&AS 141, 371}
\citebox{Gonz{\'a}lez Delgado, R.M., Garc{\'{\i}}a-Benito, R., P{\'e}rez, E., \etal\ 2015, A\&A, 581, A103 (GD15)}
\citebox{Gordon, K.D., Clayton, G.C., Misselt, K.A., Landolt, A.U., \& Wolff, M.J. 2003, ApJ 594, 279}
\citebox{Grogin, N.A., Kocevski, D.D., Faber, S.M., \etal\ 2011, ApJS 197, 35}
\citebox{Gunn, J.E., Carr, M., Rockosi, C., \etal\ 1998, AJ 116, 3040}
\citebox{Hanisch, R.J., Farris, A., Greisen, E.W., \etal\ 2001, A\&A, 376, 359}
\citebox{Hayes, D.~S., \& Latham, D.~W.\ 1975, ApJ 197, 593}
\citebox{Horner, S.D., \& Rieke, M.~J.\ 2004, SPIE 5487, 628}
\citebox{Hoyos, C., Arag{\'o}n-Salamanca, A., Gray, M.~E., \etal\ 2016, MNRAS 455, 295}
\citebox{Hubble, E.P. 1926, ApJ 64, 321}
\citebox{Humason, M.L., Mayall, N.U., \& Sandage, A.R. 1956, AJ 61, 97}
\citebox{Jansen, R.A., Franx, M., Fabricant, D., \& Caldwell, N. 2000a, ApJS 126, 271}
\citebox{Jansen, R.A., Fabricant, D., Franx, M., \& Caldwell, N. 2000b, ApJS 126, 331}
\citebox{Jarret, T.H., Cohen, M., Masci, F., Wright, E., Stern, D., \etal\ 2011, ApJ 735, 112}
\citebox{Jester, S., Schneider, D.P., Richards, G.T., \etal\ 2005, AJ 130, 873}
\citebox{Kauffmann, G., Heckman, T.M., De Lucia, G., \etal\ 2006, MNRAS 367, 1394}
\citebox{Kelvin, L.S., Driver, S.P., Robotham, A.S.G., \etal\ 2014, MNRAS 444, 1647}
\citebox{Kennicutt, R.C. Jr., Hao, C.-N., Calzetti, D., \etal\ 2009, ApJ 703, 1672}
\citebox{Kessler, M.F., Steinz, J.A., Anderegg, M.E., \etal\ 1996, A\&A 315, L27}
\citebox{Koekemoer, A.M., Faber, S.M., Ferguson, H.C., \etal\ 2011, ApJS 197, 36}
\citebox{Kong, X., Charlot, S., Brinchmann, J., \& Fall, S.M. 2004, MNRAS 349, 769}
\citebox{Kroupa, P. 2002, Science 295, 82}
\citebox{Kurucz, R.L. 1993, CDROM No.\,13, 18 (Cambridge MA: Smithsonial
  Astrophysical Observatory); \url{http://kurucz.harvard.edu}}
\citebox{L\'eger, A., \& Puget, J.~L.\ 1984, \aap, 137, L5}
\citebox{Leitherer, C., Schaerer, D., Goldader, J.D., \etal\ 1999, ApJS 123, 3 (\textsf{Starburst99})}
\citebox{Lindblad, B. 1941, Stockholms Observatoriums Annaler 13, 8}
\citebox{Ma\'{\i}z-Apell\'{a}niz, J., P\'{e}rez, E., \& Mas-Hesse, J.M. 2004, AJ 128, 1196}
\citebox{Maiolino, R., Nagao, T., Grazian, A., \etal\ 2008, A\&A 488, 463}
\citebox{Maraston, C. 2005, MNRAS 362, 799}
\citebox{Marigo, P., Girardi, L., Bressan, A., \etal\ 2008, A\&A 482, 883}
\citebox{Mathis, J.S., Rumpl, W., \& Nordsieck, K.H. 1977, ApJ 217, 425}
\citebox{Meurer, G.R., Heckman, T.M., \& Calzetti, D. 1999, ApJ 521, 64}
\citebox{Schaerer, D., Charbonnel, C., Meynet, G., Maeder, A., \& Schaller, G.\ 1993b, \aaps, 102, 339}
\citebox{Misselt, K.A., Clayton, G.C., \& Gordon, K.D. 1999, ApJ 515, 128}
\citebox{Neugebauer, G., Habing, H.J., van~Duinen, R., \etal\ 1984, ApJ 278, L1}
\citebox{Odewahn, S.~C. 1995, ApL\&C, 31, 55}
\citebox{Oke, J.B. 1974, ApJS 27, 21}
\citebox{Oke, J.B., \& Gunn, J.E. 1983, ApJ 266, 713}
\citebox{Panuzzo, P., Bressan, A., Granato, G.L., Silva, L., \& Danese, L. 2003, A\&A 409, 99}
\citebox{Petersen, L., \& Gammelgaard, P. 1997, A\&A 323, 697}
\citebox{Pilbratt, G.L. 2004, Proc.\,SPIE 5487, 401}
\citebox{Planck Collaboration, Ade, P.~A.~R., Aghanim, N., \etal\ 2016,
   A\&A 594, A13; \emph{``Planck 2015 Results. XIII. Cosmological
   Parameters''}}
\citebox{Postman, M., Coe, D., Ben{\'{\i}}tez, N., \etal\ 2012, ApJS 199, 25}
\citebox{Price, S.D., Carey, S.J., \& Egan, M.P. 2002, Adv.~Space~Research 30, 2027}
\citebox{Rela\~{n}o, M., Lisenfeld, U., Vilchez, J. M., \& Battaner, E. 2006, A\&A 452, 413}
\citebox{Rieke, M. 2011, in ASP Conf.\ Ser.\ 446, Galaxy Evolution: Infrared to Milimeter Wavelength Perspective, ed. W. Wang \etal\ (San Francisco, CA: ASP), 331}
\citebox{Roussel, H., Gil~de~Paz, A., Seibert, M., \etal\ 2005, ApJ 632, 227}
\citebox{Rudy, R.J. 1984, ApJ 284, 33}
\citebox{Salpeter, E.E. 1955, ApJ 121, 161}
\citebox{Scoville, N.Z., Polletta, M., Ewald, S., \etal\ 2001, AJ 122, 3017}
\citebox{Schaerer, D., Meynet, G., Maeder, A., \& Schaller, G.\ 1993a, \aaps, 98, 523}
\citebox{Schaerer, D., Charbonnel, C., Meynet, G., Maeder, A., \& Schaller, G.\ 1993b, \aaps, 102, 339}
\citebox{Schaller, G., Schaerer, D., Meynet, G., \& Maeder, A.\ 1992, \aaps, 96, 269}
\citebox{Scoville, N., Abraham, R.G., Aussel, H., \etal\ 2007, ApJS 172, 38}
\citebox{Tamura, K., Jansen, R.A., \& Windhorst, R.A. 2009, AJ 138, 1634 (T09)}
\citebox{Tamura, K., Jansen, R.A., Eskridge, P.B., Cohen, S.H., \& Windhorst, R.A. 2010, AJ 139, 2557}
\citebox{Taylor, V.A., Jansen, R.A., Windhorst, R.A., Odewahn, S.C., \& Hibbard, J.E. 2005, ApJ 630, 784}
\citebox{Taylor-Mager, V.A., Conselice, C.J., Windhorst, R.A., \& Jansen, R.A. 2007, ApJ 659, 162}
\citebox{Trumpler, R.J. 1930, PASP 42, 214}
\citebox{van den Bergh, S. 1998, Galaxy Morphology and Classification (Cambridge: Cambridge Univ. Press)}
\citebox{Vazdekis, A., S\'{a}nchez-Bl\'{a}zquez, P., Falc\'{o}n-Barroso, J., \etal\ 2010, MNRAS 404, 1639}
\citebox{V\'{a}zquez, G.A., \& Leitherer, C. 2005, ApJ 621, 695}
\citebox{Viallefond, F., Goss, W.M., \& Allen, R.J. 1982, A\&A 115, 373}
\citebox{Walcher, J., Groves, B., Budav{\'a}ri, T., \& Dale, D.\ 2011, \apss, 331, 1}
\citebox{Walterbos, R.A.M., \& Kennicutt, R.C. Jr. 1988, A\&A 198, 61}
\citebox{Waller, W.H., Gurwell, M., \& Tamura, M. 1992, AJ 104, 63}
\citebox{Wells, D.C., Greisen, E.W., \& Harten, R.H.\ 1981, A\&AS, 44, 363}
\citebox{Werner, M.W., Roellig, T.L., Low, F.J., \etal\ 2004, ApJS 154, 1}
\citebox{Windhorst, R.A., Taylor, V.A., Jansen, R.A., \etal\ 2002, ApJS 143, 113}
\citebox{Windhorst, R.A., Cohen, S.H., Hathi, N.P., \etal\ 2011, ApJS 193, 27}
\citebox{Witt, A.N., Thronson, H.A. Jr., \& Capuano, J.M. Jr. 1992, ApJ 393, 611}
\citebox{Wright, E.L., Eisenhardt, P.R.M., Mainzer, A.K., \etal\ 2010, AJ 140, 1868}
\citebox{Xu, C., \& Helou, G. 1996, ApJ 456, 152}
}%*%---end-of-footnotesize

%%%%%%%%%%%%%%%%%%%%%%%%%%%% A P P E N D I X %%%%%%%%%%%%%%%%%%%%%%%%%%%%%
\begin{appendix}

\section{A SET OF DISTINGUISHABLE SSPs}

\begin{figure*}[h]
\centerline{
  \includegraphics[width=0.425\txw]{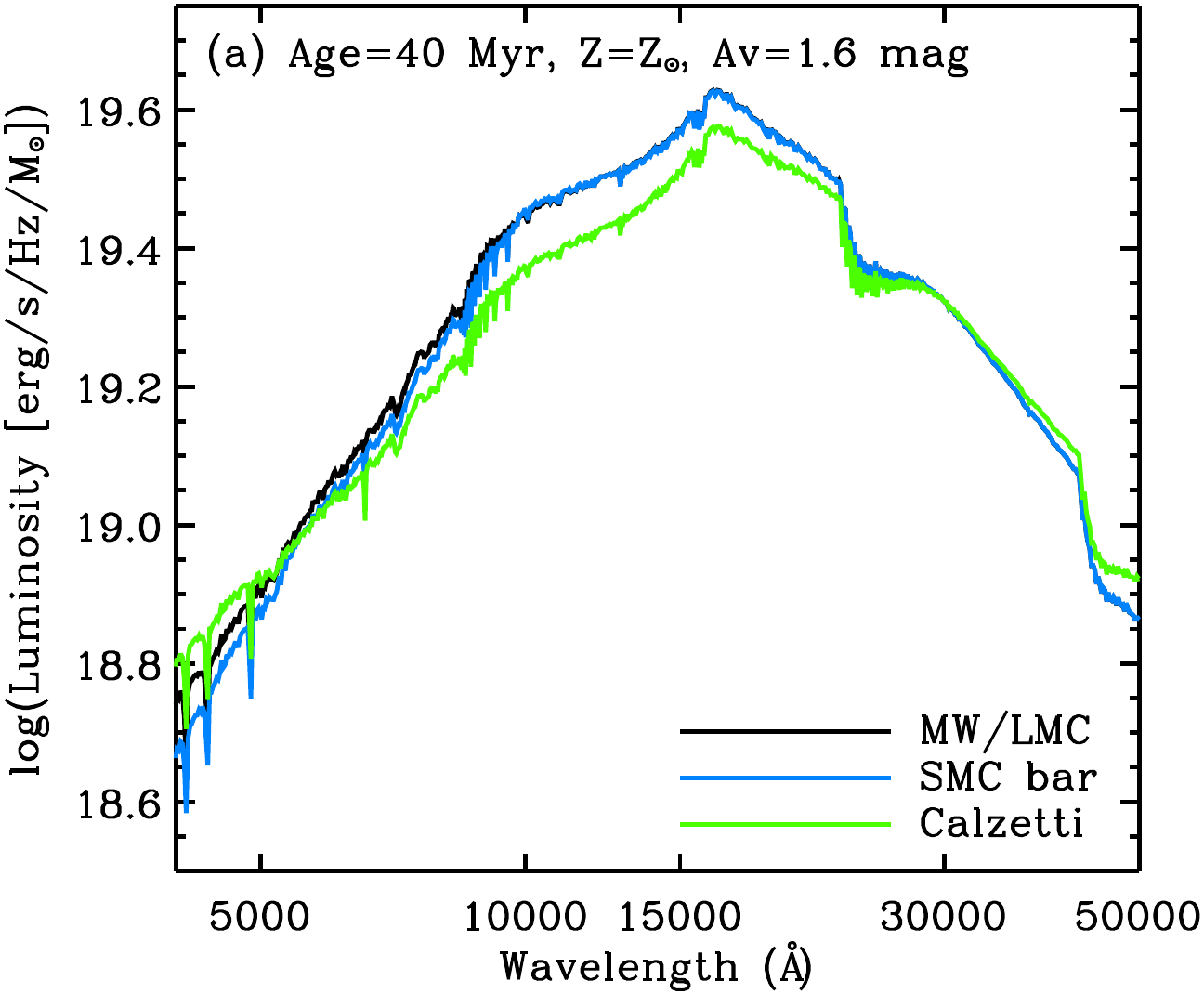}\ \ \ \ \ 
  \includegraphics[width=0.425\txw]{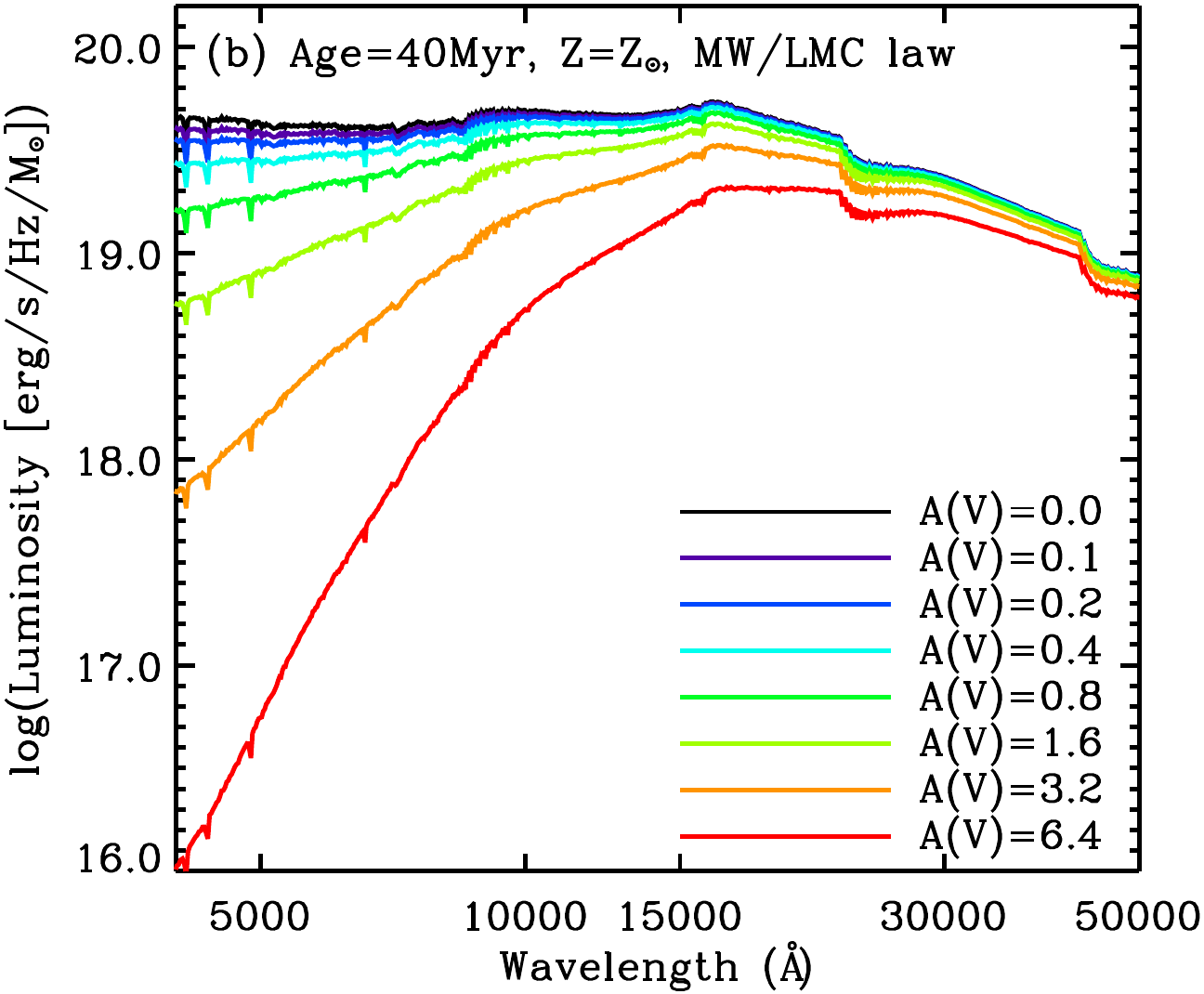}
}
\caption{\noindent\small
Examples of our adopted combined array of SSP SEDs. (a) SEDs for a 40\,Myr old
SSP with $Z$\,=\,$Z_{\odot}$, for three different extinction laws and
\AV\,=\,1.6\,mag. Note that over the wavelength range of interest
(0.4--3.75\,\micron), the Milky Way and LMC extinction laws are essentially
indistinguishable and were combined; (b) SEDs for a 40\,Myr old SSP with
$Z$\,=\,$Z_{\odot}$, for eight different extinction values and a MW/LMC
extinction law.\label{fig:figureA1}}
\end{figure*}

We construct our combined model SSP SEDs for a manageably small, yet
comprehensive set of parameters that span the range of representative stellar
population properties and dust extinction. We consider six metallicity and 16
age values of the stellar population, four extinction laws, and eight $V$-band
extinction values \AV\ (see Table~\ref{tab:tableA1}). We adopt a grid of 16
stellar population ages between 3\,Myr and 13.25\,Gyr\,\footnote{In
Planck\,2015 cosmology (Planck Collaboration \etal\ 2016), the Hubble time
$t_{H}$\,$\simeq$\,13.8\,Gyr. Hence, $t$\,=\,13.25\,Gyr corresponds to
0.55\,Gyr after the Big Bang or $\mathrm{z}_{f}$\,$\simeq$\,9, consistent with
the Planck Collaboration (Paper \,2016) results of $\tau$\,=\,0.058 and
$\mathrm{z}_{\mathrm{reion}}$\,$\simeq$\,8, so this maximum age is a deliberate
choice.} in steps that roughly double for ages between 10\,Myr and 10\,Gyr.
This allows us to follow the rapid evolution of massive stars, without
reserving too many SEDs for the slow evolution of low-mass stars. Because the
MW (Fitzpatrick 1999) and LMC (both the LMC2 Supershell and LMC Average from
Misselt \etal\ 1999) extinction laws are nearly indistinguishable in the
0.4--3.75\,\micron\ wavelength range of interest for the present study (see
Figure~\ref{fig:figureA1}(a)), we averaged both into a single ``MW/LMC''
extinction law and adopt $R_V$\,$=$\,3.1. We furthermore consider the Small
Magellanic Cloud (SMC; SMC bar of Gordon \etal\ 2003) extinction law with
$R_V$\,$=$\,2.74, and the law appropriate for starburst galaxies of Calzetti
\etal\ (2000), which has less reddening per unit extinction, for which we adopt
$R_V$\,$=$\,4.33. We substitute the MW/LMC curve for the SMC extinction law at
$\lambda$\,$>$\,1\,\micron, where no SMC data is available. The Calzetti
extinction law is also only defined to 2.2\,\micron, so we extrapolated to
3.75\,\micron, imposing a minimum residual attenuation of 5\% of that in $V$.
For each of these three extinction laws, we select 7 extinction values between
\AV\,=\,0.1 and \AV\,=\,6.4\,mag, each double the value of the previous one,
plus the trivial case of zero extinction (\AV\,=\,0\,mag), generating 22 SSP
SEDs for each of the 6 metallicity values and 16 ages, \ie\ a total of 2112
unique SSP SEDs. Figure~\ref{fig:figure2} and Figure~\ref{fig:figureA1} shows
example SEDs for different metallicities, ages, extinction laws, and extinction
values.

%%%%%%%%%%%%%%%%%%%%%%%%%%%%%%%%%  FIGURE A1  %%%%%%%%%%%%%%%%%%%%%%%%%%%%%%%%%
\noindent
%%%%%%%%%%%%%%%%%%%%%%%%%%%%%%%%%%%%%%%%%%%%%%%%%%%%%%%%%%%%%%%%%%%%%%%%%%%%%%
   %
%%%%%%%%%%%%%%%%%%%%%%%%%%%%%%%%  FIGURE A2  %%%%%%%%%%%%%%%%%%%%%%%%%%%%%%%%%%
\noindent\begin{figure}[h]
\centerline{
  \includegraphics[width=0.425\txw,height=0.35\txw]{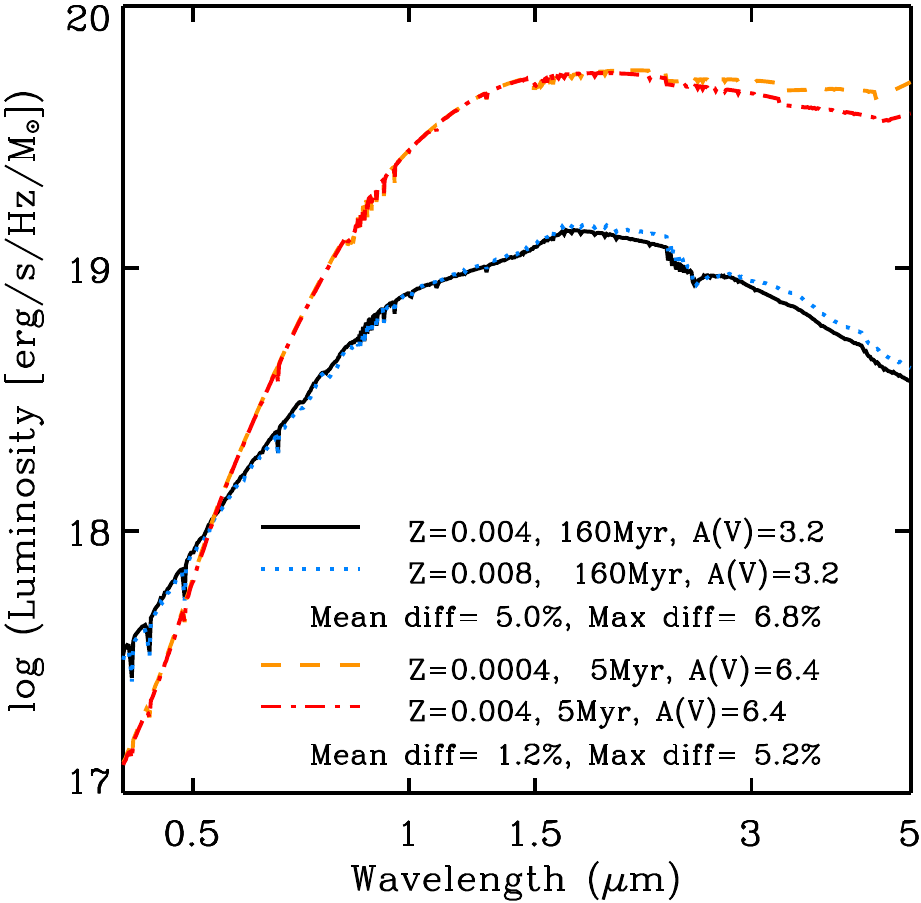} }   
\caption{\noindent\small 
Examples of two pairs of SSP model SEDs that \emph{just} meet our
$\max\{\Delta^{ij}_k\}$ or mean$\{\Delta^{ij}_k\}$ $>$ 5\% criteria for being
distinguishable over the 0.4--3.75\,\micron\ wavelength range, each normalized
at $\lambda$\,=\,1\,\micron. The lower pair (black/blue) is for
mean$\{\Delta^{ij}_k\}$ $>$ 5\% and therefore also has $\max\{\Delta^{ij}_k\}$
greater than 5\%, and the upper pair (red/orange) is for
$\max\{\Delta^{ij}_k\}$ $>$ 5\%.\label{fig:figureA2}}
\end{figure}
%%%%%%%%%%%%%%%%%%%%%%%%%%%%%%%%%%%%%%%%%%%%%%%%%%%%%%%%%%%%%%%%%%%%%%%%%%%%%%

Within typical observational uncertainties, many of these unique SSP SEDs will
be indistinguishable from one or more of the others over the wavelength range
of interest. We quantified the differences between each pair of SEDs, where we
only need to consider differences in SED shape, not in the absolute flux scale
(which depends on the total mass of stars formed). We first divide the
0.4--3.75\,\micron\ wavelength range in 10 equal linear bins, and normalize the
integrated flux in each bin to the mean integrated flux in all bins. For each
pair of SEDs, $(i,j)$, and each wavelength bin, $k$, we compare the difference
between normalized integrated fluxes as:
\begin{equation}
\Delta^{ij}_{k} = 
    \left\vert\, 
	\int_{k}{F_i(\lambda)} \middle/ \overline{\int_{k}{F_i(\lambda)}}\ - 
	\int_{k}{F_j(\lambda)} \middle/ \overline{\int_{k}{F_j(\lambda)}} 
    \,\right\vert .
\label{eq:delta}
\end{equation}

%%%%%%%%%%%%%%%%%%%%%%%%%%%%%%%%%  TABLE A1  %%%%%%%%%%%%%%%%%%%%%%%%%%%%%%%%%%
\noindent\begin{table*}[b]
\centering
\caption{\small Parameter set of our combined SSP model SEDs.\label{tab:tableA1}}
\begin{tabular}{p{0.15\txw} p{0.80\txw}}
\hline\\[-8pt]
Parameter & Values\\[2pt]
\hline\\[-8pt]
$Z$      & 0.0001, 0.0004, ~0.004, ~0.008, ~0.02, 0.05 \\[2pt]
Age (Myr)& 3, ~4, ~5, ~7, ~10, ~20, ~40, ~80, ~160, ~320, \\
         & 640, 1250, ~2500, ~5000, ~10000, ~13250                      \\[2pt]
Extinction law & MW$^{\mathrm{a}}$/LMC$^{\mathrm{b}}$, SMC$^{\mathrm{c}}$,
						Calzetti$^{\mathrm{d}}$\\[2pt]
\AV\ (mag)     & 0.0, ~0.1, ~0.2, ~0.4, ~0.8, ~1.6, ~3.2, ~6.4          \\[2pt]
\hline\\[-8pt]
\end{tabular}
\begin{minipage}{0.98\txw}{$^{\mathrm{a}}$Fitzpatrick (1999),
$^{\mathrm{b}}$Misselt \etal\ (1999), $^{\mathrm{c}}$Gordon \etal\ (2003),
$^{\mathrm{d}}$Calzetti \etal\ (2000).}\end{minipage}\vspace*{8pt}
\end{table*}
%%%%%%%%%%%%%%%%%%%%%%%%%%%%%%%%%%%%%%%%%%%%%%%%%%%%%%%%%%%%%%%%%%%%%%%%%%%%%%

We reject SED $j$ as indistinguishable from SED $i$ when both the maximum
difference, $\max\{\Delta^{ij}_k\}$, and the mean$\{\Delta^{ij}_k\}$ are
smaller than 5\%. Using this criterion, we rejected 1388 of our 2112 SSP SEDs,
leaving a set of 724 unique and distinguishable SEDs for further analysis of
the \betaV\ method and its comparison with the SED fitting method, and for
fitting observed SEDs of individual regions within galaxies (Jansen \etal, in
prep.; Kim \etal, in prep.). Figure~\ref{fig:figureA2} shows two examples of
pairs of SEDs that were deemed to be \emph{just} distinguishable ($>$5\%) over
the 0.4--3.75\micron\ interval.

\section{THE MAGNITUDE OFFSETS BETWEEN AB \& VEGA MAGNITUDE SYSTEMS FOR VARIOUS FILTERS}

Throughout this study, we use the AB magnitude system (Oke 1974; Oke \& Gunn
1983). For observers with data calibrated onto the Vega magnitude system, we
here provide magnitude offsets between AB and Vega magnitudes, derived using
the Kurucz (1993) Vega spectrum as available in \code{STSDAS}\,3.16
(\url{www.stsci.edu/institute/software\_hardware/stsdas/}). We
also list the central wavelength, $\lambda_{c}$, and bandwidth (FWHM), $\Delta\lambda$, of each filter in Table~\ref{tab:tableB1}.

%%%%%%%%%%%%%%%%%%%%%%%%%%%%%%%%%  TABLE B1  %%%%%%%%%%%%%%%%%%%%%%%%%%%%%%%%%
\noindent\begin{table}[h!]
\caption{\small Central wavelengths, bandwidths, and magnitude offsets
between AB and Vega magnitudes for each of the filters used in the present
study.\label{tab:tableB1}}
\centering{
\setlength{\tabcolsep}{5.8pt}
\begin{tabular}{cccccccc}
\hline\\[-5pt]
Filter  & $\lambda_{c}$\tablenotemark{a}(\AA) & $\Delta\lambda$\tablenotemark{b}(\AA) & $m_{\lambda,\mathrm{AB}}-m_{\lambda,\mathrm{Vega}}$\tablenotemark{c} &
Filter  & $\lambda_{c}$\tablenotemark{a}(\AA) &  $\Delta\lambda$\tablenotemark{b}(\AA) & $m_{\lambda,\mathrm{AB}}-m_{\lambda,\mathrm{Vega}}$\tablenotemark{c}\\[2pt]
\hline\\[-6pt]
$B$              & 4361\tablenotemark{d} &  890\tablenotemark{d} &$-$0.098 &
WFC3\,UVIS F475W & 4773\tablenotemark{e} & 1344\tablenotemark{e} &$-$0.088\\
$V$              & 5448\tablenotemark{d} &  840\tablenotemark{d} &  0.022 &
WFC3\,UVIS F555W & 5308\tablenotemark{e} & 1562\tablenotemark{e} &$-$0.013\\
$R_c$            & 6400\tablenotemark{f} & 1500\tablenotemark{f} &  0.213 &
WFC3\,UVIS F606W & 5887\tablenotemark{e} & 2182\tablenotemark{e} &  0.096\\
$I_c$            & 7900\tablenotemark{f} & 1500\tablenotemark{f} &  0.455 &
WFC3\,UVIS F625W & 6242\tablenotemark{e} & 1463\tablenotemark{e} &  0.163\\
$I$              & 7980\tablenotemark{d} & 1540\tablenotemark{d} &  0.436 &
WFC3\,UVIS F775W & 7647\tablenotemark{e} & 1171\tablenotemark{e} &  0.396\\
SDSS\,$g$        & 4774\tablenotemark{g} & 1377\tablenotemark{g} &$-$0.085 &
WFC3\,UVIS F814W & 8024\tablenotemark{e} & 1536\tablenotemark{e} &  0.433\\
SDSS\,$r$        & 6231\tablenotemark{g} & 1371\tablenotemark{g} &  0.167 &
WFC3\,UVIS F850LP& 9166\tablenotemark{e} & 1182\tablenotemark{e} &  0.536\\
SDSS\,$i$        & 7615\tablenotemark{g} & 1510\tablenotemark{g} &  0.394 &
WFC3\,IR F098M   & 9864\tablenotemark{e} & 1570\tablenotemark{e} &  0.576\\
SDSS\,$z$        & 9132\tablenotemark{g} &  940\tablenotemark{g} &  0.530 &
WFC3\,IR F105W   &10552\tablenotemark{e} & 2650\tablenotemark{e} &  0.660\\
ACS\,WFC F435W   & 4297\tablenotemark{h} & 1038\tablenotemark{h} &$-$0.094 &
WFC3\,IR F110W   &11534\tablenotemark{e} & 4430\tablenotemark{e} &  0.775\\
ACS\,WFC F475W   & 4760\tablenotemark{h} & 1458\tablenotemark{h} &$-$0.088 &
WFC3\,IR F125W   &12486\tablenotemark{e} & 2845\tablenotemark{e} &  0.916\\
ACS\,WFC F555W 	 & 5346\tablenotemark{h} & 1193\tablenotemark{h} &  0.006 &
WFC3\,IR F160W   &15369\tablenotemark{e} & 2683\tablenotemark{e} &  1.267\\ 
ACS\,WFC F606W 	 & 5907\tablenotemark{h} & 2342\tablenotemark{h} &  0.100 &
$J$              &12546                  & 1620\tablenotemark{i} &  0.909\\
ACS\,WFC F625W 	 & 6318\tablenotemark{h} & 1442\tablenotemark{h} &  0.178 &
$H$              &16487                  & 2510\tablenotemark{i} &  1.383\\
ACS\,WFC F775W 	 & 7764\tablenotemark{h} & 1528\tablenotemark{h} &  0.403 &
$K_S$            &21634                  & 2620\tablenotemark{i} &  1.853\\
ACS\,WFC F814W 	 & 8333\tablenotemark{h} & 2511\tablenotemark{h} &  0.438 &
$K$              &22053                  & 3889                  &  1.886\\
ACS\,WFC F850LP  & 9445\tablenotemark{h} & 1229\tablenotemark{h} &  0.536 &
$K'$             &21218                  & 3404                  &  2.800\\
WFC3\,UVIS F438W & 4325\tablenotemark{e} &  618\tablenotemark{e} &$-$0.144 &
                 &                       &                       &      \\[2pt]
\hline\\[-5pt]
$L$              &34831                  & 5103                  & 2.760 &
NIRCam F410M     &40820\tablenotemark{j} & 4380\tablenotemark{j} & 3.087\\
$L'$             &38333                  & 5880                  & 2.954 &
NIRCam F444W     &44080\tablenotemark{j} &10290\tablenotemark{j} & 3.224\\
\WISE\,$W$\textsl{1}&33836               & 7934                  & 2.681 &
NIRCam F480M     &48740\tablenotemark{j} & 3000\tablenotemark{j} & 3.423\\
\Spitzer\,$I$\textsl{1}&35466            & 7432                  & 2.797 &
MIRI F560W       &56000\tablenotemark{k} &12000\tablenotemark{k} & 3.732\\
NIRCam F200W     &19890\tablenotemark{j} & 4570\tablenotemark{j} & 1.691 &
MIRI F770W       &77000\tablenotemark{k} &22000\tablenotemark{k} & 4.352\\
NIRCam F356W     &35680\tablenotemark{j} & 7810\tablenotemark{j} & 2.793 &
MIRI F1000W      &100000\tablenotemark{k}&20000\tablenotemark{k} &4.921\\[2pt]
\hline\\[-3pt]
\end{tabular}
\begin{minipage}{0.98\txw}{(a) Central wavelength; (b) Width of the bandpass
(FWHM); (c) Magnitude offset between AB and Vega magnitude system;
$m_{\lambda,\mathrm{AB}}$$-$\,$m_{\lambda,\mathrm{Vega}}$=\,$-$2.5\,
$\log{(F_{\lambda,\mathrm{Vega}})}$$-$\,48.585 (Hayes \& Latham 1975; Bessell
\& Brett 1988; Bessell 1990; Colina \& Bohlin 1994) (d) Bessell (2005); (e)
Dressel (2015); (f) Bessell (1979); (g) Gunn \etal\ (1998); (h) Avila \etal\
(2015); (i) Cohen \etal\ (2003); (j)
\url{http://www.stsci.edu/jwst/instruments/nircam/instrumentdesign/filters/};
(k) \url{http://ircamera.as.arizona.edu/MIRI}}
\end{minipage}
}
\end{table}
%%%%%%%%%%%%%%%%%%%%%%%%%%%%%%%%%%%%%%%%%%%%%%%%%%%%%%%%%%%%%%%%%%%%%%%%%%%%%

\vfill\null

\end{appendix}

\input{figset5.tex}

\input{figset10.tex}

\end{document}

%% file: figset5.tex
%%%%%%%%%%%%%%%%%%%%%%%%% Figure Set 5 %%%%%%%%%%%%%%%%%%%%%%%%%%%%%%%%
\begin{figure*}[t!]
\centerline{
	\includegraphics[width=\txw]{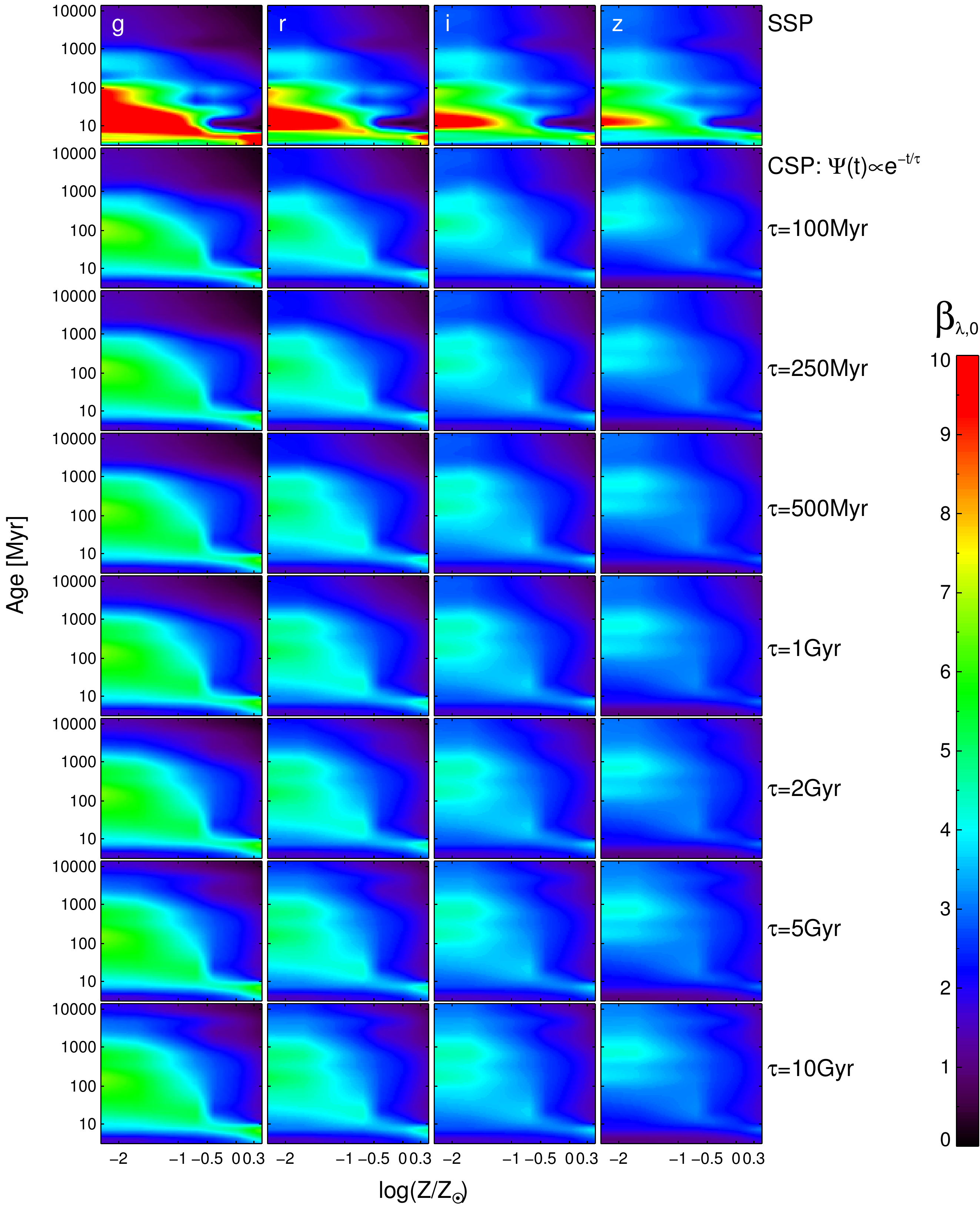}
}
\caption{\noindent\small
Same as Figure~\ref{fig:figure5} for the SDSS $g$, $r$, $i$, and $z$ filters.}
\end{figure*}

\begin{figure*}
\centerline{
	\includegraphics[width=\txw]{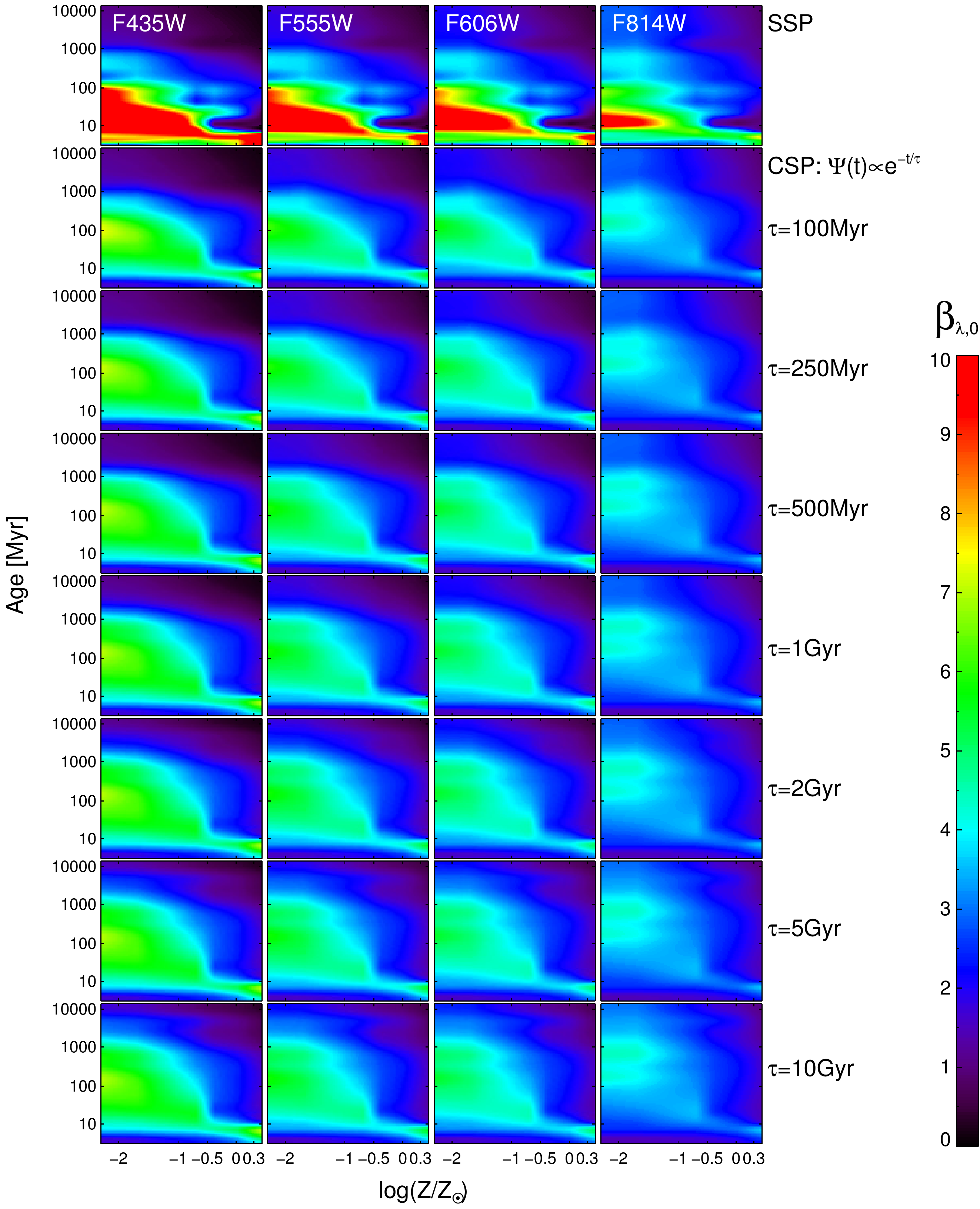}
}
\caption{\noindent\small
Same as Figure~\ref{fig:figure5} for the \HST/ACS\,WFC F435W, F555W, F606W, and
F814W filters.}
\end{figure*}

\begin{figure*}
\centerline{
	\includegraphics[width=\txw]{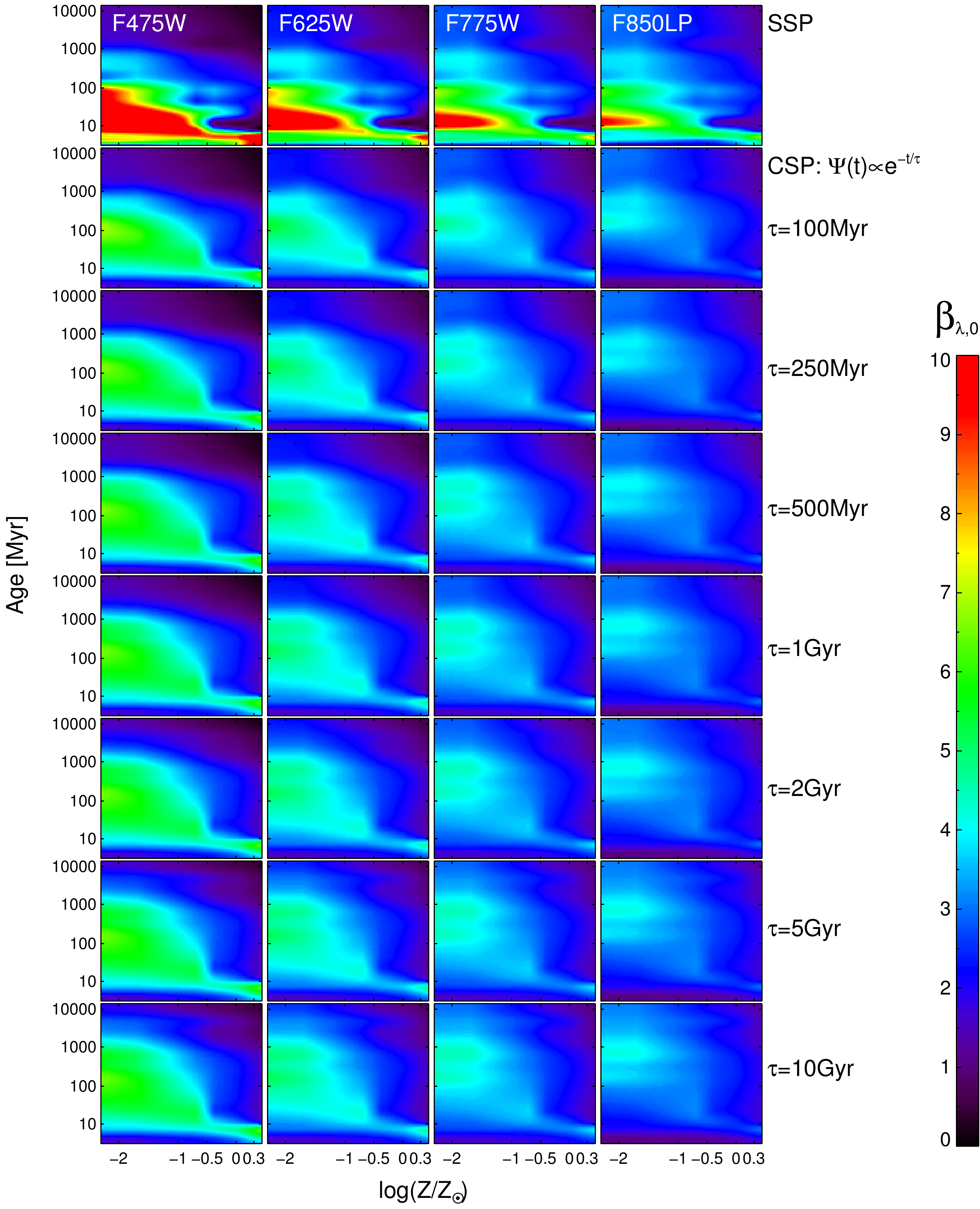}
}
\caption{\noindent\small
Same as Figure~\ref{fig:figure5} for the \HST/ACS\,WFC F475W, F625W, F775W, and
F850LP filters.}
\end{figure*}

\begin{figure*}
\centerline{
	\includegraphics[width=\txw]{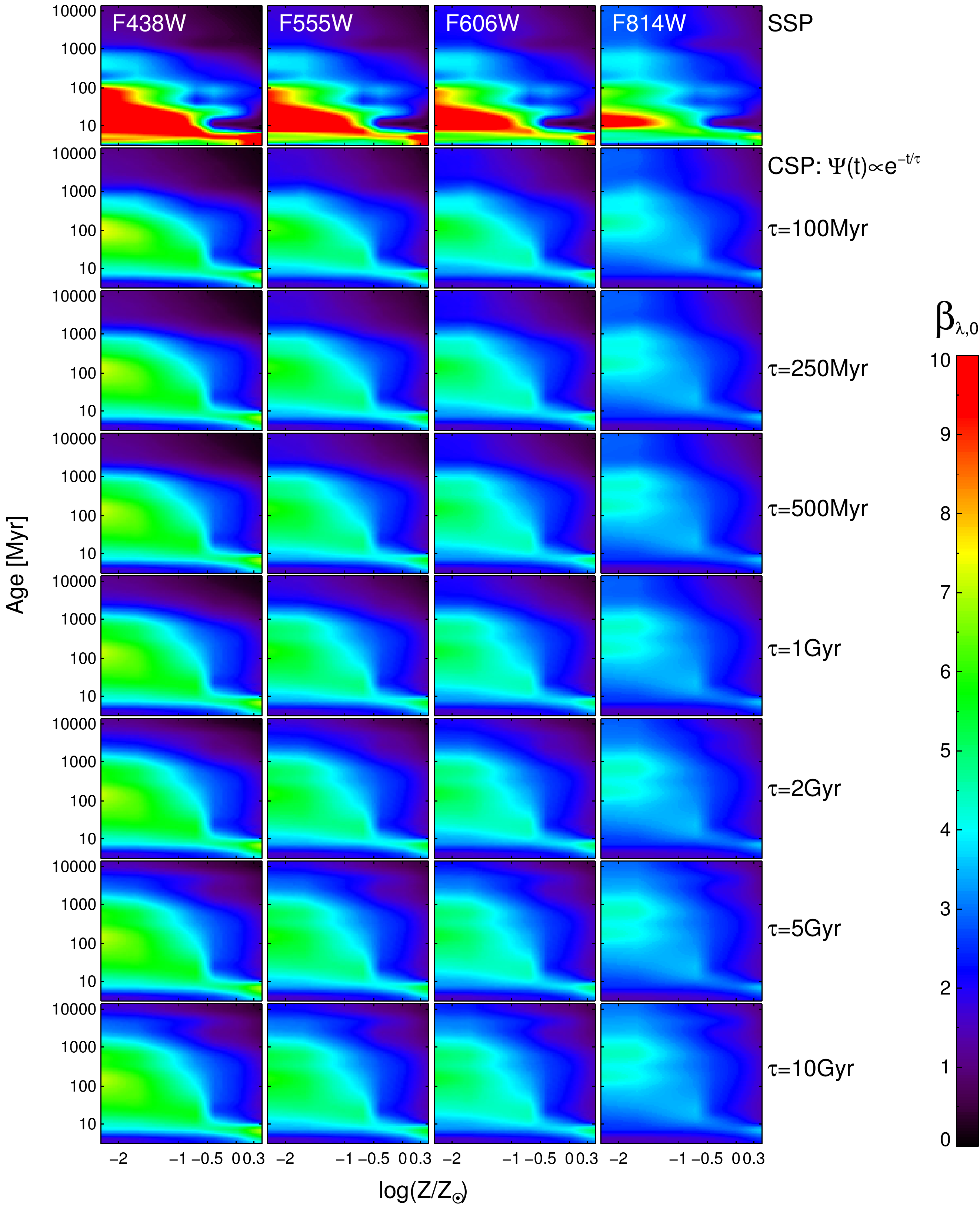}
}
\caption{\noindent\small
Same as Figure~\ref{fig:figure5} for the \HST/WFC3\,UVIS F438W, F555W, F606W, and
F814W filters.}
\end{figure*}

\begin{figure*}
\centerline{
	\includegraphics[width=\txw]{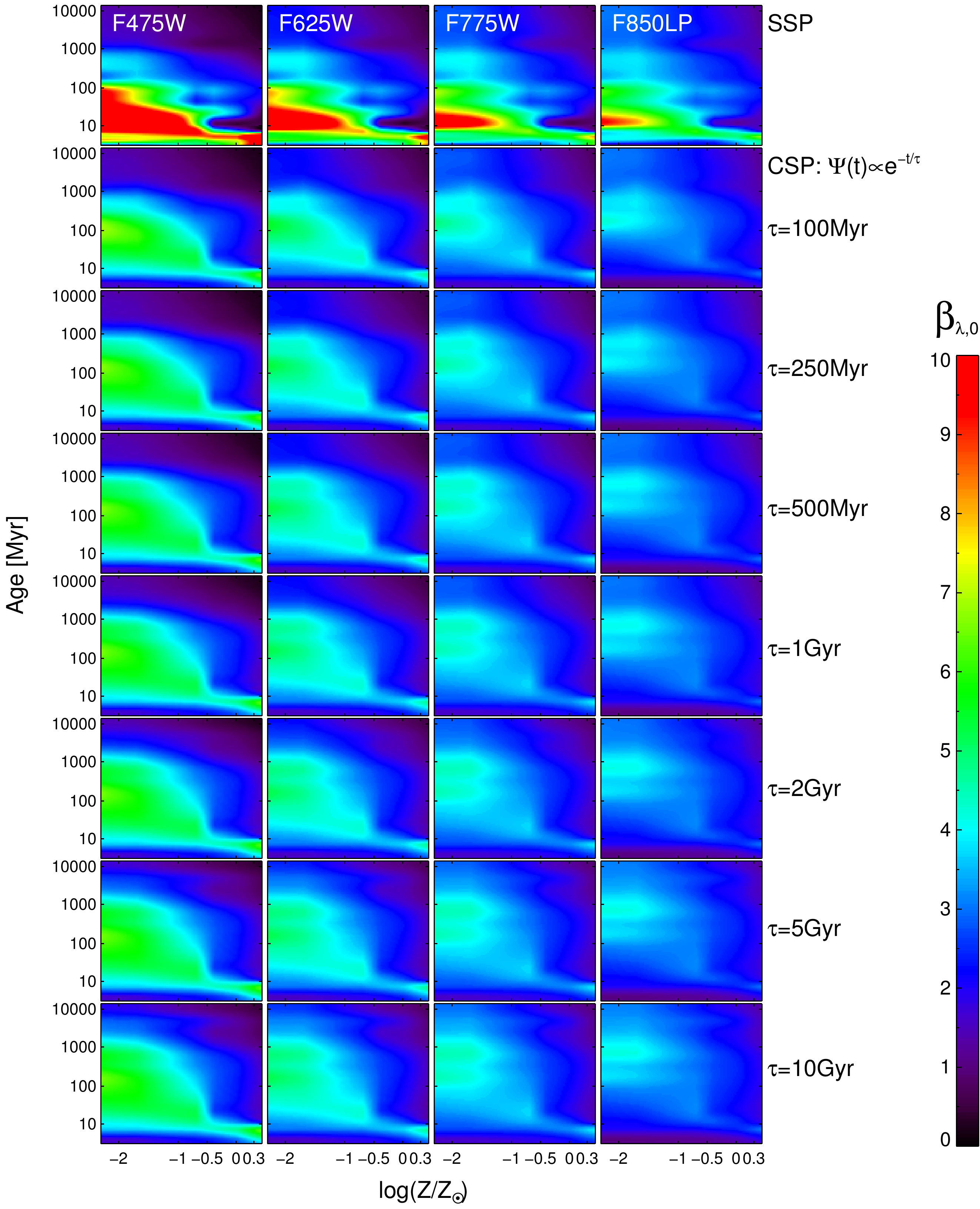}
}
\caption{\noindent\small
Same as Figure~\ref{fig:figure5} for the \HST/WFC3\,UVIS F475W, F625W, F775W, and
F850LP filters.}
\end{figure*}

\begin{figure*}
\centerline{
	\includegraphics[width=\txw]{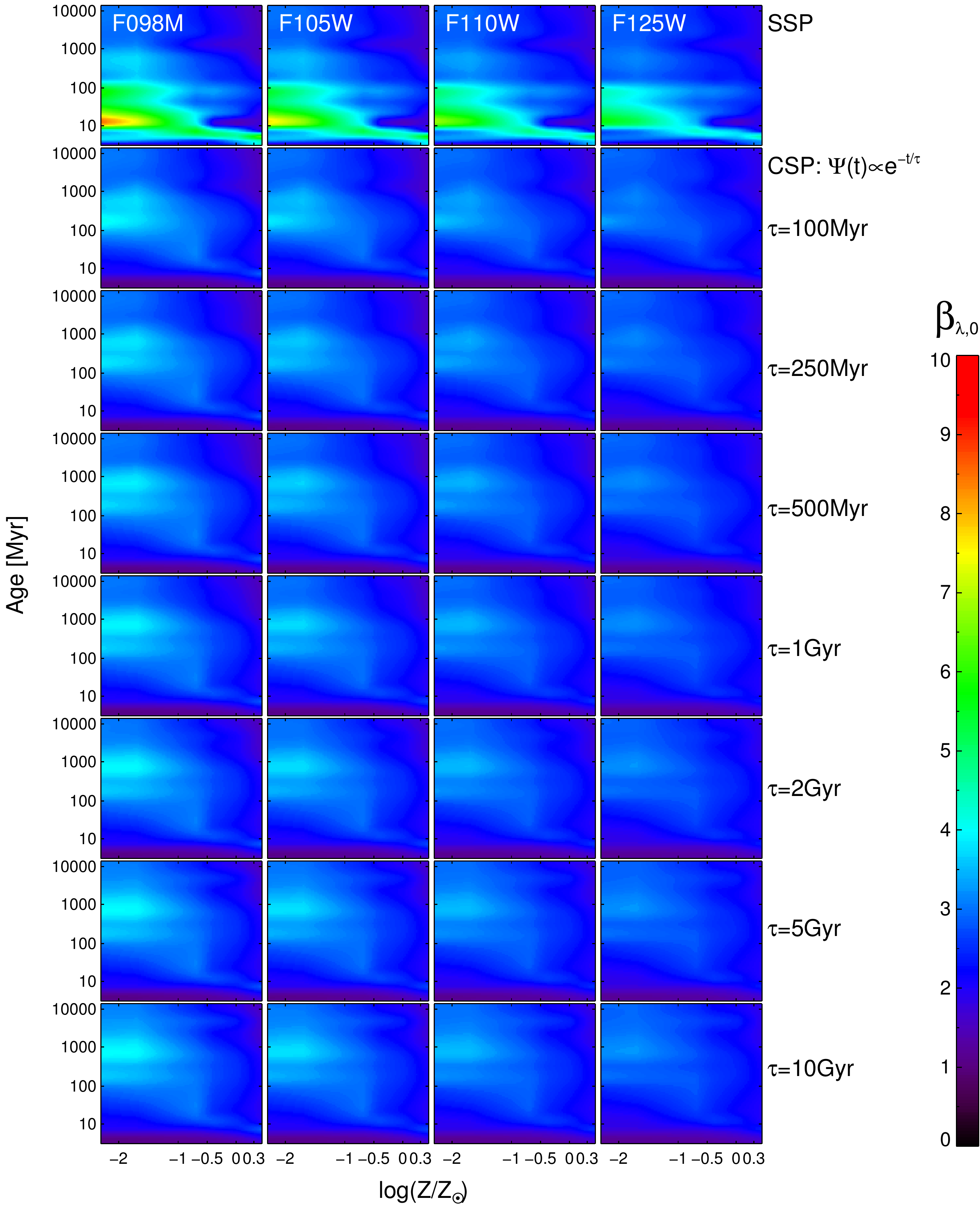}
}
\caption{\noindent\small
Same as Figure~\ref{fig:figure5} for the \HST/WFC3\,IR F098M, F105W, F110W, and
F125W filters.}
\end{figure*}
%%%%%%%%%%%%%%%%%%%%%%%%% Figure Set 5 %%%%%%%%%%%%%%%%%%%%%%%%%%%%%%%%

%% file: figset10.tex
%%%%%%%%%%%%%%%%%%%%%%%%% Figure Set 10 %%%%%%%%%%%%%%%%%%%%%%%%%%%%%%%%
\begin{figure*}
\centerline{
\parbox[b][0.440\txh][t]{0.03\txw}{\textbf{(a)}} \
	\includegraphics[height=0.45\txh]{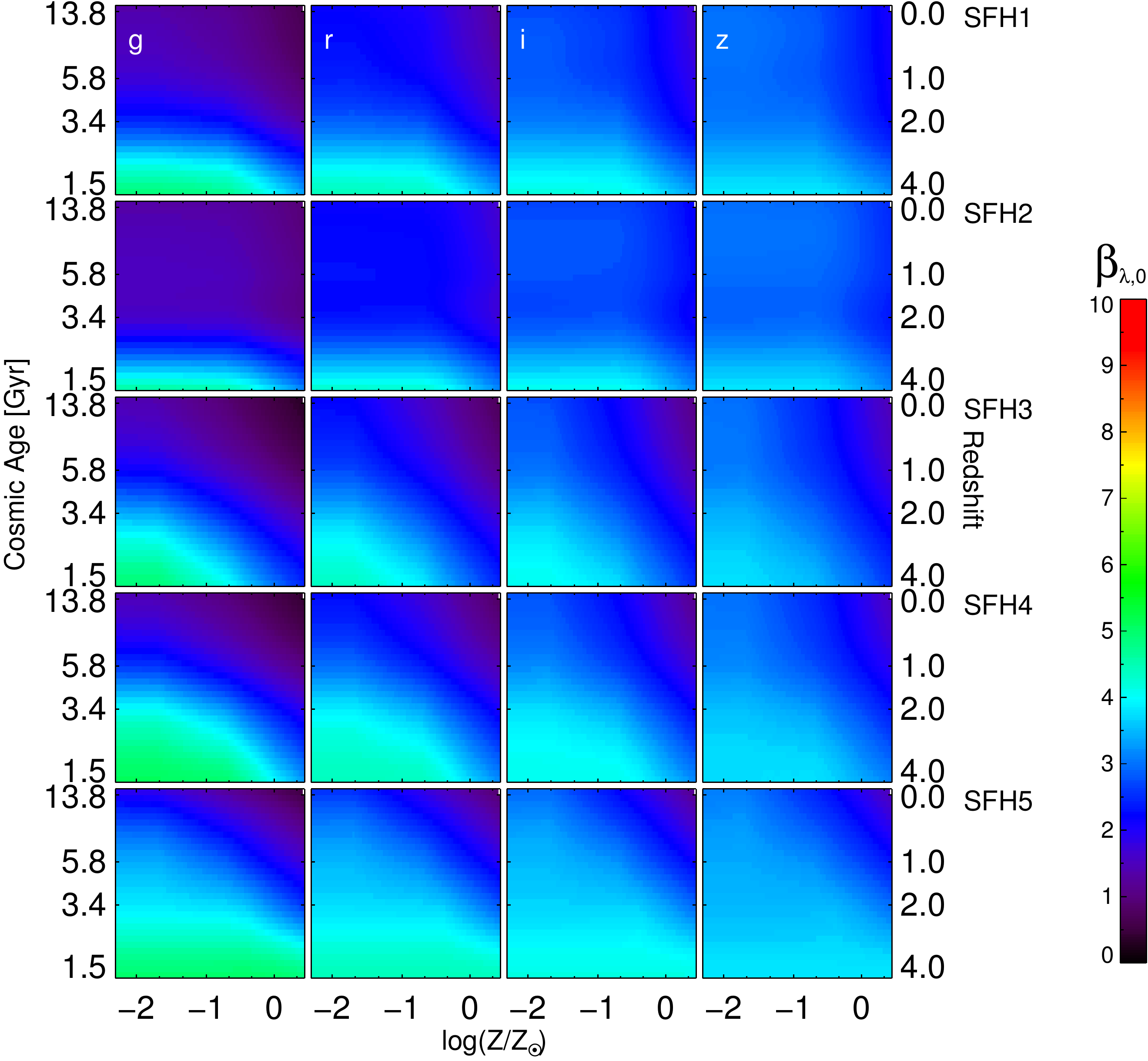}
}
\centerline{
\parbox[b][0.440\txh][t]{0.03\txw}{\textbf{(b)}} \
	\includegraphics[height=0.45\txh]{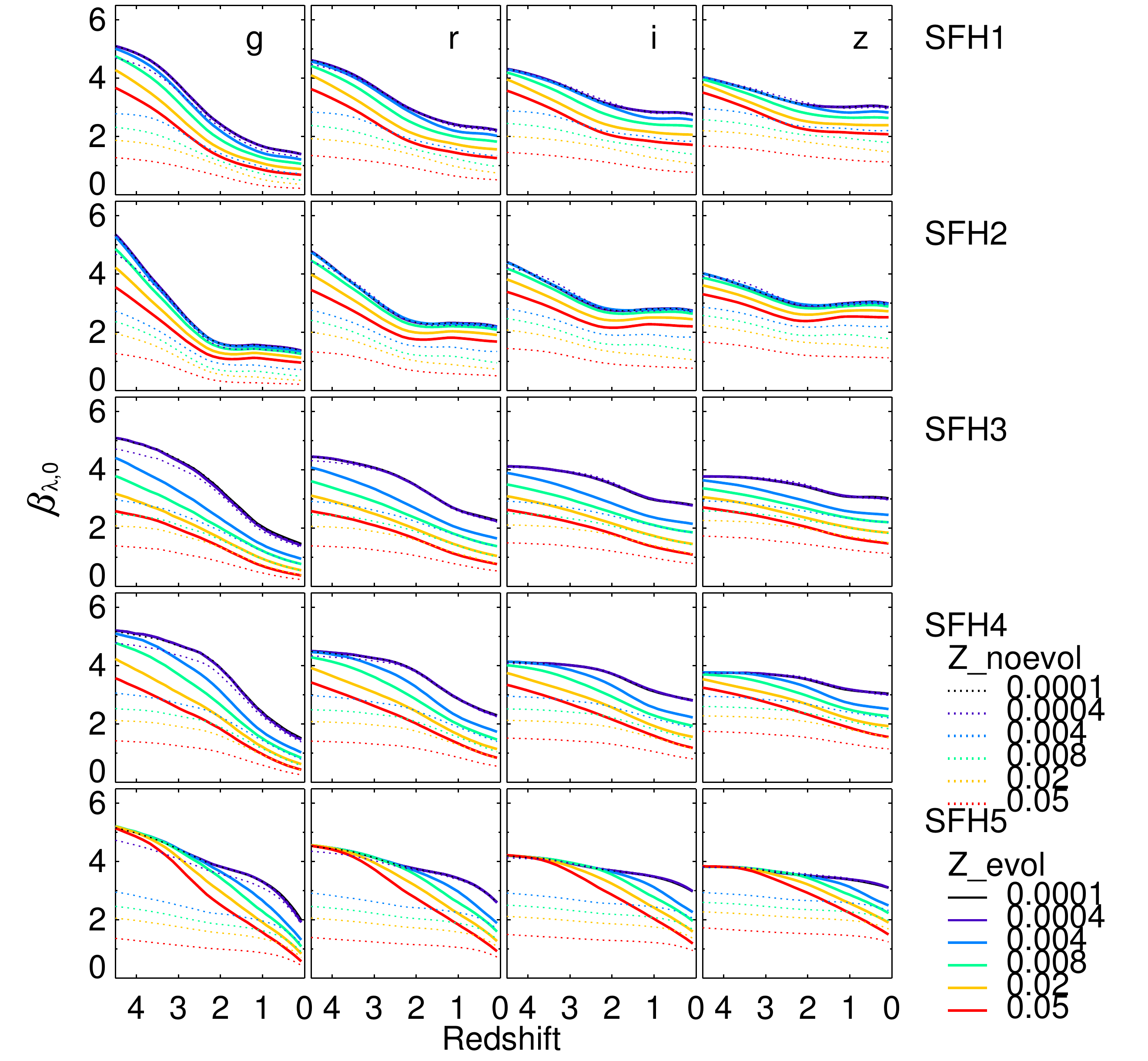}
}
\caption{\noindent\small
Same as Figure~\ref{fig:figure10} for the SDSS $g$, $r$, $i$, and
$z$ filters.}
\end{figure*}

\begin{figure*}
\centerline{
\parbox[b][0.440\txh][t]{0.03\txw}{\textbf{(a)}} \
	\includegraphics[height=0.45\txh]{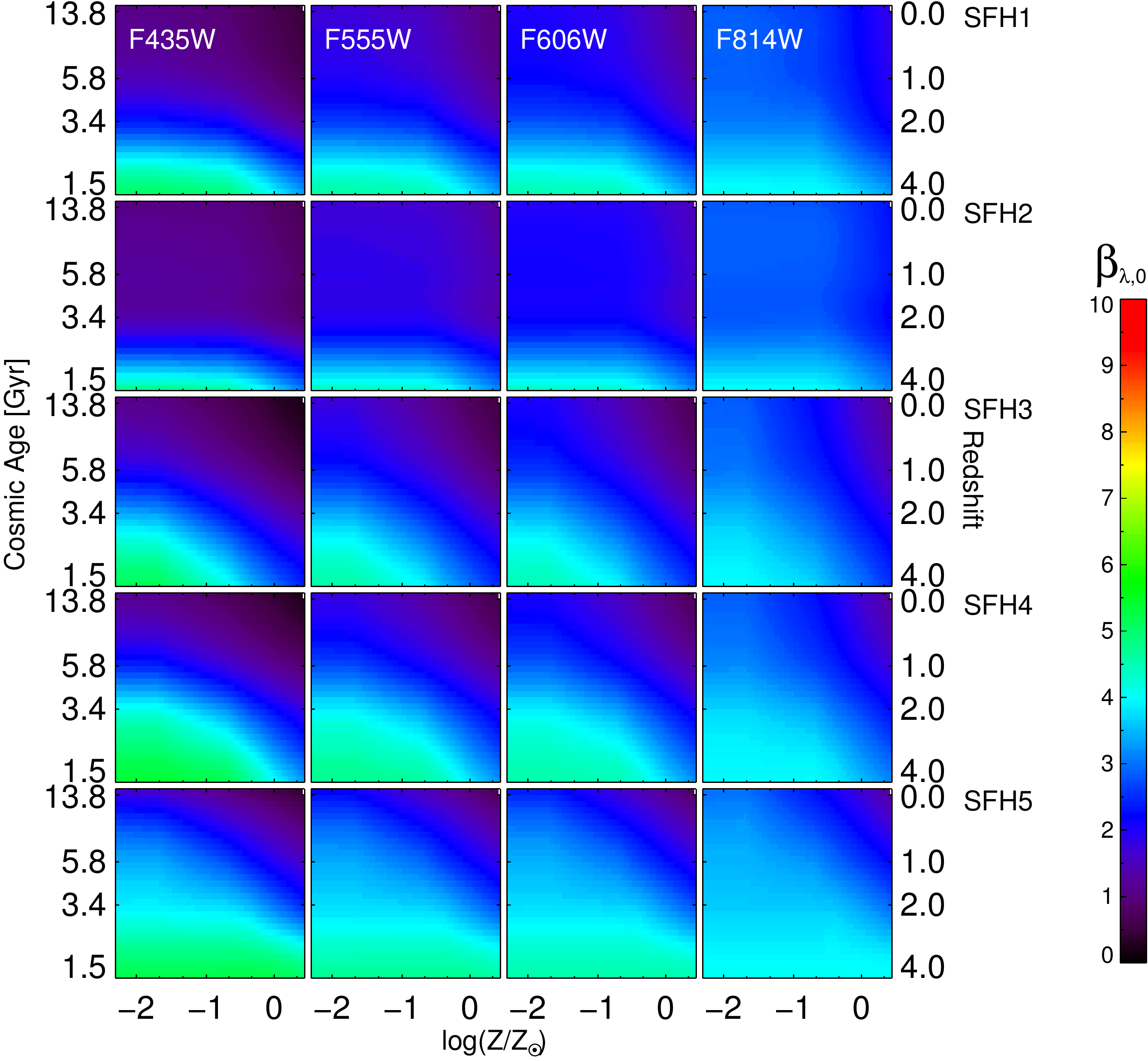}
}
\centerline{
\parbox[b][0.440\txh][t]{0.03\txw}{\textbf{(b)}} \
	\includegraphics[height=0.45\txh]{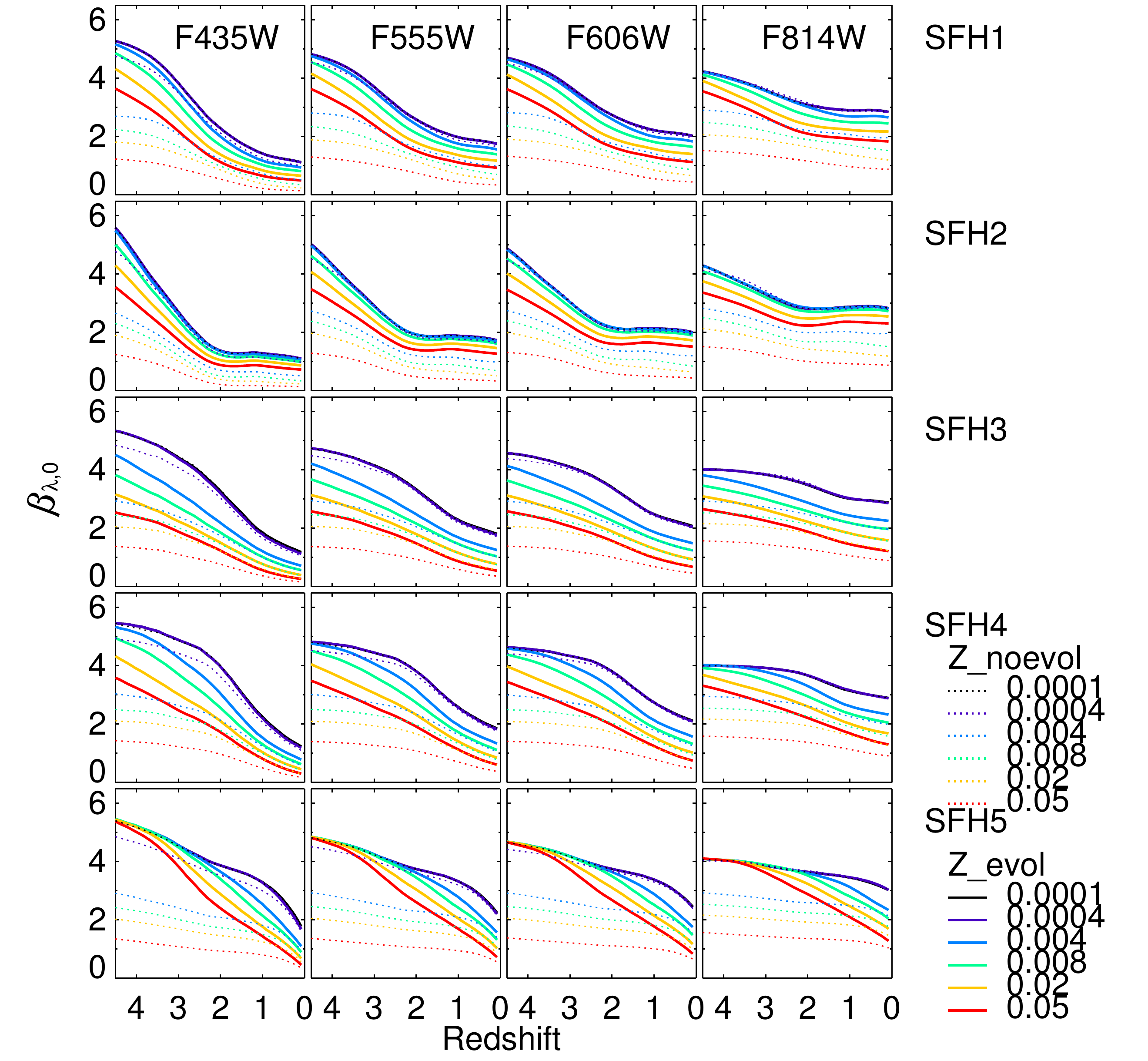}
}
\caption{\noindent\small
Same as Figure~\ref{fig:figure10} for the \HST/ACS\,WFC F435W, F555W, F606W, and
F814W filters.}
\end{figure*}

\begin{figure*}
\centerline{
\parbox[b][0.440\txh][t]{0.03\txw}{\textbf{(a)}} \
	\includegraphics[height=0.45\txh]{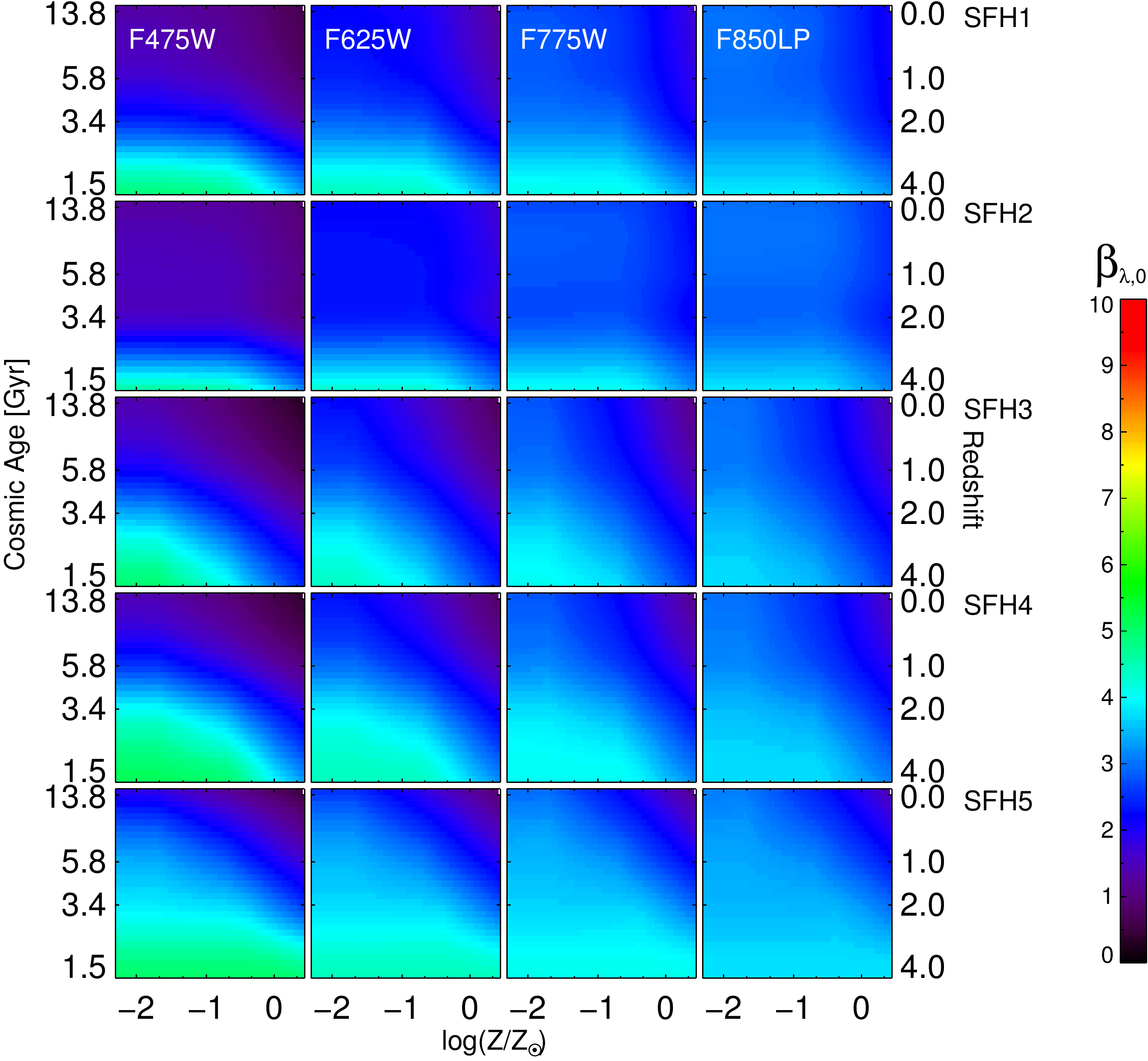}
}
\centerline{
\parbox[b][0.440\txh][t]{0.03\txw}{\textbf{(b)}} \
	\includegraphics[height=0.45\txh]{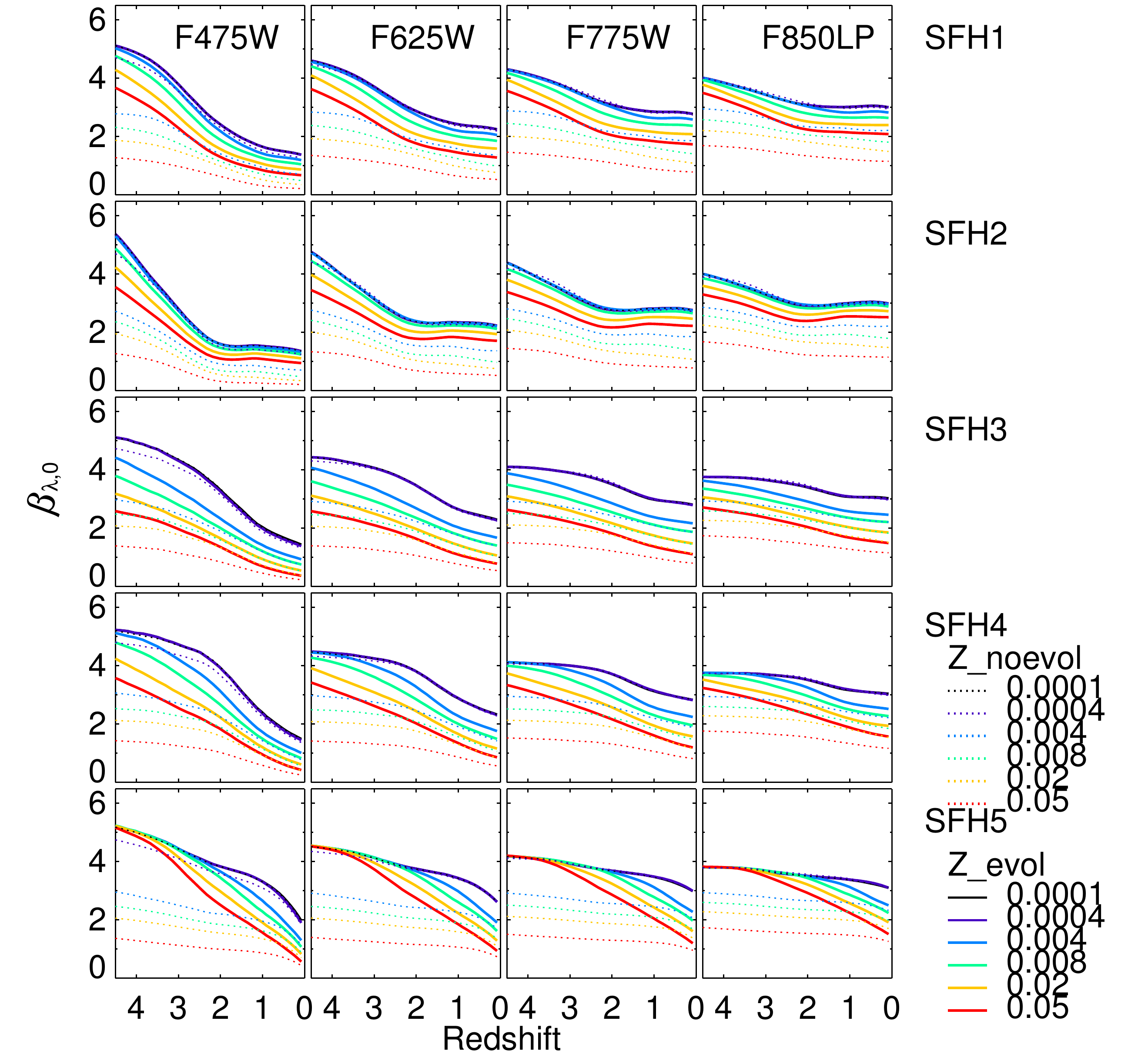}
}
\caption{\noindent\small
Same as Figure~\ref{fig:figure10} for the \HST/ACS\,WFC F475W, F625W, F775W, and
F850LP filters.}
\end{figure*}

\begin{figure*}
\centerline{
\parbox[b][0.440\txh][t]{0.03\txw}{\textbf{(a)}} \
	\includegraphics[height=0.45\txh]{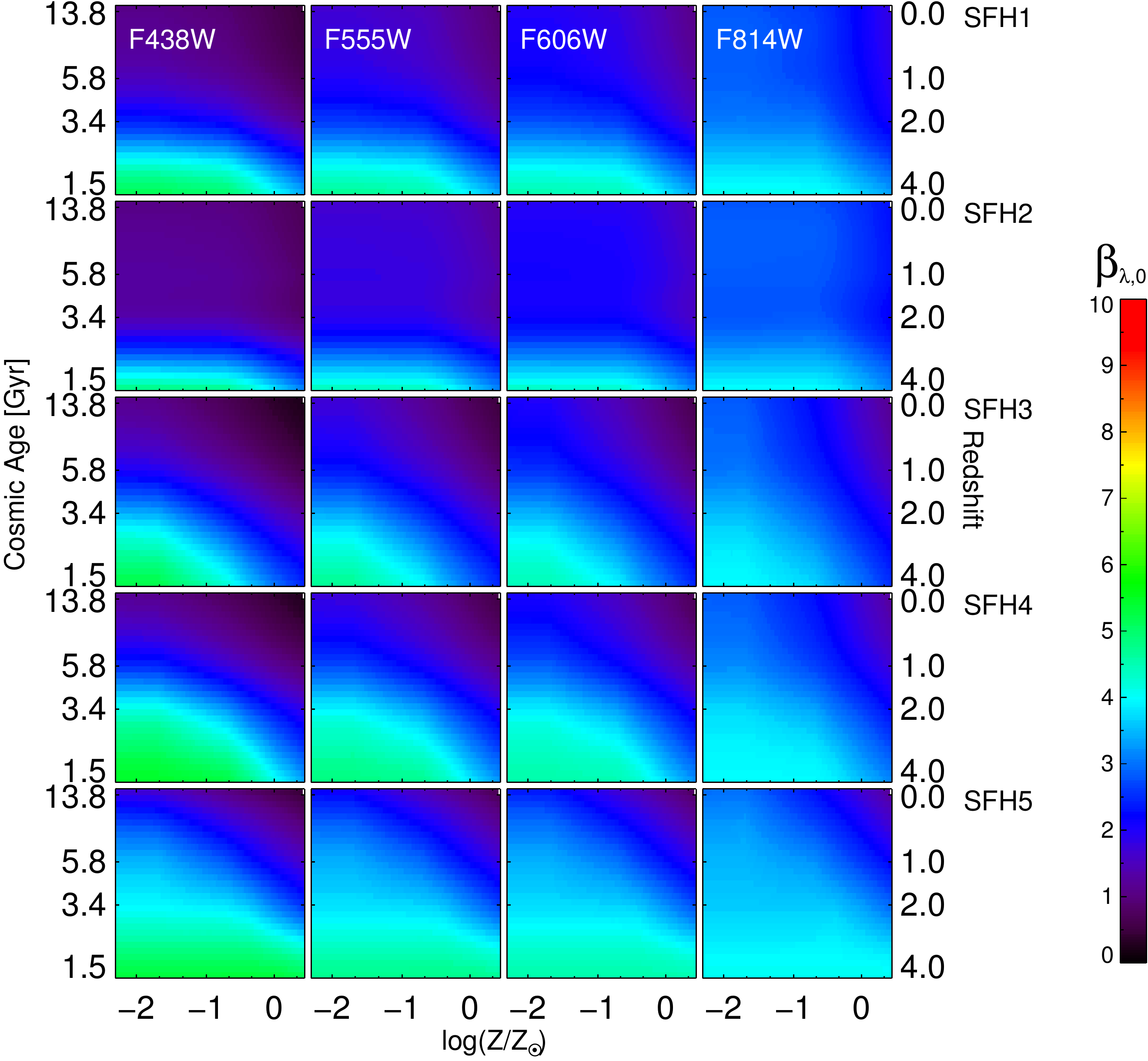}
}
\centerline{
\parbox[b][0.440\txh][t]{0.03\txw}{\textbf{(b)}} \
	\includegraphics[height=0.45\txh]{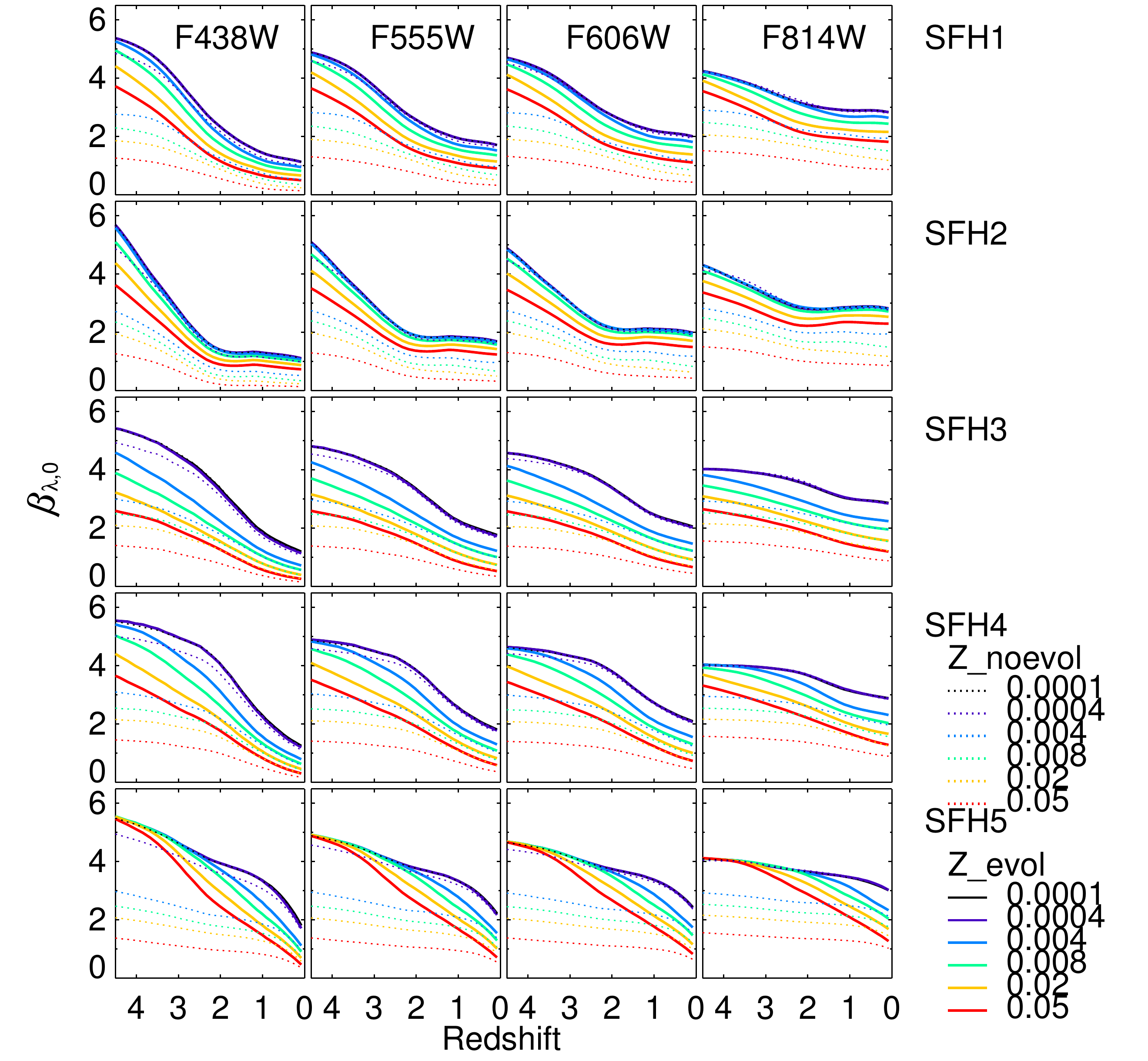}
}
\caption{\noindent\small
Same as Figure~\ref{fig:figure10} for the \HST/WFC3\,UVIS F438W, F555W, F606W, and
F814W filters.}
\end{figure*}

\begin{figure*}
\centerline{
\parbox[b][0.440\txh][t]{0.03\txw}{\textbf{(a)}} \
	\includegraphics[height=0.45\txh]{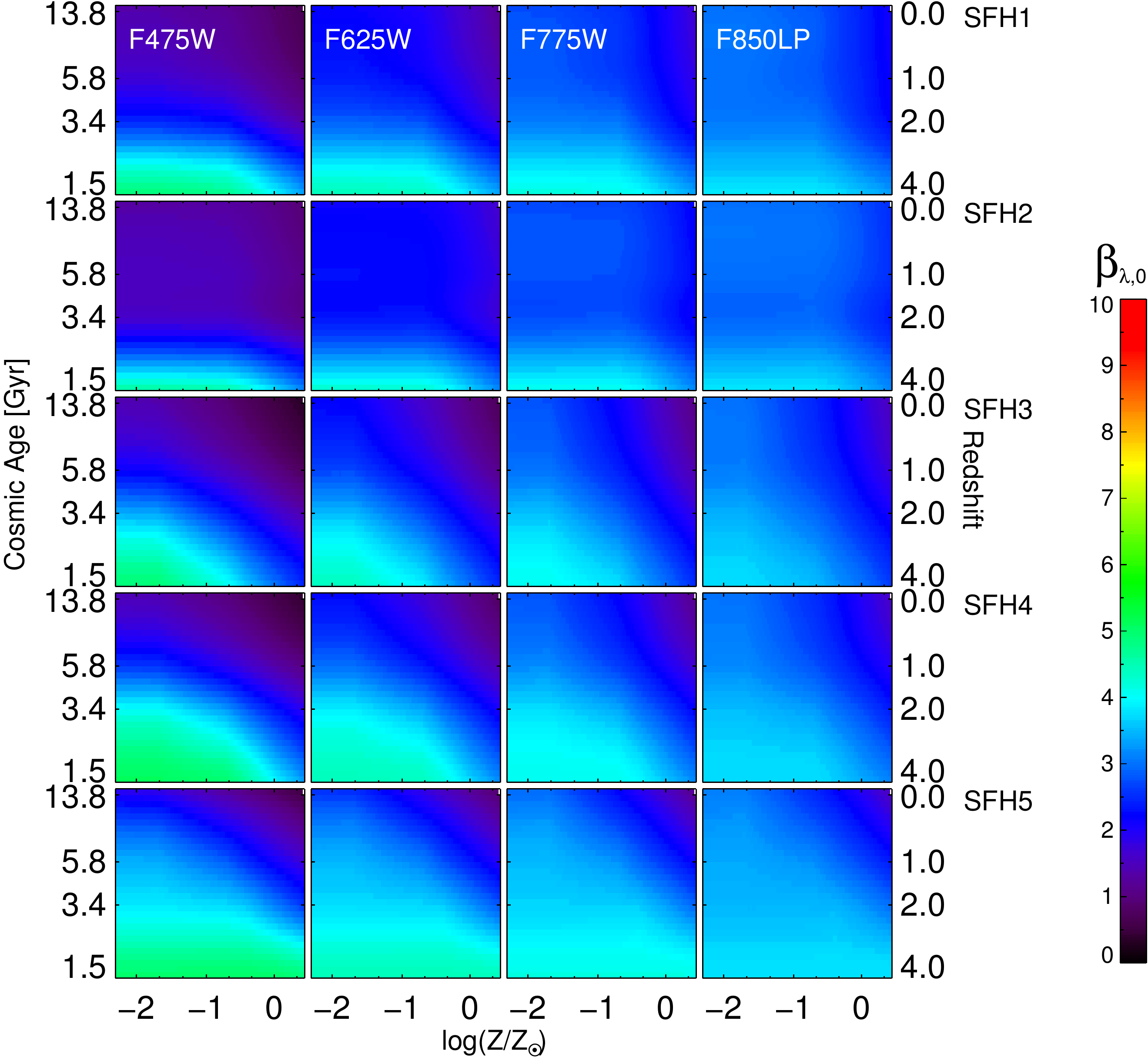}
}
\centerline{
\parbox[b][0.440\txh][t]{0.03\txw}{\textbf{(b)}} \
	\includegraphics[height=0.45\txh]{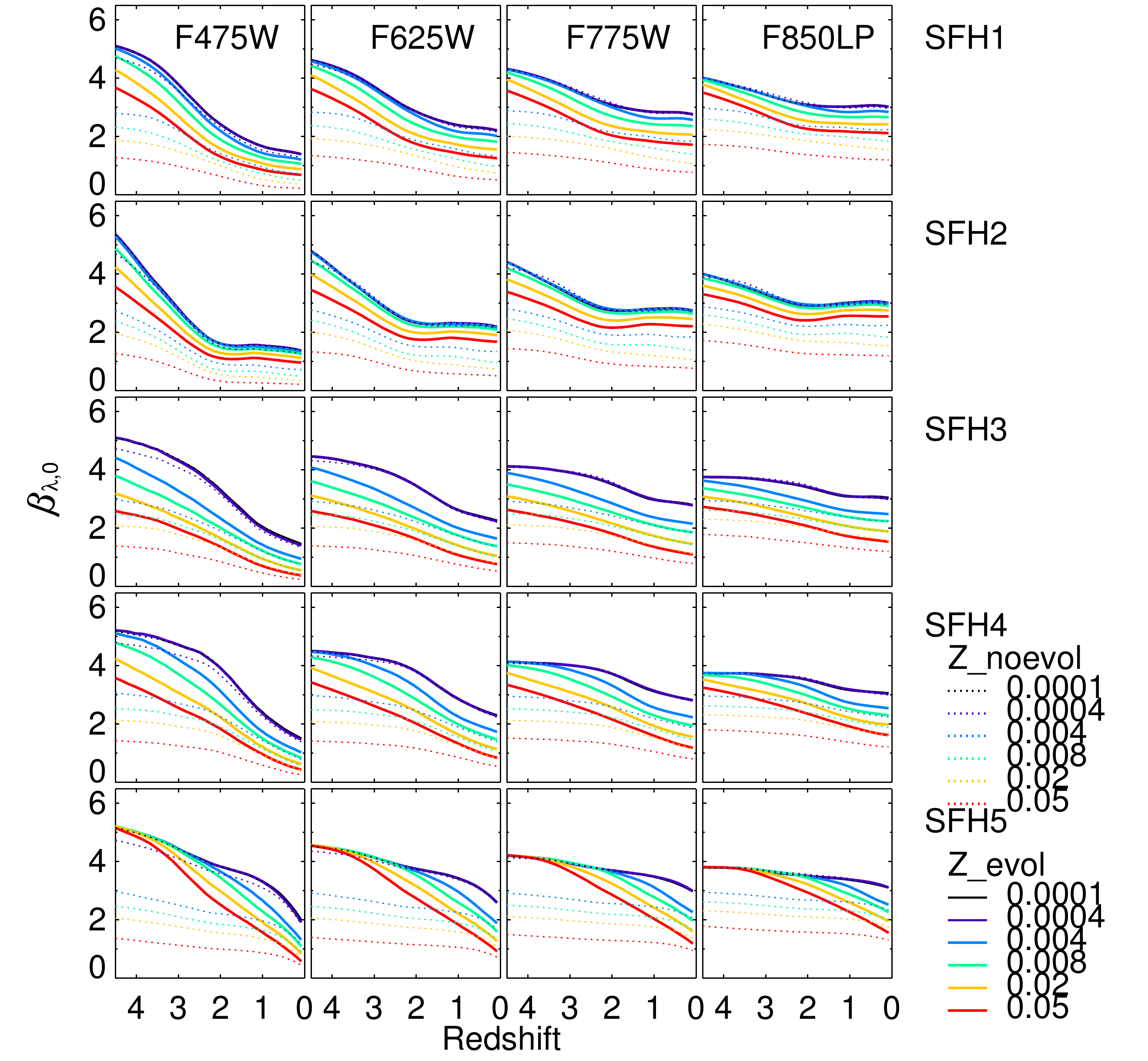}
}
\caption{\noindent\small
Same as Figure~\ref{fig:figure10} for the \HST/WFC3\,UVIS F475W, F625W, F775W, and
F850LP filters.}
\end{figure*}

\begin{figure*}
\centerline{
\parbox[b][0.440\txh][t]{0.03\txw}{\textbf{(a)}} \
	\includegraphics[height=0.45\txh]{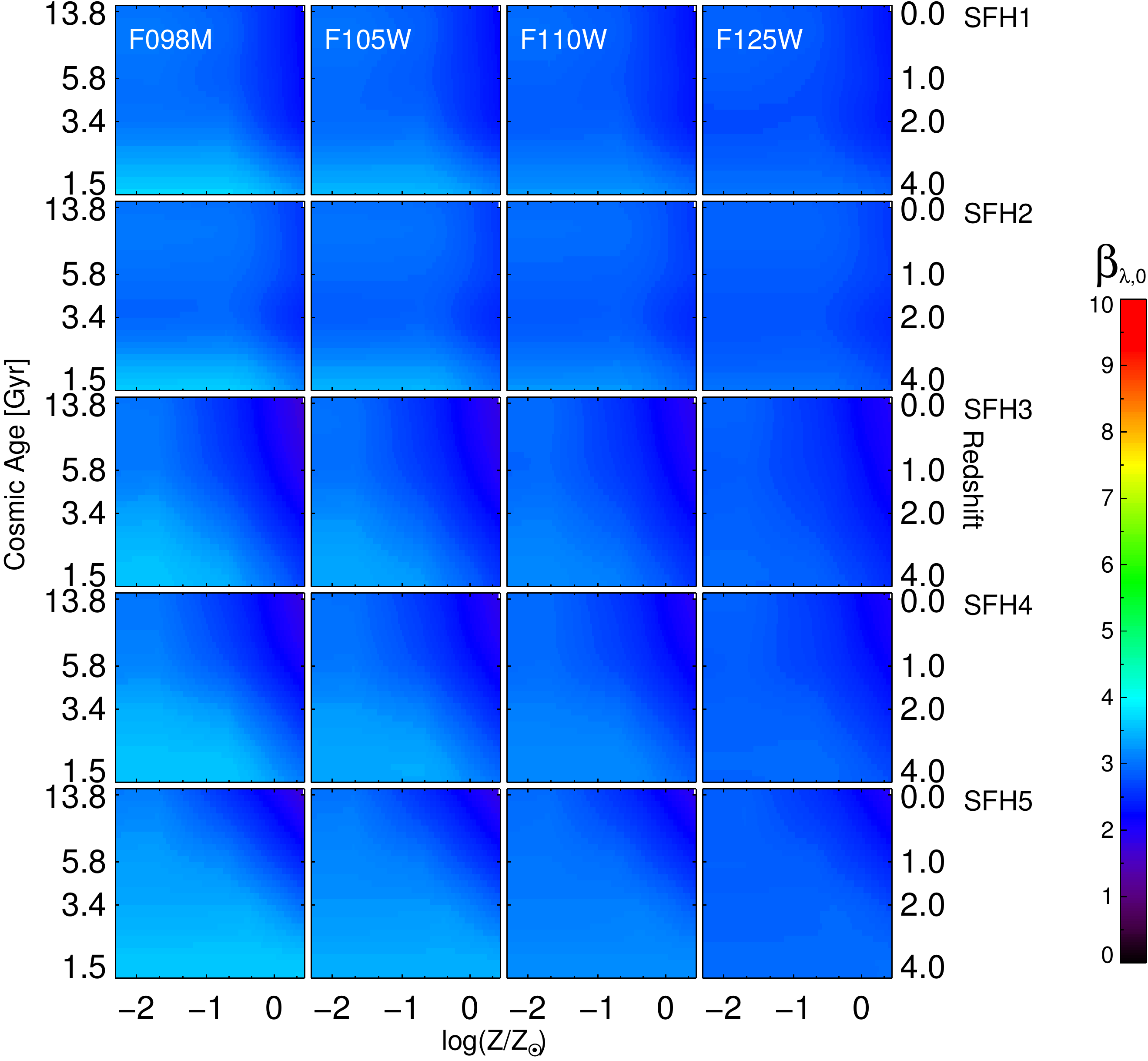}
}
\centerline{
\parbox[b][0.440\txh][t]{0.03\txw}{\textbf{(b)}} \
	\includegraphics[height=0.45\txh]{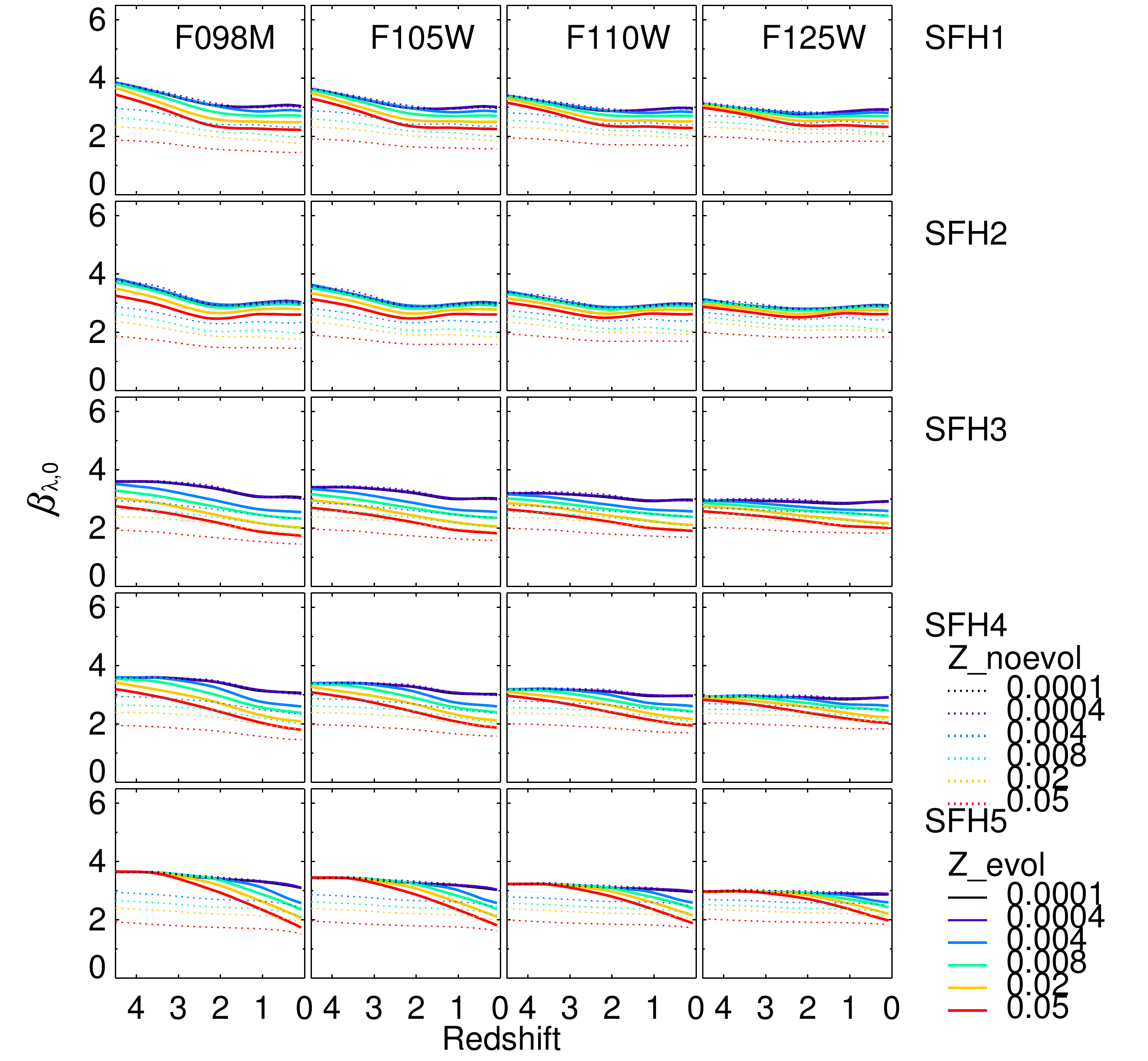}
}
\caption{\noindent\small
Same as Figure~\ref{fig:figure10} for the \HST/WFC3\,IR F098M, F105W, F110W, and
F125W filters.}
\end{figure*}
%%%%%%%%%%%%%%%%%%%%%%%%% Figure Set 10 %%%%%%%%%%%%%%%%%%%%%%%%%%%%%%%%